\documentclass[10pt]{article}
\usepackage{graphicx}
\title{Integrable nonlocal nonlinear equations$^*$}
\author{Mark J. Ablowitz$^1$ and Ziad H. Musslimani$^2$\\\\
$^1$Department of Applied Mathematics, University of Colorado\\
Campus Box 526,  Boulder, Colorado 80309-0526\\
$^2$Department of Mathematics, Florida State University, Tallahassee, FL 32306-4510}
\usepackage{color} 
\usepackage{footmisc} 
\usepackage{amsmath,epsf,citesort,times}
\usepackage{amsmath,amssymb,epsf,graphicx}
\usepackage{IEEEtrantools}
\usepackage{soul,xcolor}
\usepackage{soul}
\usepackage{mathrsfs}

\newcommand{\ba}{\begin{array}}
\newcommand{\ea}{\end{array}}

\newcommand{\ben}{\begin{enumerate}}
\newcommand{\een}{\end{enumerate}}

\date{\today}
\begin{document}
\setstcolor{red}
 \maketitle
\input{epsf.tex}
%%%%%%%%%%%%%%%%%%%%%%%%%%%%%%%%%%%%%%%%%%%%
%%%%%%%%%%%%%%%%%%%%%%%%%%%%%%%%%%%%%%%%%%%%
\begin{abstract} 
A nonlocal nonlinear Schr\"odinger  (NLS) equation was recently found by the authors 
and shown to be an integrable infinite dimensional Hamiltonian equation. Unlike the classical (local) case, here the nonlinearly induced ``potential" is $PT$ symmetric thus
the nonlocal NLS equation is also $PT$ symmetric. 
In this paper, new {\it reverse space-time} and {\it reverse time} nonlocal nonlinear integrable equations are introduced. They arise from remarkably simple symmetry reductions of general AKNS scattering problems where the nonlocality appears in both space and time or time alone. They are integrable infinite dimensional Hamiltonian dynamical systems. These include the reverse space-time, and in some cases reverse time, nonlocal nonlinear Schr\"odinger, modified 
Korteweg-deVries (mKdV), sine-Gordon, $(1+1)$ and $(2+1)$ dimensional three-wave interaction, derivative NLS, ``loop soliton", Davey-Stewartson (DS), partially $PT$ symmetric  DS and partially reverse space-time DS equations. Linear Lax pairs,  an infinite number of conservation laws, inverse scattering transforms are discussed and one soliton solutions are found. Integrable reverse space-time and reverse time nonlocal discrete 
nonlinear Schr\"odinger type equations are also introduced along with few conserved quantities. Finally, nonlocal Painlev\'e type equations are derived from the reverse space-time and reverse 
time nonlocal NLS equations.
 \end{abstract}
 \footnotetext{$^*$This paper will be published in Studies in Applied Mathematics in a special volume dedicated to Professor David J. Benny,}
%%%%%%%%%%%%%%%%%%%%%%%%%%%%%%%%%%%%%%%%%%%%%
%%%%%%%%%%%%%%%%%%%%%%%%%%%%%%%%%%%%%%%%%%%%%
%%%%%%%%%%%%%%%%%%%%%%%%%%%%%%%%%%%%%%%%%%%%%
\section{Introduction}
%%%%%%%%%%%%%%%%%%%%%%%%%%%%%%%%%%%%%%%%%%%%%
Since their fundamental discovery in 1965 by Zabusky and Kruskal, solitons have emerged as one of the most basic concepts in nonlinear sciences. Physically speaking, solitons (or previously termed solitary waves) represent robust nonlinear coherent structures that often form as a result of a delicate balance between effects of dispersion and/or diffraction and wave steepening. They have been theoretically predicted and observed in laboratory experiments in many branches of the physical, biological and chemical sciences. Examples include water waves, temporal and spatial optics, Bose-Einstein condensation, magnetics, plasma physics to name a few -- see \cite{Chen_et.al,Lederer_et.al,kivshar_et.al,kev_et.al,yangbook} and references therein for reviews discussing soliton applications. \\\\
%%%%%%%%%%%%%%%%%%%%%%%%%%%%%%%%%%%%%%%%%%%%%
From a mathematical point of view, solitons naturally arise as a special class of solutions to so-called integrable evolution equations. Such integrable systems exhibit unique mathematical structure by admitting an infinite number of constants of motion corresponding to an  infinite number of conservation laws. Furthermore, by applying the inverse scattering transform (IST;  cf. \cite{Ablowitz2,NOVIKOV,Cal-Deg}), for decaying data, one can linearize  the system and obtain significant information about  the structure of their solutions. In many situations, one can even express these solutions in closed form.\\\\
%%%%%%%%%%%%%%%%%%%%%%%%%%%%%%%%%%%%%%%%%%%%%
Historically speaking, the first integrable nonlinear evolution equation solved by the method of IST was the Korteweg-deVries (KdV) equation \cite{GGKM}. Remarkably, it was shown that solitons corresponded to eigenvalues of the time independent linear
Schr\"odinger equation. Soon thereafter, the concept of Lax pair \cite{Lax} was introduced and the KdV equation, and others, were expressed as a compatibility condition of two linear equations. A few years later, Zakharov and Shabat \cite{ZS} used the idea of Lax pair to integrate the nonlinear Schr\"odinger equation 
\begin{equation}
\label{NLS}
iq_t(x,t) = q_{xx}(x,t) - 2\sigma q^2(x,t)q^*(x,t),  ~~\sigma= \mp 1\;,
  \end{equation}
for decaying data, where $*$ is the complex conjugate, and obtain soliton solutions. Subsequently, a method to generate a class of integrable nonlinear integrable evolution equations solvable by IST was developed \cite{AKNS}. Soon thereafter interest in the theory of integrability has grown significantly  and many integrable nonlinear partial differential equations have been identified in both one and two space dimensions as well as in discrete settings. Some notable  equations include the modified KdV, sine-Gordon, sinh-Gordon, coupled NLS, Boussinesq, Kadomtsev-Petviashvili, Davey-Stewartson, 
Benjamin-Ono (BO), Intermediate Long Wave (ILW), integrable discrete NLS equations, the Toda and discrete KdV  lattices, to name a few cf. \cite{Ablowitz1}.\\\\
%%%%%%%%%%%%%%%%%%%%%%%%%%%%%%%%%%%%%%%%%%%%%
In 2013, a new nonlocal reduction of the AKNS scattering problem was 
found \cite{AblowitzMusslimani} which gave rise to an integrable nonlocal NLS equation
\begin{equation}
\label{PTNLS-1}
iq_t(x,t) = q_{xx}(x,t) - 2\sigma q^2(x,t)q^*(-x,t),\;\;\; \sigma = \mp 1 \;.
\end{equation}
Remarkably, Eq.~(\ref{PTNLS-1}) has a self-induced nonlinear ``potential", thus, it is a $PT$ symmetric equation \cite{Bender}. In other words, one can view (\ref{PTNLS-1}) as a linear Schr\"odinger equation
\begin{equation}
\label{PTNLS-1V}
iq_t(x,t) =q_{xx}(x,t)+V[q,x,t]q(x,t)\;,
\end{equation}
with a self induced potential $V[q,x,t]  \equiv - 2\sigma q(x,t)q^*(-x,t)$ satisfying the $PT$ symmetry condition $V[q,x,t]=V^*[q,-x,t].$ We point out that $PT$ symmetric systems, which allow for lossless-like propagation due to their balance of gain and 
loss \cite{RKDM2,RKDM3}, have attracted considerable attention in recent 
years -- see \cite{yang_review} and references therein for an extensive review on linear and nonlinear waves in $PT$ symmetric systems. Equation (\ref{PTNLS-1}) was derived in  \cite{AblowitzMusslimani} with physical intuition. Recently, the nonlocal nonlinear 
Schr\"odinger equation (\ref{PTNLS-1}) was derived in a physical application of magnetics \cite{ApplicNlclNLS16}. In \cite{AblowitzMusslimani2} an integrable discrete $PT$ symmetric 
``discretization" of equation (\ref{PTNLS-1}) was obtained from a new nonlocal $PT$ symmetric reduction of the Ablowitz-Ladik scattering problem \cite{Abl-ladik}. In \cite{AblowitzMusslimaniNonlinearity} the detailed IST associated with the nonlocal NLS system (\ref{PTNLS-1}) was carried out and integrable nonlocal versions of the modified KdV and sine-Gordon equations were introduced. An extension to a $(2+1)$ dimensional integrable nonlocal NLS type equations was subsequently analyzed in \cite{Fokas2016}. \\\\
%%%%%%%%%%%%%%%%%%%%%%%%%%%%%%%%%%%%%%%%%%%%%%%%
These findings have triggered renewed interest in integrable systems.
New types of soliton solutions have been also recently reported \cite{He,He2}. 
Moreover, recently, it was proposed that the integrable nonlocal (in space) NLS equation is gauge equivalent to an unconventional system of coupled 
Landau-Lifshitz equations \cite{ApplicNlclNLS16}. Possible application of the nonlocal nonlinear Schr\"odinger and mKdV equations have been discussed in \cite{Lou1,Lou2} in the context of ``Alice-Bob systems".
%\\\\
%%%%%%%%%%%%%%%%%%%%%%%%%%%%%%%%%%%%%%%%%%%%%
%%%%%%%%%%%%%%%%%%%%%%%%%%%%%%%%%%%%%%%%%%%%%
In this paper we identify new nonlocal symmetry reductions for the general AKNS scattering problem of the {\it reverse space-time} and {\it reverse time type}. 
Unlike the integrable $PT$ symmetric equation (\ref{PTNLS-1}) \cite{AblowitzMusslimani}, 
here the symmetry reductions are nonlocal both in space and time or time alone and lead to new integrable reverse space-time nonlocal evolution equations of the nonlinear Schr\"odinger, modified KdV, sine-Gordon,  (1+1) and (2+1) dimensional multi-wave interaction (including the three-wave), derivative NLS, ``loop soliton", Davey-Stewartson (DS), partially $PT$ symmetric  DS and partially reverse space-time DS equations. Furthermore, discrete-time nonlocal NLS type equations are also derived. Finally, nonlocal Painlev\'e type equations are derived from the reverse space-time and reverse time nonlocal NLS equations. \\
Next, we list some of these equations (here $\sigma=\mp 1; \gamma^2=\pm 1, \alpha, \beta,c_j, {\bf C}_j$ are constant):\newpage
%%%%%%%%%%%%%%%%%%%%%%
%%%%%%%%%%%%%%%%%%%%%%
\begin{flalign}
\label{complex-space-time-nonloc-NLS}
&iq_t(x,t) = q_{xx}(x,t) - 2\sigma q^2(x,t)q(-x,-t), &\\
&\nonumber \text{\small{Reverse space-time nonlocal NLS}}&\\ 
 \nonumber \\ 
%%%%%%%%%%%%%%%%%%%%%%
%%%%%%%%%%%%%%%%%%%%%%
&\label{Vcomplex-space-time-nonloc-NLS}
i{\bf q}_t(x,t) = {\bf q}_{xx}(x,t) - 2\sigma  [{\bf q}(x,t)\cdot {\bf q}(-x,-t)]  {\bf q}(x,t), &  \\
&\nonumber \text{\small{Reverse space-time vector nonlocal NLS}} & \\  \nonumber \\ 
%%%%%%%%%%%%%%%%%%%%%%
%%%%%%%%%%%%%%%%%%%%%%
&\label{complex-time-nonloc-NLS}
iq_t(x,t)=q_{xx}(x,t) - 2\sigma q^2(x,t)q(x,-t), &   \\
&\nonumber \text{\small{Reverse time nonlocal NLS}} & 
\\ \nonumber \\
%%%%%%%%%%%%%%%%%%%%%%
%\end{flalign}
%%%%%%%%%%%%%%%%%%%%%%%%%%%%%%%%%%%%%%%%%%
%\begin{flalign}
%%%%%%%%%%%%%%%%%%%%%%
&\nonumber
q_t(x,t) = iq_{xx}(x,t) + \alpha \sigma (q^2(x,t)q(-x,-t))_x 
+ i\beta \sigma q^2(x,t)q(-x,-t), &\\
&\label{complex-space-time-nonloc-dNLS}
\text{\small{Reverse space-time coupled nonlocal NLS -- derivative NLS $(\alpha, \beta \in \mathbb{R})$}}& 
\\ \nonumber \\
%%%%%%%%%%%%%%%%%%%%%%
%%%%%%%%%%%%%%%%%%%%%%
& \label{real-space-time-nonloc-loop}
\frac{\partial q(x,t)}{\partial t} + \frac{\partial^2}{\partial x^2} \left(  \frac{q_{x}(x,t)}{[1 - \sigma q(x,t)q(-x,-t) ]^{3/2}}   \right) =0, &\\
&\nonumber \text{\small{Real reverse space-time nonlocal nonlinear ``loop soliton"}} &
\\  \nonumber \\ 
%%%%%%%%%%%%%%%%%%%%%
%%%%%%%%%%%%%%%%%%%%%%
&\label{complex-space-time-nonloc-mkdv}
q_t(x,t)+q_{xxx}(x,t)  - 6 \sigma q(x,t)q^*(-x,-t)q_x(x,t)=0,  q \in \mathbb{C}, &\\ 
&\nonumber\text{\small{Complex reverse space-time nonlocal mKdV}} & \\  \nonumber \\ 
%%%%%%%%%%%%%%%%%%%%%
%%%%%%%%%%%%%%%%%%%%%%
&\label{real-space-time-nonloc-mkdv}
q_t(x,t)+q_{xxx}(x,t)  - 6\sigma q(x,t)q(-x,-t)q_x(x,t)=0,  q \in \mathbb{R}, &\\
&\nonumber \text{\small{Real reverse space-time nonlocal mKdV}} & \\  \nonumber \\ 
%%%%%%%%%%%%%%%%%%%%%
%%%%%%%%%%%%%%%%%%%%%%
&  \nonumber q_{xt}(x,t)+2 s (x,t) q(x,t) =0, \;\;\; q \in \mathbb{R}, & \\
&  \nonumber  s_x(x,t) = (q(x,t)q(-x,-t))_t,   &\\
&\label{real-space-time-nonloc-sG}
\text{\small{Real reverse space-time nonlocal sine-Gordon}} & \\  \nonumber \\ 
%%%%%%%%%%%%%%%%%%%%%
%%%%%%%%%%%%%%%%%%%%%
&\nonumber
iq_t({\bf x},t)+\frac{1}{2}\left[\gamma^2 q_{xx}({\bf x},t)  + q_{yy}({\bf x},t) \right]
+ \sigma  q^2({\bf x},t) q(-{\bf x},-t) 
 = \phi ({\bf x},t) q({\bf x},t), &\\ 
%%%%%%%
&\nonumber\phi_{xx} ({\bf x},t) -\gamma^2 \phi_{yy} ({\bf x},t) 
= 2\sigma \left[ q({\bf x},t) q(-{\bf x},-t)\right]_{xx}, &\\
&\label{complex-space-time-nonloc-DS}
\text{\small{ Reverse space-time nonlocal Davey-Stewartson}}& \\  \nonumber \\ 
%%%%%%%%%%%%%%%%%%%%%
%%%%%%%%%%%%%%%%%%%%%
&\nonumber
iq_t({\bf x},t)+\frac{1}{2}\left[\gamma^2 q_{xx}({\bf x},t)  + q_{yy}({\bf x},t) \right]
+ \sigma  q^2({\bf x},t) q({\bf x},-t) = \phi ({\bf x},t) q({\bf x},t) ,   &\\
%%%%%
&\nonumber\phi_{xx} ({\bf x},t) -\gamma^2 \phi_{yy} ({\bf x},t) 
= 2\sigma \left[ q({\bf x},t) q({\bf x},-t)\right]_{xx}, &\\
&\label{complex-time-nonloc-DS}
\text{\small{ Reverse time nonlocal Davey-Stewartson}} &
%%%%%%%%%%%%%%%%%%%%%
\end{flalign}
\begin{flalign}
%%%%%%%%%%%%%%%%%%%%%
%%%%%%%%%%%%%%%%%%%%%%%%%%%%%%%%%%%%%%%%%%
%%%%%%%%%%%%%%%%%%%%%%%%%%%%%%%%%%%%%%%%%
%%%%%%%%%%%%%%%%%%%%%
%%%%%%%%%%%%%%%%%%%%%
&\nonumber
iq_t({\bf x},t)+\frac{1}{2}\left[\gamma^2 q_{xx}({\bf x},t)  + q_{yy}({\bf x},t) \right]
+ \sigma  q^2({\bf x},t) q^*(-x,y,t) 
 = \phi ({\bf x},t) q({\bf x},t), &\\ 
%%%%%%%
&\nonumber\phi_{xx} ({\bf x},t) -\gamma^2 \phi_{yy} ({\bf x},t) 
= 2\sigma \left[ q({\bf x},t) q^*(-x,y,t)\right]_{xx}, &\\
&\label{complex-space-time-nonloc-DS-PT-part}
\text{\small{Partially $PT$ symmetric nonlocal Davey-Stewartson}} & \\  \nonumber \\ 
%%%%%%%%%%%%%%%%%%%%%
%%%%%%%%%%%%%%%%%%%%%
&\nonumber
iq_t({\bf x},t)+\frac{1}{2}\left[\gamma^2 q_{xx}({\bf x},t)  + q_{yy}({\bf x},t) \right]
+ \sigma  q^2({\bf x},t) q(-x,y,-t) = \phi ({\bf x},t) q({\bf x},t) ,   &\\
%%%%%
&\nonumber\phi_{xx} ({\bf x},t) -\gamma^2 \phi_{yy} ({\bf x},t) 
= 2\sigma \left[ q({\bf x},t) q(-x,y,-t)\right]_{xx}, &\\
&\label{complex-time-nonloc-DS-partially}
\text{\small{Partial reverse space-time nonlocal Davey-Stewartson}} & \\  \nonumber \\ 
%%%%%%%%%%%%%%%%%%%%%
%%%%%%%%%%%%%%%%%%%%%%%%%%%%%%%%%%%%%%%%%
%%%%%%%%%%%%%%%%%%%%%%%%%%%%%%%%%%%%%%%%%
&\nonumber
Q_{1,t} + c_{1}Q_{1,x}  = \sigma_{3}Q_{2}(-x,-t)Q_{3}(-x,-t),\;\; \sigma_{3}=\pm 1 &\\
%%%%%
&\nonumber
Q_{2,t} + c_{2}Q_{2,x}  = -\sigma_{2}Q_{1}(-x,-t)Q_{3}(-x,-t), \;\; \sigma_{2}=\pm 1& \\
%%%%%
&\nonumber
Q_{3,t} + c_{3}Q_{3,x}  = \sigma_{1}Q_{1}(-x,-t)Q_{2}(-x,-t), \;\; \sigma_{1}=\pm 1&\\  
%%%%%
&\text{\small{Reverse space-time nonlocal three wave interaction with $c_3>c_2>c_1, 
~\sigma_{1}\sigma_{3}/\sigma_{2}=1.$}} & \\ \nonumber \\
%%%%%%%%%%%%%%%%%%%%%%%
%%%%%%%%%%%%%%%%%%%%%%%%%%%%%%%%
%%%%%%%%%%%%%%%%%%%%
&\nonumber
Q_{1,t}({\bf x},t) + {\bf C}_{1}\cdot\nabla Q_{1} ({\bf x},t)
=\sigma_{3}Q^{*}_{2}(-{\bf x},-t)Q^{*}_{3}(-{\bf x},-t), \;\; \sigma_{3}=\pm 1 &\\
%%%%%%%%%%%%%%%%%%%%
&\nonumber
Q_{2,t}({\bf x},t) + {\bf C}_{2} \cdot\nabla Q_{2}({\bf x},t)
=-\sigma_{2}Q^{*}_{1}(-{\bf x},-t)Q^{*}_{3}(-{\bf x},-t),\;\; \sigma_{2}=\pm 1 & \\
%%%%%%%%%%%%%%%%%%%%%%%%
&\nonumber
Q_{3,t}({\bf x},t) + {\bf C}_{3} \cdot\nabla Q_{3}({\bf x},t)
=\sigma_{1}Q^{*}_{1}(-{\bf x},-t) Q^{*}_{2}(-{\bf x},-t), \;\; \sigma_{1}=\pm 1 &\\
%%%%%
&\nonumber
\text{\small{Multi-dimensional reverse space-time nonlocal three wave interaction}}& \\
&\text{\small{ with distinct ${\bf C}_j, j=1,2,3, 
~\sigma_{1}\sigma_{3}/\sigma_{2}=1.$}}& \\ \nonumber \\
%%%%%%%%%%%%%%%%%%%%%%%%%%%%%%%%%%%%%%%
&\nonumber
\label{dNLS-1}
i\frac{dQ_n(t)}{dt} = Q_{n+1}(t)  - 2Q_n(t) + Q_{n-1}(t)  &\\
&\nonumber
\hspace{1.5cm} -\sigma  Q_n(t)Q_{n}(-t) \left[Q_{n+1}(t) + Q_{n-1}(t) \right], &\\
&\text{\small{ Reverse time nonlocal discrete NLS}} & \\ \nonumber \\
%%%%%%%%%%%%%%%%%%%%%%%%%%%%%%%
%%%%%%%%%%%%%%%%%%%%%%%%%%%%%%%%
&\nonumber
\label{dNLS-2}
i\frac{dQ_n(t)}{dt} = Q_{n+1}(t)  - 2Q_n(t) + Q_{n-1}(t)  &\\
&\nonumber
\hspace{1.5cm} -\sigma Q_n(t)Q_{-n}(-t) \left[Q_{n+1}(t) + Q_{n-1}(t) \right],&\\
&\text{\small{ Reverse discrete-time nonlocal discrete NLS}} & \\ \nonumber
%%%%%%%%%%%%%%%%%%%%%%%%%%%%%%%
\end{flalign}
\newpage
%%%%%%%%%%%%%%%%%
In the above, ${\bf x}$ represents $(x,y)$ and $*$ denotes complex conjugation. 
Unless otherwise specified $q(x,t)$ 
or $q({\bf x},t)$ is a complex valued function of the real variables ${\bf x}$ and $t$. 
There are also nonlocal matrix and vector extensions of many of the above equations. 
In this paper, we will show how Eqns. (\ref{complex-space-time-nonloc-NLS}) -- 
(\ref{complex-time-nonloc-DS}) arise from a rather simple but extremely important symmetry reductions of various AKNS scattering problems in one and multi dimensions and show that they form a Hamiltonian integrable systems. For these equations, we provide few integrals of motions (conserved quantities) or indicate how an infinite number of them can be obtained and outline the solution strategy through the theory of inverse scattering transform. We then give a one soliton solution for a number of equations and discuss their properties.\\\\
%%%%%%%%%
In this paper we do not discuss in detail vector or matrix extensions of the integrable nonlocal NLS equations, i.e., their $PT$ - symmetric nonlocal versions, such as the equation obtained by replacing $[{\bf q}(x,t)\cdot {\bf q}(-x,-t)] $ in Eq.~ (\ref{Vcomplex-space-time-nonloc-NLS}) 
by $[{\bf q}(x,t)\cdot {\bf q^*}(-x,t)]$ in which case the resulting $PT$ symmetric 
multi-component nonlocal NLS equation reads
%%%%%%%%%%%%%%%%%%%%%%
%%%%%%%%%%%%%%%%%%%%%%
\begin{equation}
\label{VPTNLS-PT}
i{\bf q}_t(x,t) = {\bf q}_{xx}(x,t) - 2\sigma  [{\bf q}(x,t)\cdot {\bf q}^*(-x,t)]  {\bf q}(x,t)\;.
\end{equation}
%%%%%%%%%%%%%%%%%%%%%%
%%%%%%%%%%%%%%%%%%%%%%
As is the case in (\ref{Vcomplex-space-time-nonloc-NLS}), 
here dot stands for the usual vector scalar product.
We consider these equations to be direct extensions, though the IST is likely to contain novel aspects. In this regard we note that direct and inverse scattering of the AKNS $2\times 2$ and 
$n \times n$ systems have important applications in their own right.\\
%%%%%%%%%%%%%%%%%%%%%%%%%%%%%%%%%%%%%%%%%%%%%%%%%%%%%%%%%%%%%%%%%%%%%%%%%%%%%%%%%%%%%%%%%%%%%
The paper is organized as follows. In Sec. \ref{compatibility} we use the AKNS theory to derive various nonlocal reverse space-time and reverse time only NLS type equations in terms of two (complex or real) potentials: $q(x,t)$ and $r(x,t)$. In Sec. \ref{der-NLS} we show how one can derive the nonlocal analogue of the derivative NLS equation and show that it is an integrable nonlocal system. We also give few conserved quantities. The derivation of nonlocal mKdV and sine-Gordon is given in Sec.~ \ref{nonlocal-mKdv-SG}. The extension of the reverse 
space-time and the reverse time nonlocal NLS equation to the multi-dimensional case, i.e., Davey-Stewartson system is presented in Sec.~\ref{DSeqn}. The partially $PT$ symmetric and partially reverse space-time DS equations are obtained in Sec.~ \ref{PTDSeqn}. 
The derivation of the (1+1) and (2+1) dimensional nonlocal in space and time analogue of the multi wave equations is presented in Secs.~\ref{Nwave} and \ref{Nwave-2} respectively. The discrete analogues for the above mentioned nonlocal NLS equations are also derived in Sec.~\ref{discrete-sys}. For AKNS problems, the basic inverse scattering problem and reconstruction formula of the potential
is developed in Sec. \ref{ISTsoliton}. The important symmetries of the associated eigenfunctions and scattering data together with soliton solutions is presented 
in Sec.~\ref{solitons-sol}. Finally, we conclude in Sec.~\ref{conclusion} with an outlook for a future directions in the newly emerging field of integrable nonlocal equations including reverse space-time, reverse time and $PT$ symmetric nonlocal integrable systems.
%%%%%%%%%%%%%%%%%%%%%%%%%%%%%%%%%%%%%%%%%%%%%
%%%%%%%%%%%%%%%%%%%%%%%%%%%%%%%%%%%%%%%%%%%%%
\section{Linear pair and compatibility conditions: Nonlocal NLS hierarchy}
\label{compatibility}
%%%%%%%%%%%%%%%%%%%%%%%%%%%%%%%%%%%%%%%%%%%%%
%%%%%%%%%%%%%%%%%%%%%%%%%%%%%%%%%%%%%%%%%%%%%
Our starting point is the AKNS scattering problem \cite{Ablowitz2,Ablowitz3}
%%%%%%%%%%%%%%%%%%
\begin{equation}
\label{AKNS}
{\bf v}_x = \mathsf{X} {\bf v}\;,
\end{equation}
%%%%%%%%%%%%%%%%%%
where ${\bf v}={\bf v}(x,t)$ is a two-component vector, i.e., ${\bf v}(x,t)=(v_1(x,t), v_2(x,t))^T$ and $q(x,t), r(x,t)$ are (in general) complex valued functions of the real variables $x$ and $t$ that vanish rapidly as $|x|\rightarrow \infty$ and $k$ is a complex spectral parameter. The matrix $\mathsf{X}$ depends on the functions $q(x,t)$ and $r(x,t)$ as well as on the spectral parameter $k$
\begin{equation}
\label{LAX-1}
\mathsf{X}
=
 \left(\begin{array}{lc}
-ik&q(x,t) \\
r(x,t)&ik
\end{array}
\right )  \;.
\end{equation}
Associated with the scattering problem (\ref{AKNS}) is the time evolution equation of the eigenfunctions $v_j, j=1,2$ which is given by
%%%%%%%%%%%%%%%%%%%%
\begin{equation}
\label{time-AKNS}
{\bf v}_t = \mathsf{T} {\bf v}\;,
\end{equation}
where
\begin{equation}
\label{LAX-2}
\mathsf{T} =
 \left(\begin{array}{cr}
\mathsf{A}&\mathsf{B}
\\
\mathsf{C}&-\mathsf{A}
\end{array}\right) \;,
\end{equation}
and the quantities $\mathsf{A}, \mathsf{B}$ and $\mathsf{C}$ are scalar functions of 
$q(x,t), r(x,t)$ and the spectral parameter $k.$ Depending on the choice of these functions one finds an evolution equation for the potential functions $q(x,t)$ and $r(x,t)$ which, under a certain symmetry restriction, leads to a single evolution equation for either 
$q(x,t)$ or $r(x,t)$.  
%%%%%%%%%%%%%%%%%%%%%%%%%%%%%%%%%%%%%%%%%%%%%
%%%%%%%%%%%%%%%%%%%%%%%%%%%%%%%%%%%%%%%%%%%%%
In the case where the quantities $\mathsf{A}, \mathsf{B}$ and $\mathsf{C}$ are second order polynomials in the isospectral parameter $k$ with coefficients depending on $q(x,t), r(x,t)$, i.e., 
\begin{eqnarray}
\label{A0-series}
&&\mathsf{A}=2ik^2+iq(x,t)r(x,t)\;,
\\
\label{B0-series}
&&\mathsf{B}=-2kq(x,t)-iq_x(x,t)\;,
\\
\label{C0-series}
&&\mathsf{C}=-2kr(x,t)+ir_x(x,t)\;,
\end{eqnarray}
the compatibility condition of system (\ref{AKNS}) and (\ref{time-AKNS}) leads to
\begin{equation}
\label{q-r-sys-1}
iq_t(x,t)=q_{xx}(x,t) -2r(x,t)q^2(x,t)\;,
\end{equation}
\begin{equation}
\label{q-r-sys-2}
-ir_t(x,t)=r_{xx}(x,t) -2q(x,t)r^2(x,t)\;.
\end{equation}
Under the symmetry reduction 
\begin{equation}
\label{sym-reduction-intro-minus-2}
r(x,t) = \sigma q(-x,-t), ~~\sigma=\mp 1\;,
\end{equation}
the system (\ref{q-r-sys-1}) and (\ref{q-r-sys-2}) are compatible and leads to the reverse 
space-time nonlocal nonlinear Schr\"odinger 
equation (\ref{complex-space-time-nonloc-NLS}), which for convenience we rewrite 
again:
\begin{equation}
\label{PTNLS-new}
iq_t(x,t) = q_{xx}(x,t)  - 2 \sigma q^2(x,t)q(-x,-t)\;.
\end{equation}
We remark that the symmetry reduction (\ref{sym-reduction-intro-minus-2}) is new and,
since $q$ is complex valued, is different from the symmetry 
\begin{equation}
\label{PTsym2}
r(x,t)= \sigma q^*(-x,t) \;.
\end{equation}
The latter was found in \cite{AblowitzMusslimani} and leads to the $PT$ symmetric nonlocal NLS Eq.~ (\ref{PTNLS-1}). However, the new symmetry condition
(\ref{sym-reduction-intro-minus-2}) gives rise to a new class of nonlocal (in both space and time) integrable evolution equations including a nonlocal NLS hierarchy. Equation (\ref{PTNLS-new}) is another special and remarkably simple reduction of the more general $q,r$ system mentioned above. For completeness, we give the compatible pair associated with Eq.~(\ref{PTNLS-new}):
%%%%%%%%%%%%%%%
%%%%%%%%%%%%%%%
\begin{equation}
%%%%%%%
\mathsf{X}= \left(\begin{array}{lc}
-ik&q(x,t) \\
\sigma q(-x,-t)&ik
\end{array}
\right ),
\end{equation}
%%%%%%%%%%
\begin{equation}
\mathsf{T}= \left(\begin{array}{cr}
2ik^2 + i\sigma  q(x,t)q(-x,-t) &-2kq(x,t)-iq_x(x,t)
\\
-2\sigma kq(-x,-t) -\sigma i q_x(-x,-t)& -2ik^2 - i \sigma q(x,t)q(-x,-t)
\end{array}\right) 
\;.
\end{equation}
%\end{widetext}
It is well-known that the compatible pair 
(\ref{LAX-1})-(\ref{time-AKNS}) with (\ref{A0-series})-(\ref{C0-series}) lead to an infinite number of conservation laws and conserved quantities cf. \cite{Ablowitz2}. The first few  conserved quantities associated with Eq.~(\ref{PTNLS-new}) are given by
%%%%%%%%%%%%%%%%%%%%%%%%%%%%
\begin{equation}
\label{conserve-1}
\int_{\mathbb{R}}q(x,t)q(-x,-t)dx = \text{constant}\;,
\end{equation}
%%%%%%%%%%%%%%%%%%%%%%%%%%%%
\begin{equation}
\label{conserve-2}
\int_{\mathbb{R}}q_x(x,t)q(-x,-t)dx = \text{constant} \;,
\end{equation}
%%%%%%%%%%%%%%%%%%%%%%%%%%%%
\begin{equation}
\label{conserve-3}
\int_{\mathbb{R}}\left(\sigma q_x(x,t)q_x(-x,-t) + q^2(x,t)q^2(-x,-t)  \right)dx 
= \text{constant}\;.
\end{equation}
%%%%%%%%%%%%%%%%%%%%%%%%%%%%
In the context of $PT$ symmetric linear/nonlinear optics, the analogous quantity in Eq.~(\ref{conserve-1}) is referred to as the ``quasipower." We also note that 
Eq. (\ref{PTNLS-new}) is an integrable Hamiltonian system with Hamiltonian 
given by equation (\ref{conserve-3}).\\\\
%%%%%%%%%%%%%%%%
We also note that equations such as (\ref{PTNLS-new}) are nonlocal in both space and time. Alone, it is not immediately clear how  (\ref{PTNLS-new}) is an evolution equation. But with the symmetry reduction (\ref{sym-reduction-intro-minus-2}) we can consider (\ref{PTNLS-new}) as arising as the unique solution associated with the evolution system (\ref{q-r-sys-1})-(\ref{q-r-sys-2}) with initial conditions $r(x,t=0)= \sigma q^*(-x,t=0)$. All nonlocal in time equations in this paper can be considered in a similar way.\\\\
%%%%%%%%%%%%%%%%
Another interesting nonlocal symmetry reduction that system 
(\ref{q-r-sys-1}) and (\ref{q-r-sys-2}) admits is given by
 \begin{equation}
\label{sym-reduction-nls-time}
r(x,t) = \sigma q(x,-t)\;,
\end{equation}
which, in turn, leads to the following new {\it reverse-time} nonlocal nonlinear Schr\"odinger equation 
\begin{equation}
\label{TNLS-2}
iq_t(x,t) = q_{xx}(x,t) - 2 \sigma q^2(x,t)q(x,-t)\;. 
\end{equation}
Again, the condition (\ref{sym-reduction-nls-time}) is new, remarkably simple, and has not been noticed in the literature and leads to a nonlocal in time NLS hierarchy. 
Furthermore, since this equation arises from the above AKNS scattering problem, it is an integrable Hamiltonian evolution equation that admits an infinite number of conservation laws /conserved quantities. The first few are listed below:
%%%%%%%%%%%%%%%%%%%%%%%%%%%%
\begin{equation}
\label{conserve-1a}
\int_{\mathbb{R}}q(x,t)q(x,-t)dx = \text{constant}  \;,
\end{equation}
%%%%%%%%%%%%%%%%%%%%%%%%%%%%
\begin{equation}
\label{conserve-2a}
\int_{\mathbb{R}}q(x,t)q_x(x,-t)dx = \text{constant}  \;,
\end{equation}
%%%%%%%%%%%%%%%%%%%%%%%%%%%%
\begin{equation}
\label{conserve-3a}
\int_{\mathbb{R}}\left(\sigma q_x(x,t)q_x(x,-t) + q^2(x,t)q^2(x,-t)  \right)dx 
= \text{constant} \;.
\end{equation}
%%%%%%%%%%%%%%%%%%%%%%%%%%%%
The Lax pairs associated with Eq.~(\ref{TNLS-2}) are thus given by
%%%%%%%%%%%%%%%
\begin{equation}
\label{TNLS-lax-1}
\mathsf{X} = \left(\begin{array}{lc}
-ik&q(x,t) \\
\sigma q(x,-t)&ik
\end{array}
\right ),
\end{equation}
%%%%%%%%%%%%%%%
\begin{equation}
\mathsf{T} = \left(\begin{array}{cr}
2ik^2 +i \sigma  q(x,t)q(x,-t) &-2kq(x,t)-iq_x(x,t)
\\
- 2 \sigma kq(x,-t) \pm i q_x(x,-t)& -2ik^2 - i \sigma q(x,t)q(x,-t)
\end{array}\right) 
\;.
\end{equation}
%\end{widetext}
%%%%%%%%%%%%%%%%%%%%%%%%%%%%
%%%%%%%%%%%%%%%%%%%%%%%%%%%%%%%%%%%%%%%%%%%%%
The extension to the matrix or vector (multi component) reverse space-time or reverse time only nonlocal NLS system can be carried out in a similar fashion. Indeed, if we start from the matrix generalization of the AKNS scattering problem then the compatibility condition generalizing the one given in system (\ref{q-r-sys-1}) and  (\ref{q-r-sys-2}) would now read
\begin{equation}
\label{q-r-sys-1-matrix}
i{\bf Q}_t(x,t) = {\bf Q}_{xx}(x,t) - 2 {\bf Q}(x,t) {\bf R}(x,t) {\bf Q}(x,t) \;,
\end{equation}
\begin{equation}
\label{q-r-sys-2-matrix}
-i{\bf R}_t(x,t) = {\bf R}_{xx}(x,t) - 2 {\bf R}(x,t) {\bf Q}(x,t) {\bf R}(x,t) \;,
\end{equation}
where ${\bf Q}(x,t)$ is an $N\times M$ matrix; ${\bf R}(x,t)$ is an $M\times N$ matrix of the real variables $x$ and $t$ and super script $T$ denotes matrix transpose. Under the symmetry reduction 
\begin{equation}
\label{sym-reduction-matrix}
{\bf R}(x,t) = \sigma {\bf Q}^T(-x,-t)\;,\;\;\;\; \sigma = \mp 1,
\end{equation}
system (\ref{q-r-sys-1-matrix}) and (\ref{q-r-sys-2-matrix}) are compatible and this leads to the reverse space-time nonlocal matrix nonlinear Schr\"odinger equation 
\begin{equation}
\label{q-r-sys-1-matrix-eqn}
i{\bf Q}_t(x,t) = {\bf Q}_{xx}(x,t)  - 2 \sigma {\bf Q}(x,t) {\bf Q}^T(-x,-t) {\bf Q}(x,t) \;.
\end{equation}
In the special case where ${\bf Q}$ is either a column vector ($M=1$) then 
Eq.~(\ref{q-r-sys-1-matrix-eqn}) reduces to (\ref{Vcomplex-space-time-nonloc-NLS}), i.e.,
\begin{equation}
\label{PTNLS-new-1}
i{\bf q}_t(x,t) = {\bf q}_{xx}(x,t)  - 2 \sigma [{\bf q}(x,t)\cdot {\bf q}(-x,-t)]  {\bf q}(x,t),
\end{equation}
where dot stands for the vector scalar product. As in the scalar case, we can generalize Eq.~(\ref{TNLS-2}) to the matrix or vector multi component case. Indeed, we note that system (\ref{q-r-sys-1-matrix}) and  
(\ref{q-r-sys-2-matrix}) are compatible under the symmetry reduction
\begin{equation}
\label{sym-reduction-matrix-2}
{\bf R}(x,t) = \sigma {\bf Q}^T(x,-t)\;, \;\;\;\;\; \sigma = \mp 1,
\end{equation}
which in turn gives rise to the following nonlocal in time only matrix nonlinear Schr\"odinger equation 
\begin{equation}
\label{q-r-sys-1-matrix-eqn-time}
i{\bf Q}_t(x,t) = {\bf Q}_{xx}(x,t) - 2 \sigma {\bf Q}(x,t) {\bf Q}^T(x,-t) {\bf Q}(x,t) \;.
\end{equation}
To obtain the multi-component analogue of Eq.~(\ref{q-r-sys-1-matrix-eqn-time}) we restrict the matrix ${\bf Q}$ to a column vector ($N=1$) giving rise to the following nonlocal evolution equation 
\begin{equation}
\label{PTNLS-new-1-time}
i{\bf q}_t(x,t) = {\bf q}_{xx}(x,t) - 2 \sigma [{\bf q}(x,t)\cdot {\bf q}(x,-t)]  {\bf q}(x,t).
\end{equation}
%%%%%%%%%%%%%%%%%%%%%%%%%%%%%%%%%%%%%%%%%%%%%
%%%%%%%%%%%%%%%%%%%%%%%%%%%%%%%%%%%%%%%%%%%%%
\section{Reverse space-time nonlocal coupled NLS -- derivative NLS equation}
\label{der-NLS}
%%%%%%%%%%%%%%%%%%%%%%%%%%%%%%%%%%%%%%%%%%%%%
%%%%%%%%%%%%%%%%%%%%%%%%%%%%%%%%%%%%%%%%%%%%%
In this section we derive the space-time nonlocal coupled NLS-derivative NLS equation that includes the reverse space-time nonlocal NLS (as well as the reverse space-time nonlocal
derivative NLS) equations as special cases. To do so we consider a generalization to the AKNS scattering problem (\ref{AKNS}) with 
%%%%%%%%%%%%%%%%%%
\begin{equation}
\label{AKNS-der}
\mathsf{X} 
=
 \left(\begin{array}{lc}
-f(k)& g(k)q(x,t) \\
g(k)r(x,t)&f(k)
\end{array}
\right ) 
 \;,
\end{equation}
%%%%%%%%%%%%%%%%%%
where $f(k) = i\alpha k^2 -\sqrt{2\beta}k$ and $g(k)=\alpha k+i\sqrt{\beta/2}$ are functions of the complex spectral parameter $k$ and $\alpha, \beta$ are real constants.
The time evolution of the eigenfunctions ${\bf v}(x,t)$ is governed by 
Eqns.~(\ref{time-AKNS}) and (\ref{LAX-2}) where functions $\mathsf{A}, \mathsf{B}$ and $\mathsf{C}$ are now fourth order polynomials in the isospectral parameter $k$ (see \cite{Ablowitz1}). The compatibility condition of system (\ref{AKNS-der}) 
and (\ref{time-AKNS}) gives the coupled $q,r$ system
\begin{eqnarray}
\label{q-r-sys-1-der}
q_t(x,t) = iq_{xx}(x,t) + \alpha \left(r(x,t)q^2(x,t)\right)_x 
+ i\beta r(x,t)q^2(x,t) \;,
\end{eqnarray}
\begin{eqnarray}
\label{q-r-sys-2-der}
-r_t(x,t) = ir_{xx}(x,t) - \alpha \left(r^2(x,t)q(x,t)\right)_x 
+ i\beta r^2(x,t)q(x,t)\;.
\end{eqnarray}
Under the symmetry reduction (\ref{sym-reduction-intro-minus-2})
the system (\ref{q-r-sys-1-der}) and (\ref{q-r-sys-2-der}) are compatible and leads to the 
reverse space-time nonlocal coupled NLS-derivative NLS equation:
\begin{eqnarray}
\label{PTNLS-new-der}
q_t(x,t) = iq_{xx}(x,t) + \alpha\sigma \left(q(-x,-t)q^2(x,t)\right)_x 
+ i\beta\sigma q(-x,-t)q^2(x,t) \;.
\end{eqnarray}
In the special case where $\alpha =0$ and $\beta =2$ we recover Eq.~(\ref{PTNLS-new}). On the other hand, if we choose $\alpha =1$ and $\beta =0$ then we find the reverse space-time nonlocal version of the ``classical" derivative NLS equation:
\begin{equation}
\label{PTNLS-new-der-only}
iq_t(x,t) = -q_{xx}(x,t) + i\sigma \left(q(-x,-t)q^2(x,t)\right)_x  \;.
\end{equation}
\\
The linear Lax pairs associated with Eq.~(\ref{PTNLS-new-der-only}) are given by
%%%%%%%%%%%%%%%%%%
\begin{equation}
\mathsf{X} 
=
 \left(\begin{array}{lc}
-ik^2& kq(x,t) \\
k\sigma q(-x,-t)&ik^2
\end{array}
\right ) \;,
\end{equation}
%%%%%%
%\begin{widetext}
\begin{equation}
\mathsf{T} = \left(\begin{array}{cr}
\mathsf{A}^{nonloc}_{dNLS}  & \mathsf{B}^{nonloc}_{dNLS}
\\ \\
\mathsf{C}^{nonloc}_{dNLS}
& -\mathsf{A}^{nonloc}_{dNLS}
\end{array}\right) 
\;,
\end{equation}
where
%%%%%%%%%%%%%%%%%%
\begin{equation}
\label{A-nonloc-dNLS}
\mathsf{A}^{nonloc}_{dNLS} = -2ik^4  - i\sigma q(-x,-t)q(x,t)k^2\;,
\end{equation}
%%%%%%%%%%%%%%%%%%%%
%%%%%%%%%%%%%%%%%%
\begin{equation}
\label{B-nonloc-dNLS}
\mathsf{B}^{nonloc}_{dNLS} = 2q(x,t)k^3 + (iq_x(x,t)+\sigma q(-x,-t)q^2(x,t))k \;,
\end{equation}
%%%%%%%%%%%%%%%%%%%%
%%%%%%%%%%%%%%%%%%
\begin{equation}
\label{C-nonloc-dNLS}
\mathsf{C}^{nonloc}_{dNLS}
=
2\sigma kq(-x,-t)k^3 + (i\sigma q_x(-x,-t)+ q^2(-x,-t)q(x,t))k\;.
\end{equation}
%%%%%%%%%%%%%%%%%%%%
In \cite{KaupNewell} it was shown that the general $q,r$ system (\ref{q-r-sys-1-der}) 
and (\ref{q-r-sys-2-der}) for $\alpha =1$ and $\beta =0$ is integrable and admits infinitely many conservation laws. Since the new nonlocal equation (\ref{PTNLS-new-der-only}) comes out of a new symmetry reduction it is also an infinite dimensional integrable Hamiltonian system. The first two conserved quantities associated with Eq.~(\ref{PTNLS-new-der-only}) are
%%%%%%%%%%%%%%%%%%%%%%%%%%%%%%%%%%%%%%%
%%%%%%%%%%%%%%%%%%%%%%%%%%%%
\begin{equation}
\label{conserve-1a-der}
\int_{\mathbb{R}}q(x,t)q(-x,-t)dx = \text{constant}  \;,
\end{equation}
%%%%%%%%%%%%%%%%%%%%%%%%%%%%
\begin{equation}
\label{conserve-2a-der}
\int_{\mathbb{R}}q(x,t)\left[
\frac{i}{2} q^2(-x,-t)q(x,t) -  \sigma q_x(-x,-t)\right]
      dx = \text{constant}
      \;.
\end{equation}
%%%%%%%%%%%%%%%%%%%%%%%%%%%%
%%%%%%%%%%%%%%%%%%%
Another interesting nonlocal in both space and time integrable evolution equation can be obtained from the scattering problem (\ref{AKNS-der}) if one chooses the functional dependence of $f$ and $g$ on $k$ to be linear, i.e., $f(k)=g(k)=k$ with suitable functions
$\mathsf{A}, \mathsf{B}$ and $\mathsf{C}$ (see \cite{KJ84,Ablowitz1}). Following the same procedure as above, the compatibility condition gives rise to the following system of $q,r$
equations: 
%%%%%%%%%%%%%%%%%%%%%
\begin{eqnarray}
\label{q-eqn-loop}
\frac{\partial q(x,t)}{\partial t} + \frac{\partial^2}{\partial x^2} \left[ \frac{q_{x}(x,t)}{(1 - r(x,t) q(x,t))^{3/2}}\right] =0\;,
\end{eqnarray}
%%%%%%%%%%%%%%%%%%%%
\begin{eqnarray}
\label{r-eqn-loop}
\frac{\partial r(x,t)}{\partial t} + \frac{\partial^2}{\partial x^2} \left[ \frac{r_{x}(x,t)}{(1 - r(x,t) q(x,t))^{3/2}}\right] =0\;.
\end{eqnarray}
%%%%%%%%%%%%%%%%%%%%%
Now, under the symmetry reduction (\ref{sym-reduction-intro-minus-2}), i.e., 
$r(x,t)=\sigma q(-x,-t), \sigma = \mp 1$, Eqns.~(\ref{q-eqn-loop}) and (\ref{r-eqn-loop}) are compatible and leads to the reverse space-time nonlocal ``loop soliton" equation
%%%%%%%%%%%%%%%%%%%%%
\begin{eqnarray}
\label{q-eqn-loop-1}
\frac{\partial q(x,t)}{\partial t} + \frac{\partial^2}{\partial x^2} \left[ \frac{q_{x}(x,t)}{(1 - \sigma q(x,t) q(-x,-t))^{3/2}}\right] =0\;,
\end{eqnarray}
with $\sigma = \mp 1.$
%%%%%%%%%%%%%%%%%%%%
The conservation laws for this `loop soliton' system can be obtained by standard methods; cf. \cite{KSI74,WSK75}
%%%%%%%%%%%%%%%%%%%%%%%%%%%%%%%%%%%%%%%%%%%
\section{Complex and real reverse  space-time nonlocal mKdV and sine-Gordon Equations}
\label{nonlocal-mKdv-SG}
%%%%%%%%%%%%%%%%%%%%%%%%%%%%%%%%%%%%%%%%%%%
Returning to the $2 \times 2$ Lax pair given by equations (\ref{AKNS}) - (\ref{LAX-2}) we can find other integrable nonlocal equations depending on the functional form 
of $\mathsf{A}, \mathsf{B}$ and $\mathsf{C}$ on the spectral parameter $k.$ In the following few sections, we will derive the space-time nonlocal versions of the ``classical" mKdV and 
sine-Gordon equations and provide the IST formulation as well as one soliton solution. Contrary to the $PT$ symmetric nonlocal NLS case where the one soliton solution can develop a singularity in fine time \cite{AblowitzMusslimani,AblowitzMusslimaniNonlinearity}, 
here the reverse space-time nonlocal mKdV soliton can be generically regular and does not develop a singularity.
%%%%%%%%%%%%%%%%%%%%%%%%%%%%%%%%%%%%
\subsection{The complex reverse space-time nonlocal mKdV}
%%%%%%%%%%%%%%%%%%%%%%%%%%%%%%%%%%%%%
If we take
 \begin{eqnarray}
\label{A0-series-mkdv}
&&\mathsf{A_3} = -4ik^3 - 2iq(x,t)r(x,t)k + r(x,t)q_x(x,t) - q(x,t)r_x(x,t)\;,
\nonumber \\
\label{B0-series-mkdv}
&&\mathsf{B_3} = 4k^2q(x,t) + 2iq_x(x,t)k +2q^2(x,t)r(x,t) - q_{xx}(x,t)\;,
\nonumber \\
\label{C0-series-mkdv}
&&\mathsf{C_3} = 4k^2r(x,t) - 2ir_x(x,t)k + 2q(x,t)r^2(x,t) - r_{xx}(x,t) \;,
\nonumber 
\end{eqnarray}
the compatibility condition of system (\ref{AKNS}) and (\ref{time-AKNS}) yields
\begin{equation}
\label{q-r-sys-1-mkdv}
q_t(x,t) + q_{xxx}(x,t) - 6q(x,t)r(x,t)q_x(x,t) = 0\;,
\end{equation}
\begin{equation}
\label{q-r-sys-2-mkdv}
r_t(x,t) + r_{xxx}(x,t) - 6q(x,t)r(x,t)r_x(x,t) = 0\;.
\end{equation}
Under the symmetry reduction 
\begin{equation}
\label{PTsymt-mKdV}
r(x,t)=\sigma q^*(-x,-t), ~~\sigma= \mp 1 \;,
\end{equation}
the system (\ref{q-r-sys-1-mkdv}) and (\ref{q-r-sys-2-mkdv}) are compatible and leads to the  complex reverse space-time nonlocal complex mKdV equation
\begin{equation}
\label{PTNLS-r-mkdv-2}
q_t(x,t) + q_{xxx}(x,t) - 6\sigma q(x,t)q^*(-x,-t)q_x(x,t) = 0\;,
\end{equation}
where again $*$ denotes complex conjugation and $q(x,t)$ is a complex valued function of the real variables $x$ and $t$. On the other hand, using the symmetry 
(\ref{sym-reduction-intro-minus-2}) yields the real reverse space-time equation
\begin{equation}
\label{PTNLS-r-mkdv-3}
q_t(x,t) + q_{xxx}(x,t) - 6\sigma q(x,t)q(-x,-t)q_x(x,t) = 0\;,
\end{equation}
which for real initial conditions is the real mKdV equation. We also point out that under space and time even initial conditions, the nonlocal mKdV equation reduces to its classical (local) counterpart. Furthermore, when using the symmetry reduction 
$r(x,t)= \sigma q(-x,-t)$ for the NLS or mKdV case, one need not to specify whether $q$ is real or complex valued. However, if one further restricts $q$ to be real then additional  symmetry conditions on the underlying eigenfunctions and scattering data are required, beyond those that come out of the symmetry reduction $r(x,t)=\sigma q^*(-x,-t)$. 
The compatible pair associated 
with Eq.~(\ref{PTNLS-r-mkdv-2}) now is
%%%%%%%%%%%%%%%
\begin{equation}
\label{AKNS-sin}
{\bf v}_x=\left(\begin{array}{lc}
-ik&q(x,t) \\
\sigma q^*(-x,-t)&ik
\end{array}
\right ) {\bf v},
\end{equation}
\\
%%%%%%%%%%%%%%%
%%%%%%%%%%%%%%%%%%%%
\begin{equation}
{\bf v}_t=\left(\begin{array}{cr}
\mathsf{A}_{3,nonloc}&\mathsf{B}_{3,nonloc}
\\ \\
\mathsf{C}_{3,nonloc}&-\mathsf{A}_{3,nonloc}
\end{array}\right) {\bf v}\;,
\end{equation}
where
%\begin{widetext}
 \begin{eqnarray}
\label{A0-series-mkdv-1}
\mathsf{A}_{3,nonloc} &=& -4ik^3 - 2i \sigma q(x,t)q^*(-x,-t)k 
\nonumber \\
&+& \sigma q^*(-x,-t)q_x(x,t) + \sigma q(x,t) q^*_x(-x,-t),
\end{eqnarray}
\begin{eqnarray}
\label{B0-series-mkdv-1}
\mathsf{B}_{3,nonloc} &=& 4k^2q(x,t) + 2iq_x(x,t)k 
\nonumber \\
&+& 2\sigma q^2(x,t) q^*(-x,-t) - q_{xx}(x,t)\;,
\end{eqnarray}
\begin{eqnarray}
\label{C0-series-mkdv-1}
\mathsf{C}_{3,nonloc} &=& 4k^2 \sigma q^*(-x,-t) + 2i\sigma q^*_x(-x,-t)k 
\nonumber \\ 
&+& 2q(x,t) q^{*2}(-x,-t) - \sigma q^*_{xx}(-x,-t) \;.
\end{eqnarray}
%%%%%%%%%%%%%%%%%%%%%%%%%%%%%%%%%%%%%%%%%%%%%
\subsection{The reverse space-time nonlocal sine-Gordon equation}
%%%%%%%%%%%%%%%%%%%%%%%%%%%%%%%%%%%%%%%%%%%%%
If on the other hand one makes the ansatz $\mathsf{A}=\mathsf{A}_1/k, \mathsf{B}=\mathsf{B}_1/k$ and $\mathsf{C}=\mathsf{C}_1/k$ then after some algebra the compatibility condition $v_{jxt}=v_{jtx}, j=1,2$ with $k$ being the time independent
isospectral parameter leads to
\begin{equation}
\label{q-sG}
q_{xt}(x,t) + 2s(x,t)q(x,t) =0\;,
\end{equation}
\begin{equation}
\label{r-sG}
r_{xt}(x,t) + 2s(x,t)r(x,t) =0\;,
\end{equation}
\begin{equation}
\label{s-sG}
s_{x}(x,t) + (q(x,t)r(x,t))_t =0\;,
\end{equation}
where we have defined $\mathsf{A}_1=-is/2.$  Also for completeness: 
$\mathsf{B}_1=q_t/(2i), \mathsf{C}_1=-r_t/(2i)$. Under the symmetry condition 
\begin{equation}
\label{PTsymtr-SG}
r(x,t)=-q(-x,-t)\;,
\end{equation}
with $q \in \mathbb{R}$ the system of equations (\ref{q-sG}) -- (\ref{s-sG}) are compatible and give rise to the real nonlocal sine-Gordon (sG) equation
\begin{equation}
\label{q-sG-final}
q_{xt}(x,t) + 2s(x,t)q(x,t) =0\;,\;\; s(-x,-t)=s(x,t)\;.
\end{equation}
We also fix the boundary condition of $s$ as $x \rightarrow \infty$ consistent with the classical sine-Gordon equation to be:
\begin{equation}
\label{sGinfty}
s(x,t)= s(-\infty) - \int_{-\infty}^x(q(x,t)q(-x,-t))_t(x',t)dx', ~~  s(-\infty)= i/4. 
\end{equation}
We note that we could have also generated a complex form of the (sG) equation following the previous discussion.  But for simplicity, here we only give the real nonlocal (SG) equation.
%%%%%%%%%%%%%%%%%%%%%%%%%%%%%%%%%%%%%%%%%%%%%
\subsection{Overview}
%%%%%%%%%%%%%%%%%%%%%%%%%%%%%%%%%%%%%%%%%%%%%
In summary, system (\ref{q-r-sys-1}) and (\ref{q-r-sys-2}) admits six symmetry reductions. The first four of which give rise to an integrable nonlocal NLS-type equation and the last two of which yield a nonlocal integrable mKdV-type evolution equation (below $\sigma = \mp 1$): 
\begin{enumerate}
%%%%%%%%%%
\item Standard AKNS symmetry:
 $$r(x,t) = \sigma q^*(x,t),$$ which has been known in the literature for more than four decades \cite{AKNS}; the paradigm is the NLS equation (\ref{NLS});\\
%%%%%%%%%%
\item  Reverse time AKNS symmetry 
$$r(x,t) = \sigma q(x,-t),$$ 
see the NLS-type equation (\ref{complex-time-nonloc-NLS});\\
%%%%%%%%%%
\item $PT$ preserving symmetry 
$$r(x,t) = \sigma q^*(-x,t),$$ 
which was found in 2013 \cite{AblowitzMusslimani}; see the NLS-type equation (\ref{PTNLS-1});\\
%%%%%%%%%%
\item Reverse space-time symmetry 
$$r(x,t) = \sigma q(-x,-t), ~q\in \mathbb{C},$$  
see the NLS-type equation (\ref{complex-space-time-nonloc-NLS});\\
%%%%%%%%%%
\item  $PT$ reverse time symmetry 
$$r(x,t) = \sigma q^*(-x,-t),$$ 
see the complex mKdV-type equation (\ref{complex-space-time-nonloc-mkdv});\\
%%%%%%%%%%
\item Real reverse space-time symmetry 
$$r(x,t) = \sigma q(-x,-t), ~q\in \mathbb{R},$$ 
see the real mKdV-type equation (\ref{real-space-time-nonloc-mkdv}).
\end{enumerate}
In Sec.~\ref{solitons-sol} we find soliton solutions of nonlocal NLS, mKdV and sine-Gordon type equations with these symmetries.
%%%%%%%%%%%%%%%%%%%%%%%%%%%%%%%%%%%%%%%%%%%%%
%%%%%%%%%%%%%%%%%%%%%%%%%%%%%%%%%%%%%%%%%%%%%
\section{Reverse space-time and reverse time nonlocal Davey-Stewartson system}
\label{DSeqn}
%%%%%%%%%%%%%%%%%%%%%%%%%%%%%%%%%%%%%%%%%%%%%
%%%%%%%%%%%%%%%%%%%%%%%%%%%%%%%%%%%%%%%%%%%%%
The integrable  two spatial dimensional extension of the NLS equation was obtained from a $2 \times 2$  compatible linear pair in \cite{AblHa75}. The IST was carried out later--cf. \cite{Ablowitz1}. The spatial part of the linear pair generalizes the operator $X$ in (\ref{AKNS}) - (\ref{LAX-1}) where the eigenvalue $k$ is replaced by an operator in the transverse spatial variable $y$. This new operator still contains the potentials $q,r$ which now depend on $x,y$ and $t$. The general Davey-Stewartson (DS) $(q,r)$ system 
is given by
%%%%%%%%%%%%%%%
%\begin{widetext}
\begin{eqnarray}
\label{q-DS}
iq_t({\bf x},t) &+& \frac{1}{2}\left[\gamma^2 q_{xx}({\bf x},t)  + q_{yy}({\bf x},t) \right]
+ q^2({\bf x},t) r({\bf x},t)
\nonumber \\
&=& \phi ({\bf x},t) q({\bf x},t) ,   
\end{eqnarray}
%%%%%%%%%%%%%%%
\begin{eqnarray}
\label{r-DS}
-ir_t({\bf x},t) &+& \frac{1}{2}\left[\gamma^2 r_{xx}({\bf x},t)  + r_{yy}({\bf x},t) \right]
 + r^2({\bf x},t)q({\bf x},t) 
\nonumber \\
&=& \phi ({\bf x},t) r({\bf x},t) ,   
\end{eqnarray}
%%%%%%%%%%%%%%%
%%%%%%%%%%%%%%%
\begin{equation}
\label{phi-DS}
\phi_{xx} ({\bf x},t) - \gamma^2 \phi_{yy} ({\bf x},t) = 2\left[ q({\bf x},t) r({\bf x},t)\right]_{xx},
\end{equation}
%\end{widetext}
%%%%%%%%%%%%%%%%
where $\gamma^2 = \pm 1$ and ${\bf x}=(x,y)$ is the transverse plane. In \cite{AblHa75} it was shown that the system of equations (\ref{q-DS}) and (\ref{r-DS}) are consistent under the symmetry reduction $r({\bf x},t) = \sigma q^*({\bf x},t)$ and leads to the ``classical" DS equation and in \cite{Fokas2016} a $PT$ symmetric reduction in the form 
$r({\bf x},t) = \sigma q^*(-{\bf x},t)$ was also reported. In this paper, we shall identify two new nonlocal symmetry reductions to the above DS system: 
$r({\bf x},t) = \sigma q(-{\bf x},-t)$ and $r({\bf x},t) = \sigma q({\bf x},-t)$ each of which leads to a new DS system.
%%%%%%%%%%%%%%%%%%%%%%%%%%%%%%%%%%%%%%%%%%%
\subsection{Reverse space-time nonlocal Davey-Stewartson equation}
%%%%%%%%%%%%%%%%%%%%%%%%%%%%%%%%%%%%%%%%%%
Under the symmetry reduction 
\begin{equation}
\label{sym-reduction-DS}
r({\bf x},t) = \sigma q(-{\bf x},-t)\;,
\end{equation}
it can be shown that system (\ref{q-DS}) and (\ref{r-DS}) are compatible and lead to the reverse space-time nonlocal Davey-Stewartson equation (\ref{complex-space-time-nonloc-DS}) 
which, for the convenience of the reader we rewrite again:
%%%%%%%%%%%%%%%%%%%%%
%%%%%%%%%%%%%%%%%%%%%%
%%%%%%%%%%%
\begin{eqnarray}
\label{complex_DS-q}
iq_t({\bf x},t) &+& \frac{1}{2}\left[\gamma^2 q_{xx}({\bf x},t)  + q_{yy}({\bf x},t) \right]
+\sigma  q^2({\bf x},t) q(-{\bf x},-t)
\nonumber \\ 
&=&  \phi ({\bf x},t) q({\bf x},t)\;,
\end{eqnarray}
%%%%%%%%%%%%%%%
\begin{eqnarray}
\label{complex_DS-phi}
\phi_{xx} ({\bf x},t) - \gamma^2 \phi_{yy} ({\bf x},t) = 2\sigma \left[ q({\bf x},t) q(-{\bf x},-t)\right]_{xx} \;.
\end{eqnarray}
%%%%%%%%%%%%%%%%
Note that from Eq.~(\ref{complex_DS-phi}) it follows that the potential $\phi$ has a solution that satisfies the relation $\phi (-{\bf x},-t) = \phi ({\bf x},t).$ The solution $\phi$ can, in principle, have boundary conditions that do not allow 
$\phi (-{\bf x},-t) = \phi ({\bf x},t).$ For the decaying infinite space problem we are considering here, one can expect the symmetry relation for $\phi ({\bf x},t)$ to hold. The elliptic case in the $\phi$ equation is easier than the hyperbolic one. In general, to prove 
$\phi (-{\bf x},-t) = \phi ({\bf x},t)$ one need to study the Greens function and see if this symmetry reduction holds. For the two-dimensional elliptic case with $\gamma^2=-1$ this condition appears to be true. Thus, the existence of the symmetry property for
$\phi ({\bf x},t)$ is {\it necessary} for the $(q,r)$ DS system to be compatible. 
Any solution $\phi ({\bf x},t)$ that breaks the symmetry $\phi (-{\bf x},-t) = \phi ({\bf x},t)$, 
would force the $(q,r)$ system to be become inconsistent. As such, the proposed new nonlocal DS equations are valid so long $\phi ({\bf x},t)$ satisfy the necessary underlying symmetry induced from the nonlocal AKNS symmetry reduction.
%%%%%%%%%%%%%%%%%%%%%%%%%%%%%%%%%%%%%%%%%%%%
%%%%%%%%%%%%%%%%%%%%%%%%%%%%%%%%%%%%%%%%%%%%%
\subsection{Reverse time nonlocal Davey-Stewartson equation}
%%%%%%%%%%%%%%%%%%%%%%%%%%%%%%%%%%%%%%%%%%%%%
Another interesting and new symmetry reduction which was not noticed in the literature so far is the time only nonlocal reduction given by
\begin{equation}
\label{sym-reduction-DS-time}
r({\bf x},t) = \sigma q({\bf x},-t)\;,\;\;\;\;\; \sigma = \mp 1.
\end{equation}
With this symmetry condition, system (\ref{q-DS}) and (\ref{r-DS}) are consistent and give rise to the following reverse time-only nonlocal DS system of equation
%%%%%%%%%%%%%%%%%%%%%
%%%%%%%%%%%%%%%%%%%%%%
%%%%%%%%%%%
\begin{eqnarray}
\label{complex_DS-q-time}
iq_t({\bf x},t) &+& \frac{1}{2}\left[\gamma^2 q_{xx}({\bf x},t)  + q_{yy}({\bf x},t) \right]
+\sigma  q^2({\bf x},t) q({\bf x},-t) 
\nonumber \\
 &=&  \phi ({\bf x},t) q({\bf x},t) \;,
\end{eqnarray}
%%%%%%%%%%%%%%%
\begin{eqnarray}
\label{complex_DS-phi-time}
\phi_{xx} ({\bf x},t) - \gamma^2 \phi_{yy} ({\bf x},t) 
= 2 \sigma \left[ q({\bf x},t) q({\bf x},-t)\right]_{xx} \;.
\end{eqnarray}
%%%%%%%%%%%
Note that from Eqns.~(\ref{complex_DS-q-time}) and (\ref{complex_DS-phi-time}) it follows that the potential $\phi$ has a solution that satisfies 
$\phi ({\bf x},t) = \phi ({\bf x},-t).$
In this paper, we will not go into further detail regarding the integrability properties of the above systems nor will we construct soliton solutions or an inverse scattering theory. 
This will be left for future work. 
%%%%%%%%%%%%%%%%%%%%%%%%%%%%%%%%%%%%%%%%%%%%%%%
%%%%%%%%%%%%%%%%%%%%%%%%%%%%%%%%%%%%%%%%%%%%%%%
%%%%%%%%%%%%%%%%%%%%%%%%%%%%%%%%%%%%%%%%%%%%%
%%%%%%%%%%%%%%%%%%%%%%%%%%%%%%%%%%%%%%%%%%%%%
\section{Fully $PT$ symmetric, partially $PT$ symmetric and partial reverse space-time nonlocal Davey-Stewartson system}
\label{PTDSeqn}
%%%%%%%%%%%%%%%%%%%%%%%%%%%%%%%%%%%%%%%%%%%%%
%%%%%%%%%%%%%%%%%%%%%%%%%%%%%%%%%%%%%%%%%%%%%
In this section we show that the (DS) system (\ref{q-DS}) and (\ref{r-DS}) admit yet other types of symmetry reductions. These new symmetry reductions fall into three distinct categories: (i) full $PT$ symmetry, (ii) partial $PT$ symmetry and (iii) partial reverse 
space-time symmetry. Generally speaking, a linear or nonlinear partial differential equation (PDE) is said to be $PT$ symmetric if it is invariant under the combined action of the (not necessarily linear) $PT$ operator. In $(1+1)$ dimensions, this amounts to invariance under the joint transformation $x\rightarrow -x, t\rightarrow -t$ and complex conjugation. The situation for the $(2+1)$ case is more rich. Here, one can talk about two different types of $PT$ symmetries: full and partial. 
If we denote by ${\bf x}\equiv (x,y),$ then a linear or nonlinear PDE in $(2+1)$ dimensions is said to be fully $PT$ symmetric if it is invariant under the combined 
transformation ${\bf x} \rightarrow -{\bf x}$ (parity operator $P$), $t\rightarrow -t$ plus complex conjugation ($T$ operator). Note that the space reflection is performed in both space coordinates. On the other hand, a linear or nonlinear PDE in $(2+1)$ dimensions is said to be partially $PT$ symmetric if it is invariant under the combined 
transformation $(x,y) \rightarrow (-x,y)$ or $(x,y) \rightarrow (x,-y)$, $t\rightarrow -t$ plus complex conjugation. Partially $PT$ symmetric optical potentials have been studied 
in \cite{yang-partialPT} and shown that such potentials exhibit pure real spectra and can support (in the presence of cubic type nonlinearity) continuous families of solitons. Below, we use these new symmetry reductions to derive the corresponding Davey-Stewartson like equations.
%%%%%%%%%%%%%%%%%%%%%%%%%%%%%%%%%%%%%%%%%%%%%
%%%%%%%%%%%%%%%%%%%%%%%%%%%%%%%%%%%%%%%%%%%%%
%%%%%%%%%%%%%%%%%%%%%%%%%%%%%%%%%%%%%%%%%%%
\subsection{Partially $PT$ symmetric nonlocal Davey-Stewartson equation}
%%%%%%%%%%%%%%%%%%%%%%%%%%%%%%%%%%%%%%%%%%
Under the symmetry reduction 
\begin{equation}
\label{sym-reduction-DS-PT-part}
r(x,y,t) = \sigma q^*(-x,y,t)\;,
\end{equation}
it can be shown that system (\ref{q-DS}) and (\ref{r-DS}) are compatible and lead to the partially $PT$ symmetric nonlocal Davey-Stewartson 
equation (\ref{complex-space-time-nonloc-DS-PT-part}) which, for the convenience of the reader we rewrite again:
%%%%%%%%%%%%%%%%%%%%%
%%%%%%%%%%%%%%%%%%%%%%
%%%%%%%%%%%
\begin{eqnarray}
\label{complex_DS-q-PT-part}
iq_t(x,y,t) &+& \frac{1}{2}\left[\gamma^2 q_{xx}(x,y,t)  + q_{yy}(x,y,t) \right]
\nonumber \\ 
&+&\sigma  q^2(x,y,t) q^*(-x,y,t) = \phi (x,y,t) q(x,y,t)\;,
\end{eqnarray}
%%%%%%%%%%%%%%%
\begin{eqnarray}
\label{complex_DS-phi-PT-part}
\phi_{xx} (x,y,t) - \gamma^2 \phi_{yy} (x,y,t) 
= 2\sigma \left[ q(x,y,t) q^*(-x,y,t)\right]_{xx} \;.
\end{eqnarray}
%%%%%%%%%%%%%%%%
Note that from Eq.~(\ref{complex_DS-phi-PT-part}) it follows that the potential $\phi$ has a solution that satisfies the relation $\phi (x,y,t) = \phi^* (-x,y,t),$ in other words, the potential satisfies the partial $PT$ symmetry requirement. Note, here and below we could have also considered the partial $PT$ reduction:
\begin{equation}
\label{sym-reduction-DS-PT-part-y}
r(x,y,t) = \sigma q^*(x,-y,t)\;,
\end{equation}
which would lead to another DS type equation. 
%%%%%%%%%%%%%%%%%%%%%%%%%%%%%%%%%%%%%%%%%%%%
%%%%%%%%%%%%%%%%%%%%%%%%%%%%%%%%%%%%%%%%%%%%%
\subsection{Partial reverse space-time nonlocal Davey-Stewartson equation}
%%%%%%%%%%%%%%%%%%%%%%%%%%%%%%%%%%%%%%%%%%%%%
Another new symmetry reduction which was not noticed in the literature so far is the partially reverse space-time nonlocal reduction given by
\begin{equation}
\label{sym-reduction-DS-time-part}
r(x,y,t) = \sigma q(-x,y,-t)\;,\;\;\;\;\; \sigma = \mp 1.
\end{equation}
With this symmetry condition, system (\ref{q-DS}) and (\ref{r-DS}) are consistent and give rise to the following partially reverse space-time nonlocal DS system of equation
%%%%%%%%%%%%%%%%%%%%%
%%%%%%%%%%%%%%%%%%%%%%
%%%%%%%%%%%
\begin{eqnarray}
\label{complex_DS-q-time-part}
iq_t(x,y,t) &+& \frac{1}{2}\left[\gamma^2 q_{xx}(x,y,t)  + q_{yy}(x,y,t) \right]
\nonumber \\
&+& \sigma  q^2(x,y,t) q(-x,y,-t) =  \phi (x,y,t) q(x,y,t) \;,
\end{eqnarray}
%%%%%%%%%%%%%%%
\begin{eqnarray}
\label{complex_DS-phi-time-part}
\phi_{xx} (x,y,t) - \gamma^2 \phi_{yy} (x,y,t) 
= 2 \sigma \left[ q(x,y,t) q(-x,y,-t)\right]_{xx} \;.
\end{eqnarray}
%%%%%%%%%%%
Note that from Eqns.~(\ref{complex_DS-q-time-part}) and (\ref{complex_DS-phi-time-part}) it follows that the potential $\phi$ has a solution that satisfies 
$\phi (x,y,t) = \phi (-x,y,-t).$ Again, one can consider the partial reverse space-time reduction:
\begin{equation}
\label{sym-reduction-DS-reverse-part-y}
r(x,y,t) = \sigma q(x,-y,-t)\;,
\end{equation}
and obtain the corresponding DS equation. In this paper, we will not go into further detail regarding the integrability properties of the above systems nor will we construct soliton solutions or an inverse scattering theory. 
This will be left for future work. We close this section by mentioning that the fully $PT$ symmetric nonlocal Davey-Stewartson equation was obtained by Fokas in \cite{Fokas2016}. Indeed, under the symmetry condition $r({\bf x},t) = \sigma q^*(-{\bf x},t)$ the system (\ref{q-DS}) and (\ref{r-DS}) are again compatible and lead to the following $PT$ symmetric nonlocal DS equation \cite{Fokas2016}:
%%%%%%%%%%%
\begin{eqnarray}
\label{complex_DS-q-space}
iq_t({\bf x},t) &+& \frac{1}{2}\left[\gamma^2 q_{xx}({\bf x},t)  + q_{yy}({\bf x},t) \right]
+\sigma  q^2({\bf x},t) q^*(-{\bf x},t)  
\nonumber \\
&=&  \phi ({\bf x},t) q({\bf x},t) = 0\;,
\end{eqnarray}
%%%%%%%%%%%%%%%
\begin{eqnarray}
\label{complex_DS-phi-space}
\phi_{xx} ({\bf x},t) - \gamma^2 \phi_{yy} ({\bf x},t) = 2\sigma \left[ q({\bf x},t) q^*(-{\bf x},t)\right]_{xx} \;,
\end{eqnarray}
%%%%%%%%%%%
with the potential $\phi ({\bf x},t)$ satisfying the $PT$ symmetry condition: 
$\phi^* (-{\bf x},t) = \phi ({\bf x},t).$\\\\
%%%%%%%%%%%%%%%%%%%%%%%%%
%%%%%%%%%%%%%%%%%%%%%%%%%%%%%%%%%%%%%%%%%%%%%%
%%%%%%%%%%%%%%%%%%%%%%%%%%%%%%%%%%%%%%%%%%%%%%%
In summary, like the integrable NLS-type equations, the Davey-Stewartson system (\ref{q-DS}), (\ref{r-DS}) and (\ref{phi-DS}) admit six different symmetry reductions which we list below: 
%%%%%%%%%%
\begin{enumerate}
%%%%%%%%%%
\item
Classical $r({\bf x},t) = \sigma q^*({\bf x},t)$ observed in \cite{AblHbmn}, \\
%%%%%%%%%
\item
Fully $PT$ symmetric:  $r({\bf x},t) = \sigma q^*(-{\bf x},t)$ reported in \cite{Fokas2016},\\
%%%%%%%%%%
\item
Partially $PT$ symmetric:  $r(x,y,t) = \sigma q^*(-x,y,t)$ or $r(x,y,t) = \sigma q^*(x,-y,t)$
found in this paper,\\
%%%%%%%%%%
\item
Reverse space-time symmetry $r({\bf x},t) = \sigma q(-{\bf x},-t)$ found in this paper,\\
%%%%%%%%%%
\item
Partial reverse space-time symmetry $r(x,y,t) = \sigma q(-x,y,-t)$ or $r(x,y,t) = \sigma q(x,-y,-t)$
found in this paper,\\
%%%%%%%%%%
\item
Reverse time symmetry $r({\bf x},t) = \sigma q({\bf x},-t)$ found in this paper.\\
\end{enumerate}
%%%%%%%%%%%%%%%%%%%%%%%%%%%%%%%%%%%%%%%%%
It would be interesting for future research direction to study the solutions and possible wave collapse properties (or lack of) for each of the reported new reductions.
%%%%%%%%%%%%%%%%%%%%%%%%%%%%%%%%%%%%%%%%%%%%%%%%%%%%%%%%%%%%%%%%%%%%%%%%%%%%%%%%%%%%%%%%%%
\section{(1+1) dimensional reverse space-time nonlocal multi-wave and three-wave equations}
\label{Nwave}
%%%%%%%%%%%%%%%%%%%%%%%%%%%%%%%%%%%%%%%%%%%%
In this section we derive the reverse space-time and reverse time nonlocal multi-wave equation and its physically important reduction to three wave equations. The idea is to generalize the 
$2\times 2$ scattering AKNS scattering problem (\ref{AKNS}) and its associated time evolution (\ref{time-AKNS}) to an $n\times n$ matrix form and obtain, after following similar compatability procedure, the corresponding multi interacting nonlinear, nonlocal (in space and time) wave equation. A physically relevant reduction of the more general case, i.e., three-wave equation will be also derived. Our approach follows that given by Ablowitz and Haberman \cite{AblHbmn}. 
An $n\times n$ generalization of the scattering problem (\ref{AKNS} - \ref{LAX-1}) 
is given by
%%%%%%%%%%%%%%%%%%
\begin{equation}
\label{AKNS-N-by-N}
{\bf v}_x = i k {\mathsf D} {\bf v} + {\mathsf N} {\bf v}\;,
\end{equation}
%%%%%%%%%%%%%%%%%%
where ${\bf v}$ is a column vector of length $n$, i.e., ${\bf v}=(v_1, v_2, \cdots, v_n)^T$  where as before, $T$ denotes matrix transpose. Furthermore, ${\mathsf D}$ 
and ${\mathsf N}$ are $n\times n$ matrices with ${\mathsf D}$ being a diagonal constant, matrix, i.e., ${\mathsf D} \equiv \text{diag} (d_1, d_2, \cdots, d_n)$, ~$d_n>d_{n-1} \cdots >d_1,$ and ${\mathsf N}$ has zero entries on the main diagonal; i.e. in matrix element form $N_{\ell, \ell}=0$. 
The time evolution associated with (\ref{AKNS-N-by-N}) is given by
%%%%%%%%%%%%%%%%%%%%
\begin{equation}
\label{time-AKNS-N-by-N}
{\bf v}_t = \mathsf{Q} {\bf v}\;,
\end{equation}
%%%%%%%%%%%%%%%%%%%%
with $\mathsf{Q}$ being an $n\times n$ matrix which depends on the components of the ``potential" matrix $\mathsf{N}$ and the assumed time-independent 
spectral parameter $k.$ As in the $2\times 2$ case, the compatibility 
condition ${\bf v}_{xt} = {\bf v}_{tx}$ yields the matrix equation
%%%%%%%%%%%%%%%%%%%%
\begin{equation}
\label{AKNS-N-by-N-comp}
\mathsf{Q}_x  - \mathsf{N}_t =  ik  [{\mathsf D}, \mathsf{Q}] + [{\mathsf N}, \mathsf{Q}] \;,
\end{equation}
%%%%%%%%%%%%%%%%%%%%
where $[{\mathsf A}, \mathsf{B}] \equiv {\mathsf A}\mathsf{B} - {\mathsf B}\mathsf{A}.$ If one now expands the matrix $\mathsf{Q}$ in a first order polynomial in the spectral parameter $k$, $\mathsf{Q} = \mathsf{Q}_0 + k \mathsf{Q}_1$ then, after some algebra,  one finds $\mathsf{Q}_{1\ell j} \equiv q_\ell \delta_{\ell j}$ and 
$\mathsf{Q}_{0\ell j} = a_{\ell j} \mathsf{N}_{\ell j}$ %and zero otherwise 
and $a_{\ell j}  = -\frac{i(q_\ell - q_j)}{(d_\ell - d_j)} = a_{j\ell}.$ We want $a_{\ell j} \in \mathbb{R}$ hence $q_j, j=1, \cdots n$ are purely imaginary.
The time evolution of the matrix elements 
$\mathsf{N}_{\ell j}, \ell , j = 1,2, \cdots, n$ is found at order $k^0$ and given by
%%%%%%%%%%%%%%%%%%%%
\begin{equation}
\label{E:k^{0}}
\mathsf{N}_{\ell j,t}(x,t) - a_{\ell j} \mathsf{N}_{\ell j,x} (x,t)
=
 \sum_{m=1}^n (a_{\ell m} - a_{m j}) \mathsf{N}_{\ell m}(x,t) \mathsf{N}_{m j}(x,t) \;.
\end{equation}
%%%%%%%%%%%%%%%%%%%%
Equation (\ref{E:k^{0}}) was derived in \cite{AblHbmn} and governs the time evolution of generic ``potential" matrix elements $\mathsf{N}_{\ell j}.$ 
%%%%%%%%%%%%%%%%%%%%%%%%%%%%%%%%%%%%%%%%%%%%
\subsection{Classical multi-wave reduction: 
$\mathsf{N}_{\ell j}(x,t) = \sigma_{\ell j} \mathsf{N}^*_{j \ell} (x,t)$}
%%%%%%%%%%%%%%%%%%%%%%%%%%%%%%%%%%%%%%%%%%%
In \cite{AblHbmn} it was shown that the system of equations (\ref{E:k^{0}}) admits the following symmetry reduction
%%%%%%%%%%%%%%%%%%%%
\begin{equation}
\label{N-wave-sym-loc}
\mathsf{N}_{\ell j}(x,t) = \sigma_{\ell j} \mathsf{N}^*_{j \ell} (x,t)  \;,
\end{equation}
%%%%%%%%%%%%%%%%%%%%%
where $\sigma_{\ell j}$ are constants satisfying 
\[ \sigma^2_{\ell j} =1,  ~~\sigma_{\ell m} \sigma_{m j} = - \sigma_{\ell j}, \]
for all $m, \ell , j = 1,2, \cdots, n$ and real $a_{\ell m}.$ 
That is to say, $\mathsf{N}_{\ell j}(x,t)$ and $\mathsf{N}^*_{j \ell} (x,t)$ satisfy the same equation (\ref{E:k^{0}}) (and its complex conjugate) thus reducing the number of equation by half; there are $n(n-1)/2$ interacting nonlinear wave equations. 
%%%%%%%%%%%%%%%%%%%%%%%%%%%%%%%%%%%%%%%%%
\subsection{Classical three wave interaction equations}
%%%%%%%%%%%%%%%%%%%%%%%%%%%%%%%%%%%%%%%%%
The physically relevant and important local three wave interaction system is next derived.  
In this case $n=3$ and the ``nonlinear" matrix $\mathsf{N}$ is assumed to have the generic 
form (note that $\mathsf{N}_{j j} =0, j=1,2,3$)
%%%%%%%%%%%%%%%%%%%%
\begin{equation}
\label{N-matrix-1}
\mathsf{N} (x,t) =
\left( {\begin{array}{ccc}
   0 &\mathsf{N}_{12}(x,t)&\mathsf{N}_{13}(x,t)\\       
   \mathsf{N}_{21}(x,t) &0&\mathsf{N}_{23}(x,t)\\  
   \mathsf{N}_{31}(x,t) &\mathsf{N}_{32}(x,t)&0
    \end{array} } \right)
\;.
\end{equation}
With the symmetry \ref{N-wave-sym-loc}, one can reduce the number of independent variables in (\ref{N-matrix-1}) and write 
\begin{equation}
\label{N-matrix-2}
\mathsf{N} (x,t) =
\left( {\begin{array}{ccc}
   0 &\mathsf{N}_{12}(x,t)&\mathsf{N}_{13}(x,t)\\       
  \sigma_1 \mathsf{N}_{12}^*(x,t) &0&\mathsf{N}_{23}(x,t)\\  
  \sigma_2  \mathsf{N}_{13}^*(x,t) &\sigma_3 \mathsf{N}_{23}^*(x,t)&0
    \end{array} } \right)
\;,
\end{equation}
where 
\[\frac{\sigma_1 \sigma_1}{\sigma_2}=1, ~~\sigma_j= \pm 1, ~j=1,2,3. \]
Thus, the number of nonlinear equations is reduced from 6 to 3. Next, we consider the 
following transformation of variables, 
\begin{equation*}
\mathsf{N}_{12}(x,t) = -i\frac{Q_{3}(x,t)}{\sqrt{\beta_{13}\beta_{23}}},
\end{equation*}
%%%%%%%%%
\begin{equation*}
\mathsf{N}_{31}(x,t) = -i\frac{Q_{2}(x,t)}{\sqrt{\beta_{12}\beta_{23}}}, 
\end{equation*}
%%%%%%%%%
\begin{equation*}
\mathsf{N}_{23}(x,t) =  i\frac{Q_{1}(x,t)}{\sqrt{\beta_{12}\beta_{13}}},
%%%%%%%%%
\end{equation*}
%%%%%%%%
\begin{equation*}
\mathsf{N}_{13}(x,t)=-\gamma_1\gamma_3 \mathsf{N}^{*}_{31}(x,t), 
\end{equation*}
%%%%%%%%%
\begin{equation*}
\mathsf{N}_{32}(x,t)= \gamma_3\gamma_2 \mathsf{N}^{*}_{23}(x,t),
\end{equation*}
%%%%%%%%
\begin{equation*}
\mathsf{N}_{21}(x,t)=\gamma_1\gamma_2\mathsf{N}^{*}_{12}(x,t),
\end{equation*}
where
\begin{equation*}
\begin{split}
&\beta_{lj}:=d_{l}-d_{j}=-c_{l} + c_{j}, \Rightarrow d_j = -c_j 
\Rightarrow ~ c_3 > c_2 > c_1\\
&\gamma_j=-i\frac{c_1c_2c_3}{c_j}.
\end{split}
\end{equation*}
This results in the classical (local) three wave interaction equations
\begin{equation}
\begin{split}
&Q_{1,t}(x,t) + c_{1}Q_{1,x}(x,t) = i \gamma_1Q^{*}_{2}(x,t)  Q^{*}_{3}(x,t),\\
&Q_{2,t}(x,t) + c_{2}Q_{2,x}(x,t) = i \gamma_2Q^{*}_{1}(x,t) Q^{*}_{3}(x,t),\\
&Q_{3,t}(x,t) + c_{3}Q_{3,x}(x,t) = i \gamma_3Q^{*}_{1}(x,t) Q^{*}_{2}(x,t),
\end{split}
\end{equation}
where
\[ c_3>c_2>c_1, ~~\gamma_1 \gamma_2 \gamma_3=-1, \gamma_j= \pm 1, ~j=1,2,3.\]
From these equations, we can derive the conserved quantities
\begin{equation}
\begin{split}
%%%%%%%%%%%%%%%%%%
&\gamma_1\int_{-\infty}^{\infty}|Q_{1}(x,t)|^2dx
- \gamma_2 \int_{-\infty}^{\infty}|Q_{2}(x,t)|^2dx=\text{constant},\\
%%%%%%%%%%%%%%%%%%
&\gamma_2\int_{-\infty}^{\infty}|Q_{2}(x,t)|^2dx
-\gamma_3 \int_{-\infty}^{\infty}|Q_{3}(x,t)|^2dx=\text{constant},\\
%%%%%%%%%%%%%%%%%%
&\gamma_1\int_{-\infty}^{\infty}|Q_{1}(x,t)|^2dx
-\gamma_3 \int_{-\infty}^{\infty}|Q_{3}(x,t)|^2dx=\text{constant}.
\end{split}
\end{equation}
Positive definite energy occurs when we take two $\gamma_j$'s of different sign. This results in the `decay instability' case. If all three $\gamma_j=-1$ then the above does not lead to a positive definite energy -- this is the  `explosive instability' case.
Next we show that the system (\ref{E:k^{0}}) admits {\it new space-time} nonlocal symmetry reductions leading to nonlocal multi-wave equations. We will discuss two reductions.
%%%%%%%%%%%%%%%%%%%%%%%%%%%%%%%%%%%%%%%%%
\subsection{The complex reverse space-time multi-wave reduction: 
$\mathsf{N}_{\ell j}(x,t)= \sigma_{\ell j} \mathsf{N}_{j \ell}^*(-x,-t)$}
%%%%%%%%%%%%%%%%%%%%%%%%%%%%%%%%%%%%%%%%%
In this section we show that the system of multi-interacting waves admits a new nonlocal symmetry reduction. Later, we shall derive a simple model of a nonlocal three-wave equation. We substitute in Eq.~(\ref{E:k^{0}}) the new symmetry relation
\begin{equation}
\mathsf{N}_{\ell j}(x,t) = \sigma_{\ell j} \mathsf{N}_{j \ell}^*(-x,-t),
\end{equation} 
and call $x'=-x, t'=-t$ and find:
\begin{equation}
\label{ECC2}
-(N^*_{j \ell,t'}-a_{lj}N^*_{j \ell,x'})(x',t')= \sum_{m=1}^{n}(a_{\ell m}-a_{mj}) \frac{\sigma_{\ell m} \sigma_{mj}}{\sigma_{\ell j}} N^*_{m \ell}(x',t')N^*_{jm}(x',t')
\end{equation}
Under the condition
\[ \frac{\sigma_{\ell m} \sigma_{mj}}{\sigma_{\ell j}} =+1, \]
equation (\ref{ECC2}) agrees with the complex conjugate of equation (\ref{E:k^{0}})
 with interchanged indices.
%%%%%%%%%%%%%%%%%%%%%%%%%%%%%%%%%%%%%%%%
\subsection{Complex reverse space-time three wave equations}
%%%%%%%%%%%%%%%%%%%%%%%%%%%%%%%%%%%%%%%
With the symmetry reduction $\mathsf{N}_{21}=\sigma_{1}\mathsf{N}_{12}^{*}(-x,-t)$, 
$\mathsf{N}_{31}=\sigma_{2}\mathsf{N}_{13}^{*}(-x,-t)$ 
and $\mathsf{N}_{32}=\sigma_{3}\mathsf{N}_{23}^{*}(-x,-t)$ and assuming that $a_{lj}$ are real, where $\sigma_{1}$, $\sigma_{2}$ and $\sigma_{3}$ are chosen as real numbers with
\begin{equation}
\frac{\sigma_{1}\sigma_{3}}{\sigma_{2}}=1,  ~~\sigma_j= \pm 1, ~j=1,2,3,
\end{equation}
equation (\ref{E:k^{0}}) may be put into a set of nonlocal three-wave interaction equations by a suitable scaling of variables. For example, we find the system
\begin{equation}
\begin{split}
&Q_{1,t}(x,t) + c_{1}Q_{1,x}(x,t) =~~\sigma_{3}Q^{*}_{2}(-x,-t)Q^{*}_{3}(-x,-t),\\
&Q_{2,t}(x,t) +c_{2}Q_{2,x}(x,t) =-\sigma_{2}Q^{*}_{1}(-x,-t)Q^{*}_{3}(-x,-t),\\
&Q_{3,t}(x,t) +c_{3}Q_{3,x}(x,t) =~~\sigma_{1}Q^{*}_{1}(-x,-t)Q^{*}_{2}(-x,-t),
\end{split}
\end{equation}
if we take
%%%%%%%%%%%
\begin{equation*}
\mathsf{N}_{12}(x,t) = -\frac{Q_{3}(x,t)}{\sqrt{\beta_{13}\beta_{23}}}, 
\end{equation*}
%%%%%%%%%%%
\begin{equation*}
\mathsf{N}_{31}(x,t) = -\frac{Q_{2}(x,t)}{\sqrt{\beta_{12}\beta_{23}}}, 
\end{equation*}
%%%%%%%%%%%
\begin{equation*}
\mathsf{N}_{23}(x,t) = -\frac{Q_{1}(x,t)}{\sqrt{\beta_{12}\beta_{13}}},
\end{equation*}
%%%%%%%%%
\begin{equation*}
\mathsf{N}_{13}(x,t) = \sigma_{2} \mathsf{N}^{*}_{31}(-x,-t), 
\end{equation*}
%%%%%%%%%%
\begin{equation*}
\mathsf{N}_{32}(x,t) = \sigma_{3} \mathsf{N}^{*}_{23}(-x,-t), 
\end{equation*}
%%%%%%%%%%%
\begin{equation*}
\mathsf{N}_{21}(x,t) = \sigma_{1} \mathsf{N}^{*}_{12}(-x,-t),
\end{equation*}
%%%%%%%%%%%%
where
\begin{equation*}
\begin{split}
& \beta_{lj}:=d_{l}-d_{j}=-c_{l}+c_{j}, \Rightarrow 
~d_j=-c_j \Rightarrow ~d_{1}=-c_{1}, \\
&
d_{2}=-c_{2}, ~~d_{3}=-c_{3} \\
&q_{1}=-ic_{2}c_{3}, \ q_{2}=-ic_{1}c_{3}, 
\ q_{3}=-ic_{1}c_{2}, \\
&\ a_{12}=-c_{3}, \ a_{13}=-c_{2}, \ a_{23}=-c_{1}, ~c_3>c_2>c_1.
\end{split}
\end{equation*}
Directly from the equations, we can derive the conserved quantities
\begin{equation}
\begin{split}
&\sigma_{2}\int_{-\infty}^{\infty}Q_{1}(x,t)Q_{1}^{*}(-x,-t)dx+\sigma_{3}\int_{-\infty}^{\infty}Q_{2}(x,t)Q_{2}^{*}(-x,-t)dx=\text{constant},\\
&\sigma_{2}\int_{-\infty}^{\infty}Q_{3}(x,t)Q_{3}^{*}(-x,-t)dx+\sigma_{1}\int_{-\infty}^{\infty}Q_{2}(x,t)Q_{2}^{*}(-x,-t)dx=\text{constant},\\
&\sigma_{1}\int_{-\infty}^{\infty}Q_{1}(x,t)Q_{1}^{*}(-x,-t)dx-\sigma_{3}\int_{-\infty}^{\infty}Q_{3}(x,t)Q_{3}^{*}(-x,-t)dx=\text{constant}.
\end{split}
\end{equation}
Thus there appears to be no positive definite conserved quantities in the above equations; in the general case there likely will be blowup solutions.
%%%%%%%%%%%%%%%%%%%%%%%%%%%%%%%%%%%%%%%%%%%%
\subsection{The reverse space-time multi-wave reduction: $\mathsf{N}_{\ell j}(x,t)= \sigma_{\ell j} \mathsf{N}_{j \ell}(-x,-t)$}
%%%%%%%%%%%%%%%%%%%%%%%%%%%%%%%%%%%%%%%%%%%
If we substitute in Eq.~(\ref{E:k^{0}}) 
\begin{equation}
\mathsf{N}_{\ell j}(x,t) = \sigma_{\ell j} \mathsf{N}_{j \ell}(-x,-t)\;,
\end{equation}
and let $x'=-x, t'=-t$ then we find
\begin{equation}
\label{ECC12}
-(N_{j \ell ,t'}-a_{lj}N_{j \ell,x'})(x',t')= \sum_{m=1}^{n}(a_{\ell m}-a_{mj}) \frac{\sigma_{\ell m} \sigma_{mj}}{\sigma_{\ell j}} N_{m \ell}(x',t')N_{jm}(x',t')
\end{equation}
Under the condition
\[ \frac{\sigma_{\ell m} \sigma_{mj}}{\sigma_{\ell j}} =1, \]
Eq.~(\ref{ECC12}) agrees with Eq.~(\ref{E:k^{0}}) by interchanging the indices and without taking the complex conjugate.
%%%%%%%%%%%%%%%%%%%%%%%%%%%%%%%%%%%%%%%%%%%
\subsection{Reverse space-time three wave equations}
%%%%%%%%%%%%%%%%%%%%%%%%%%%%%%%%%%%%%%%%%%%
Under the symmetry reduction $\mathsf{N}_{21}=\sigma_{1} \mathsf{N}_{12}(-x,-t)$, $\mathsf{N}_{31}=\sigma_{2}\mathsf{N}_{13}(-x,-t)$ and $\mathsf{N}_{32}=\sigma_{3}\mathsf{N}_{23}(-x,-t)$, where $\sigma_{1}$, $\sigma_{2}$ and $\sigma_{3}$ are chosen as real numbers, we have
\begin{equation}
\frac{\sigma_{1}\sigma_{3}}{\sigma_{2}}=1,  ~~\sigma_j= \pm 1, ~j=1,2,3.
\end{equation} 
As above, Eq.~(\ref{E:k^{0}}) may be put into a standard set of nonlocal three-wave interaction equations by a suitable scaling of variables. For example, we find the system
\begin{equation}
\begin{split} 
&Q_{1,t}(x,t) + c_{1}Q_{1,x}(x,t) = \sigma_{3}Q_{2}(-x,-t)Q_{3}(-x,-t),\\
&Q_{2,t}(x,t) + c_{2}Q_{2,x}(x,t) =-\sigma_{2}Q_{1}(-x,-t)Q_{3}(-x,-t),\\
&Q_{3,t}(x,t) + c_{3}Q_{3,x}(x,t) =\sigma_{1}Q_{1}(-x,-t)Q_{2}(-x,-t),
\end{split}
\end{equation}
if we take 
%%%%%%%%% 
\begin{equation*}
\mathsf{N}_{12}(x,t) = -\frac{Q_{3}(x,t)}{\sqrt{\beta_{13}\beta_{23}}}, 
\end{equation*}
%%%%%%%%%%
\begin{equation*}
 \mathsf{N}_{31}(x,t) =-\frac{Q_{2}(x,t)}{\sqrt{\beta_{12}\beta_{23}}}, 
 \end{equation*}
 %%%%%%%%%%
 \begin{equation*}
  \mathsf{N}_{23}(x,t) =-\frac{Q_{1}(x,t)}{\sqrt{\beta_{12}\beta_{13}}},
  \end{equation*}
  %%%%%%%%%%%%%
  \begin{equation*}
\mathsf{N}_{13}=\sigma_{2}N_{31}(-x,-t), 
\end{equation*}
%%%%%%%%%%%%%
\begin{equation*}
\mathsf{N}_{32}=\sigma_{3}N_{23}(-x,-t), 
\end{equation*}
%%%%%%%%%%%%%%
\begin{equation*}
\mathsf{N}_{21}=\sigma_{1}N_{12}(-x,-t),
\end{equation*}
%%%%%%%%%%%%%%%
where 
\begin{equation*}
\begin{split}
& \beta_{lj}:=d_{l}-d_{j}=-c_{l}+c_{j} \Rightarrow d_{1}=-c_{1},
\ \ \  d_{2}=-c_{2}, \ \ \ d_{3}=-c_{3} \\
&q_{1}=-ic_{2}c_{3}, \ q_{2}=-ic_{1}c_{3}, 
\ q_{3}=-ic_{1}c_{2}, \\
&\ a_{12}=-c_{3}, \ a_{13}=-c_{2}, \ a_{23}=-c_{1}, ~c_3>c_2>c_1.
\end{split}
\end{equation*}
Directly from the equations, we can derive the conserved quantities
\begin{equation}
\begin{split}
&\sigma_{2}\int_{-\infty}^{\infty}Q_{1}(x,t)Q_{1}(-x,-t)dx+\sigma_{3}\int_{-\infty}^{\infty}Q_{2}(x,t)Q_{2}(-x,-t)dx=\text{constant},\\
&\sigma_{2}\int_{-\infty}^{\infty}Q_{3}(x,t)Q_{3}(-x,-t)dx+\sigma_{1}\int_{-\infty}^{\infty}Q_{2}(x,t)Q_{2}(-x,-t)dx=\text{constant},\\
&\sigma_{1}\int_{-\infty}^{\infty}Q_{1}(x,t)Q_{1}(-x,-t)dx-\sigma_{3}\int_{-\infty}^{\infty}Q_{3}(x,t)Q_{3}(-x,-t)dx=\text{constant}.
\end{split}
\end{equation}
From the above there appears to be no positive definite conserved quantities; it is expected that this set of equations will have blowup solutions. In future work, we aim to study the integrability properties of this nonlocal three-wave equation and construct soliton solutions.
%%%%%%%%%%%%%%%%%%%%%%%%%%%%%%%%%%%%%%%%%%%%
%%%%%%%%%%%%%%%%%%%%%%%%%%%%%%%%%%%%%%%%%%%%
%%%%%%%%%%%%%%%%%%%%%%%%%%%%%%%%%%%%%%%%%%%%%%%%%%%%%%%%%%%%%%%%%%%%%%%%%%%%%%%%%%%%%%%%%%
\section{(2+1) dimensional space-time nonlocal multi-wave and three-wave equations}
\label{Nwave-2}
%%%%%%%%%%%%%%%%%%%%%%%%%%%%%%%%%%%%%%%%%%%%
In this section we extend the analysis presented in Sec.~\ref{Nwave} to two space dimensions and derive the classical (local) multi-wave interaction equations and the space-time (as well as the time only) nonlocal multi-wave equations. The idea is to generalize the matrix scattering problem (\ref{AKNS-N-by-N}) by replacing the eigenvalue term by a derivative in the transverse $y$ coordinate. Thus, we start from the multi-dimensional generalized scattering problem
%%%%%%%%%%%%%
\begin{equation}
\label{multi-scat-space}
{\bf v}_{x} = {\bf B} {\bf v}_{y} + {\bf N}{\bf v},
\end{equation}
%%%%%%%%%%%%%
\begin{equation}
\label{multi-scat-time}
{\bf v}_t = {\bf C} {\bf v}_y + {\bf Q} {\bf v},
\end{equation}
%%%%%%%%%%%%%%
where ${\bf v}$ is a column vector of length $n, {\bf B}, {\bf N}, {\bf C}$ and ${\bf Q}$ 
are $n\times n$ matrices with ${\bf B}$ being a real constant diagonal matrix 
given by ${\bf B} = {\text diag}(b_1, b_2, \cdots, b_n)$ and ${\bf N}$ is such 
that ${\bf N}_{jj}=0, j=1,2, \cdots, n.$ 
From the compatibility condition ${\bf v}_{xt} = {\bf v}_{tx}$ one finds expressions for the mixed derivatives ${\bf v}_{yt}$ and ${\bf v}_{yx}.$ After setting the coefficients of the independent terms ${\bf v}_{yy}, {\bf v}_y$ and ${\bf v}$ to zero one finds
%%%%%%%%%%%%%%%%%
\begin{equation}
\label{E:system 2}
[{\bf C}, {\bf B}] = 0,
\end{equation}
%%%%%%%%%%%%%%%%
\begin{equation}
\label{E:system 3}
[{\bf Q}, {\bf B}] + [{\bf C}, {\bf N}] + {\bf C}_{x} - {\bf B} {\bf C}_{y} = 0,
\end{equation}
%%%%%%%%%%%%%%%%
\begin{equation}
\label{E:system 4}
{\bf N}_{t} = [{\bf Q}, {\bf N}] + {\bf Q}_{x} - {\bf B} {\bf Q}_{y} + {\bf C} {\bf N}_{y}.
\end{equation}
%%%%%%%%%%%%%%%%%%
With the choice
%%%%%%%%%%%%%%%%
\begin{equation}
{\bf B}_{lj} = b_{l}\delta_{lj}, \;\;\;\; {\bf C}_{lj} = c_{l}\delta_{lj} ,
\end{equation}
where $b_{l}$ and $c_{l}$ are taken to be real constants then Eq.~(\ref{E:system 2}) is satisfied. In this case, (\ref{E:system 3}) yields ${\bf Q}_{lj} = \alpha_{lj} {\bf N}_{lj}\ (l\neq l)$, 
where $\alpha_{lj}=\frac{c_{l}-c_{j}}{b_{l}-b_{j}}=\alpha_{jl}$. 
Moreover, ${\bf Q}_{ll}=q_{l}$, $q_{l}-q_{j}=ik(d_{l}-d_{j})\alpha_{lj}$ and $\beta_{lj}=c_{l}-\alpha_{lj}b_{l}=(c_lb_j-c_jb_l)/(b_j-b_l)=\beta_{jl}$. Hence, we have the compatible 
two-dimensional nonlinear wave equation
%%%%%%%%%%%%%%%%%%%%%%%%
\begin{equation}
\label{E:compatible 2 dimensional}
{\bf N}_{lj,t} - \alpha_{lj} {\bf N}_{lj,x} - \beta_{lj} {\bf N}_{lj,y} 
= 
\sum_{m=1}^{n}(\alpha_{lm}-\alpha_{mj}) {\bf N}_{lm} {\bf N}_{mj}.
\end{equation}
%%%%%%%%%%%%%%%%%%%%%%%%
%%%%%%%%%%%%%%%%%%%%%%%%%%%%%%%%%%%%%%%%%%%%%
%%%%%%%%%%%%%%%%%%%%%%%%%%%%%%%%%%%%%%%%%%%%%
\subsection{Classical multi-wave reduction: 
${\bf N}_{lj}({\bf x},t) = \sigma_{lj} {\bf N}_{jl}^*({\bf x},t)$}
%%%%%%%%%%%%%%%%%%%%%%%%%%%%%%%%%%%%%%%%%%%%%
%%%%%%%%%%%%%%%%%%%%%%%%%%%%%%%%%%%%%%%%%%%%%
For the ease of presentation we shall use the notation ${\bf x}\equiv (x,y).$ Under the classical symmetry reduction
\begin{equation}
\label{SymGen}
{\bf N}_{lj}({\bf x},t) = \sigma_{lj} {\bf N}_{jl}^*({\bf x},t),
\end{equation}
Ablowitz and Haberman showed that the $(2+1)$ dimensional system of equations 
(\ref{E:compatible 2 dimensional}) are compatible with its complex conjugate (recall that 
the $\alpha$ and $\beta$ coefficients are all real) so long the ``$\sigma$" coefficients 
satisfy the constraint
%%%%%%%%%%%%%%%%%%%%%%%%%%
\[ \frac{\sigma_{lm} \sigma_{mj}}{\sigma_{lj}} =-1. \]
%%%%%%%%%%%%%%%%%%%%%%%%%%
Thus,  the symmetry condition (\ref{SymGen}) reduces the number of independent equations from $n(n-1)$ to $n(n-1)/2$. Next we show that system (\ref{E:compatible 2 dimensional}) admits novel nonlocal reductions that were not reported so far in the literature. They are of the
reverse space-time nonlocal type. In the next two sections, we outline their derivations and give some conservation laws.
%%%%%%%%%%%%%%%%%%%%%%%%%%%%%%%%%%%%%%%%%%%%%
%%%%%%%%%%%%%%%%%%%%%%%%%%%%%%%%%%%%%%%%%%%%
\subsection{(2+1) dimensional complex reverse space-time multi-wave reduction:
${\bf N}_{lj}({\bf x}, t) = \sigma_{lj} {\bf N}_{jl}^*(-{\bf x}, -t)$}
%%%%%%%%%%%%%%%%%%%%%%%%%%%%%%%%%%%%%%%%%%%%
%%%%%%%%%%%%%%%%%%%%%%%%%%%%%%%%%%%%%%%%%%%%%
If one substitutes the symmetry condition
\begin{equation}
\label{sym-2d-n-wave}
{\bf N}_{lj}({\bf x}, t) = \sigma_{lj} {\bf N}_{jl}^*(-{\bf x}, -t),
\end{equation}
in Eq.~(\ref{E:compatible 2 dimensional}) then with the help of change of variables 
${\bf x}' = -{\bf x}, t'=-t$ one can show, after interchange of indices, that the system (\ref{E:compatible 2 dimensional}) is consistent with its complex conjugate (since 
all $\alpha_{lj}, \beta_{lj}$ are real) provided
\[ \frac{\sigma_{lm} \sigma_{mj}}{\sigma_{lj}} =+1. \]
The new symmetry reduction (\ref{sym-2d-n-wave}) is new and, as we shall next see, leads to a new set of $(2+1)$ dimensional interacting nonlinear waves. For simplicity, we shall derive the simple and physically important case of three interacting waves.
%%%%%%%%%%%%%%%%%%%%%%%%%%%%%%%%%%%%%%%%%%%%%%
\subsection{(2+1) dimensional complex reverse space-time three-wave equations}
\label{21-nw}
%%%%%%%%%%%%%%%%%%%%%%%%%%%%%%%%%%%%%%%%%%%%%
Here we derive the dynamical equations governing the evolution of an interacting $(2+1)$ dimensional space-time nonlocal nonlinear waves. To do so, we explicitly write down the symmetry reduction for the case $n=3.$ They are given by
%%%%%%%%%%%%%
\begin{equation}
\label{sym-1-w}
{\bf N}_{21} ({\bf x},t) = \sigma_{1} {\bf N}_{12}^{*}(-{\bf x},-t), 
\end{equation}
%%%%%%%%%%%%%%
\begin{equation}
\label{sym-2-w}
{\bf N}_{31} ({\bf x},t) = \sigma_{2} {\bf N}_{13}^{*}(-{\bf x},-t),
\end{equation}
%%%%%%%%%%%%%%
\begin{equation}
\label{sym-3-w}
{\bf N}_{32} ({\bf x},t) = \sigma_{3} {\bf N}_{23}^{*}(-{\bf x},-t),
\end{equation}
%%%%%%%%%%%%%%
where, as before, all the $\alpha_{lj}$ and $\beta_{lj}$ for $l,j = 1,2,\cdots, n$ are real, and
$\sigma_j, j=1,2,3$ are chosen as real numbers satisfying the relation
\begin{equation}
\frac{\sigma_{1}\sigma_{3}}{\sigma_{2}}=1,\;\;\; \sigma_{j}^{2}=1 \ (j=1, 2, 3).
\end{equation}
Equation (\ref{E:compatible 2 dimensional}) may be put into a standard set of space-time nonlocal nonlinear interacting three-wave system by a suitable scaling of variables. With the definition
%%%%%%%%%%%%
\begin{equation}
\label{Q1-w}
{\bf N}_{12} ({\bf x},t) =-\frac{Q_{3} ({\bf x},t)}{\sqrt{(-D_{1}+D_{3})(-D_{2}+D_{3})}},
\end{equation}
%%%%%%%%%%%%
\begin{equation}
\label{Q2-w}
 {\bf N}_{31} ({\bf x},t) =-\frac{Q_{2}({\bf x},t)}{\sqrt{(-D_{1}+D_{2})(-D_{2}+D_{3})}},
 \end{equation}
 %%%%%%%%%%%%
\begin{equation}
\label{Q3-w}
{\bf N}_{23} ({\bf x},t) =-\frac{Q_{1}({\bf x},t)}{\sqrt{(-D_{1}+D_{2})(-D_{1}+D_{3})}},
 \end{equation}
 %%%%%%%%%%%%
 where
\begin{equation*}
\begin{split}
&D_{3}>D_{2}>D_{1}>0, \ c_{1}=-D_{2}D_{3}, \ c_{2}=-D_{1}D_{3}, \ c_{3}=-D_{1}D_{2},\\
&b_{1}=-D_{1}, \ b_{2}=-D_{2}, \ b_{3}=-D_{3}, \ \alpha_{12}=-D_{3}, \ \alpha_{13}=-D_{2}, \ \alpha_{23}=-D_{1},\\
&\beta_{12}=-D_{3}(D_{1}+D_{2}), \ \beta_{13}=-D_{2}(D_{1}+D_{3}), \ \beta_{23}=-D_{1}(D_{2}+D_{3}).
\end{split}
\end{equation*}
we obtain the following system of three reverse space-time nonlocal interacting waves:
\begin{equation}
\label{3wnonloc-2d}
\begin{split}
%%%%%%%%%%%%%%%%%%%%
&Q_{1,t}({\bf x},t) + {\bf C}_{1}\cdot\nabla Q_{1} ({\bf x},t)
=\sigma_{3}Q^{*}_{2}(-{\bf x},-t)Q^{*}_{3}(-{\bf x},-t),\\
%%%%%%%%%%%%%%%%%%%%
&Q_{2,t}({\bf x},t) + {\bf C}_{2} \cdot\nabla Q_{2}({\bf x},t)
=-\sigma_{2}Q^{*}_{1}(-{\bf x},-t)Q^{*}_{3}(-{\bf x},-t),\\
%%%%%%%%%%%%%%%%%%%%%%%%
&Q_{3,t}({\bf x},t) + {\bf C}_{3} \cdot\nabla Q_{3}({\bf x},t)
=\sigma_{1}Q^{*}_{1}(-{\bf x},-t) Q^{*}_{2}(-{\bf x},-t),
%%%%%%%%%%%%%%%%%%%%
\end{split}
\end{equation}
where $\nabla$ is the two dimensional gradient, ${\bf C}_{j} = (C^{(x)}_j, C^{(y)}_j), j=1,2,3$ and
\begin{equation*}
\begin{split}
&C_{1}^{(x)}=D_{1}, \ C_{1}^{(y)}=D_{1}(D_{2}+D_{3}), \ C_{2}^{(x)}=D_{2}, \ C_{2}^{(y)}=D_{2}(D_{1}+D_{3}),\\
&C_{3}^{(x)}=D_{3}, \ C_{3}^{(y)}=D_{3}(D_{1}+D_{2}).
\end{split}
\end{equation*}
From the above set of dynamical equations, one can derive the following conserved quantities:
%%%%%%%%%%%%%%%%%%%
\begin{eqnarray}
\label{3w-cons-1}
&\sigma_{2}&\int\int_{\mathbb{R}^2} Q_{1}({\bf x},t) Q_{1}^{*}(-{\bf x},-t)dxdy
\nonumber \\
&+&
\sigma_{3}\int\int_{\mathbb{R}^2}  Q_{2}({\bf x},t)Q_{2}^{*}(-{\bf x},-t)dxdy
=\text{constant},
\end{eqnarray}
%%%%%%%%%%%%%%%%%%%
\begin{eqnarray}
\label{3w-cons-2}
&\sigma_{2}&\int\int_{\mathbb{R}^2} Q_{3}({\bf x},t) Q_{3}^{*}(-{\bf x},-t) dxdy
\nonumber \\
&+&
\sigma_{1}\int\int_{\mathbb{R}^2}  Q_{2}({\bf x},t) Q_{2}^{*}(-{\bf x},-t) dxdy
= \text{constant},
\end{eqnarray}
%%%%%%%%%%%%%%%%%%%
\begin{eqnarray}
\label{3w-cons-3}
&\sigma_{1}&\int\int_{\mathbb{R}^2}  Q_{1}({\bf x},t) Q_{1}^{*}(-{\bf x},-t) dxdy
\nonumber \\
&-&
\sigma_{3}\int\int_{\mathbb{R}^2} Q_{3}({\bf x},t) Q_{3}^{*}(-{\bf x},-t) dxdy
=
\text{constant}.
\end{eqnarray}
%%%%%%%%%%%%%%%%%%%
Since none of the above conserved quantities is guaranteed to be positive definite, it is likely that in the general case the solution will blowup in finite time. This would be an interesting future direction to consider. 
%%%%%%%%%%%%%%%%%%%%%%%%%%%%%%%%%%%%%%%%%%%%%%
%%%%%%%%%%%%%%%%%%%%%%%%%%%%%%%%%%%%%%%%%%%%%%
\subsection{(2+1) dimensional reverse space-time multi-wave reduction:
${\bf N}_{lj}({\bf x},t) = \sigma_{lj} {\bf N}_{jl}(-{\bf x},-t)$}
%%%%%%%%%%%%%%%%%%%%%%%%%%%%%%%%%%%%%%%%%%%%%%
%%%%%%%%%%%%%%%%%%%%%%%%%%%%%%%%%%%%%%%%%%%%%%
Another interesting symmetry reduction that Eq.~(\ref{E:compatible 2 dimensional}) admits is given by
%%%%%%%%%%%%%%%%%%%%%%
\begin{equation}
\label{sym-3w-21}
{\bf N}_{lj}({\bf x},t) = \sigma_{lj} {\bf N}_{jl}(-{\bf x},-t),
\end{equation}
%%%%%%%%%%%%%%%%%%%%%
which would result in a reduction of the number of equations from $n(n-1)$ to $n(n-1)/2.$ Indeed, substituting (\ref{sym-3w-21}) into (\ref{E:compatible 2 dimensional}); make the change of variables ${\bf x}'=-{\bf x}, t'=-t$ and upon rearrangement of indices, one obtain the same Eq.~(\ref{E:compatible 2 dimensional}) provided
%%%%%%%%%%%%%%%%%%%%%%
\begin{equation}
\label{sym-3w-21-sigmas}
\frac{\sigma_{lm}\sigma_{mj}}{\sigma_{lj}} =1.
\end{equation}
With the help of the symmetry condition (\ref{sym-3w-21}) we will next derive the reverse 
space-time nonlocal interacting three-wave system following the same idea we 
outlined in Sec.~\ref{21-nw}.
%%%%%%%%%%%%%%%%%%%%%%%%%%%%%%%%%%%%%%%%%%%%%
%%%%%%%%%%%%%%%%%%%%%%%%%%%%%%%%%%%%%%%%%%%%
\subsection{(2+1) dimensional reverse space-time three-wave equations}
\label{21-nw-2}
%%%%%%%%%%%%%%%%%%%%%%%%%%%%%%%%%%%%%%%%%%%%%
%%%%%%%%%%%%%%%%%%%%%%%%%%%%%%%%%%%%%%%%%%%%%%%
Under the symmetry reduction ${\bf N}_{21}({\bf x},t)=\sigma_{1}{\bf N}_{12}(-{\bf x},-t)$, 
${\bf N}_{31}({\bf x},t)=\sigma_{2}{\bf N}_{13}(-{\bf x},-t)$ and 
${\bf N}_{32}({\bf x},t)=\sigma_{3}{\bf N}_{23}(-{\bf x},-t)$, where $\sigma_{1}$, $\sigma_{2}$ and $\sigma_{3}$ are chosen as real numbers, we have
$\sigma_{1}\sigma_{3}/\sigma_{2}=1$ with $\sigma_{j}^{2}=1 \ (j=1, 2, 3)$. Equation (\ref{E:compatible 2 dimensional}) may be put into a standard set of nonlocal three-wave interaction equations by a suitable scaling of variables. For example, we find the system
\begin{equation}
\begin{split}
&Q_{1,t}({\bf x},t) + {\bf C}_{1} \cdot\nabla Q_{1}({\bf x},t)
=\sigma_{3}Q_{2}(-{\bf x},-t) Q_{3}(-{\bf x},-t),
%%%%%%%
\\
&Q_{2,t}({\bf x},t) + {\bf C}_{2} \cdot\nabla Q_{2}({\bf x},t)
=-\sigma_{2}Q_{1}(-{\bf x},-t) Q_{3}(-{\bf x},-t),
%%%%%%%
\\
&Q_{3,t}({\bf x},t) + {\bf C}_{3} \cdot\nabla Q_{3}({\bf x},t)
=\sigma_{1}Q_{1}(-{\bf x},-t) Q_{2}(-{\bf x},-t),
\end{split}
\end{equation}
%%%%%%%%%%%%
if we define the following new functions
%%%%%%%%%%%
\begin{equation}
{\bf N}_{12}({\bf x},t) = -\frac{Q_{3}({\bf x},t)}{\sqrt{(-D_{1}+D_{3})(-D_{2}+D_{3})}}, 
\end{equation}
%%%%%%%%%%%
\begin{equation}
{\bf N}_{31}({\bf x},t) = -\frac{Q_{2}({\bf x},t)}{\sqrt{(-D_{1}+D_{2})(-D_{2}+D_{3})}},
\end{equation}
%%%%%%%%%%%
\begin{equation}
{\bf N}_{23}=-\frac{Q_{1}}{\sqrt{(-D_{1}+D_{2})(-D_{1}+D_{3})}},
\end{equation}
%%%%%%%%%%%
\begin{equation}
{\bf N}_{13}({\bf x},t) = \sigma_{2}N^{*}_{31}(-{\bf x},-t),
\end{equation}
%%%%%%%%%%%
\begin{equation}
{\bf N}_{32}{\bf x},t) = \sigma_{3}N^{*}_{23}{-\bf x},-t), 
 \end{equation}
%%%%%%%%%%%
\begin{equation}
{\bf N}_{21}{\bf x},t) =\sigma_{1}N^{*}_{12}{-\bf x},-t) ,
 \end{equation}
%%%%%%%%%%%
where we have defined ${\bf C}_j\equiv (C_j^{(x)}, C_j^{(y)}), j=1,2,3$ and
\begin{equation*}
\begin{split}
&C_{1}^{(x)}=D_{1}, \ C_{1}^{(y)}=D_{1}(D_{2}+D_{3}), \ C_{2}^{(x)}=D_{2}, \ C_{2}^{(y)}=D_{2}(D_{1}+D_{3}),\\
&C_{3}^{(x)}=D_{3}, \ C_{3}^{(y)}=D_{3}(D_{1}+D_{2}),
\end{split}
\end{equation*}
%%%%%%%%%%%%%
\begin{equation*}
\begin{split}
&D_{3}>D_{2}>D_{1}>0, \ c_{1}=-D_{2}D_{3}, \ c_{2}=-D_{1}D_{3}, \ c_{3}=-D_{1}D_{2},\\
&b_{1}=-D_{1}, \ b_{2}=-D_{2}, \ b_{3}=-D_{3}, \ \alpha_{12}=-D_{3}, \ \alpha_{13}=-D_{2}, \ \alpha_{23}=-D_{1},\\
&\beta_{12}=-D_{3}(D_{1}+D_{2}), \ \beta_{13}=-D_{2}(D_{1}+D_{3}), \ \beta_{23}=-D_{1}(D_{2}+D_{3}).
\end{split}
\end{equation*}
As was done before, we can derive the following conserved quantities:
%%%%%%%%%%%%%%%%%%%%%%%%
\begin{eqnarray}
&\sigma_{2}&\int\int_{\mathbb{R}^2}   Q_{1}({\bf x},t) Q_{1}(-{\bf x},-t) dxdy
\nonumber \\
&+&
\sigma_{3} \int \int_{\mathbb{R}^2} Q_{2}({\bf x},t) Q_{2}(-{\bf x},-t) dxdy
= \text{constant},
\end{eqnarray}
%%%%%%%%%%%%%%%%%%%%%%%
\begin{eqnarray}
&\sigma_{2}&  \int\int_{\mathbb{R}^2}  Q_{3}({\bf x},t) Q_{3}(-{\bf x},-t) dxdy
\nonumber \\
&+&
\sigma_{1}\int\int_{\mathbb{R}^2}  Q_{2}({\bf x},t) Q_{2}(-{\bf x},-t) dxdy
=\text{constant},
\end{eqnarray}
%%%%%%%%%%%%%%%%%%%%%%%%%%
\begin{eqnarray}
&\sigma_{1}&\int\int_{\mathbb{R}^2}  Q_{1}({\bf x},t) Q_{1}(-{\bf x},-t) dxdy
\nonumber \\
&-&
\sigma_{3}\int\int_{\mathbb{R}^2}  Q_{3}({\bf x},t) Q_{3}(-{\bf x},-t) dxdy
=\text{constant}.
\end{eqnarray}
%%%%%%%%%%%%%%%%%%%%%%%%%%
As with the complex reverse space-time nonlocal three wave system, none of the above conserved quantity appears to be positive definite. It would be interesting to see if the above three wave system can develop a finite time singularity.
%%%%%%%%%%%%%%%%%%%%%%%%%%%%%%%%%%%%%%%%%%%%
%%%%%%%%%%%%%%%%%%%%%%%%%%%%%%%%%%%%%%%%%%%%
\section{Integrable nonlocal discrete NLS models: 
Reverse discrete-time, reverse time and $PT$ preserved symmetries}
\label{discrete-sys}
%%%%%%%%%%%%%%%%%%%%%%%%%%%%%%%%%%%%%%%%%%%%
%%%%%%%%%%%%%%%%%%%%%%%%%%%%%%%%%%%%%%%%%%%%
In this section we derive discrete analogues to the nonlocal NLS equations 
(\ref{complex-space-time-nonloc-NLS}) and (\ref{complex-time-nonloc-NLS}). The resulting models are integrable and admit infinite number of conserved quantities. 
Our approach is based on the integrable discrete scattering problem \cite{AL1}
%%%%%%%%%%%%%% 
\begin{equation}
\label{AL-Scattering}
v_{n+1}=\left(\begin{array}{cc}z&Q_n\\R_n&z^{-1}
\end{array}\right)v_n\;,
\end{equation}
%%%%%%%%%%%%%%
\begin{equation}
\label{AL-Time-evolv}
\frac{dv_n}{dt}=\left(\begin{array}{cr}
\mathsf{A}_n & \mathsf{B}_n
\\
\mathsf{C}_n & \mathsf{D}_n
\end{array}\right)v_n\;,
\end{equation}
%%%%%%%%%%%%%%%
where $v_n=(v_n^{(1)}, v_n^{(2)})^T$, $Q_n$ and $R_n$ vanish rapidly 
as $n\rightarrow \pm\infty$ and $z$ is a complex spectral parameter. Here,
\begin{equation}
\label{An}
\mathsf{A}_n=iQ_nR_{n-1}-\frac{i}{2}\left(z-z^{-1}\right)^2\;,
\end{equation}
%%%%%%%%%%%%%%
\begin{equation}
\label{Bn}
\mathsf{B}_n=-i\left(zQ_n-z^{-1}Q_{n-1}\right),
\end{equation}
%%%%%%%%%%%%%%%
%%%%%%%%%%%%%%
\begin{equation}
\label{Cn}
\mathsf{C}_n=i\left(z^{-1}R_{n}-zR_{n-1}\right)
\end{equation}
%%%%%%%%%%%%%%%
\begin{equation}
\label{Dn}
\mathsf{D}_n=-iR_{n}Q_{n-1}+\frac{i}{2}\left(z-z^{-1}\right)^2.
\end{equation}
%%%%%%%%%%%%%%%%
The discrete compatibility condition $\frac{d}{dt}v_{n+1}=\left(\frac{d}{dt}v_{m} \right)_{m=n+1}$ yields
\begin{equation}
\label{PT-IDNLS1}
i\frac{d}{dt}Q_n(t) = \Delta_nQ_n (t)
-Q_n(t) R_n(t)\left[Q_{n+1}(t) + Q_{n-1}(t) \right]\;,
\end{equation}
\begin{equation}
\label{PT-IDNLS2}
-i\frac{d}{dt}R_n(t) = \Delta_nR_n(t)
-Q_n(t) R_n(t)\left[R_{n+1}(t) + R_{n-1}(t) \right]\;,
\end{equation}
where 
%%%%%%%%%%%%%%%
\begin{equation}
\label{Discrete_Lap}
\Delta_nF_n \equiv F_{n+1} -2F_n+F_{n-1}.
\end{equation}
%%%%%%%%%%%%%%%
In \cite{AL1}, it was shown that the system of equations (\ref{PT-IDNLS1}) 
and (\ref{PT-IDNLS2}) are compatible under the symmetry reduction
\begin{equation}
\label{sym-reduction}
R_n(t) = \sigma Q_{n}^*(t)\;, \;\;\; \sigma = \mp 1\;,
\end{equation}
and gives rise to the Ablowitz-Ladik model \cite{AL1,AL2}
\begin{equation}
\label{AL-eqn}
i\frac{dQ_n(t)}{dt} = \Delta_n Q_n - \sigma   |Q_n(t)|^2 \left[Q_{n+1}(t) + Q_{n-1}(t) \right].
\end{equation}
%%%%%%%%%%%%%%%%%%%%%%%%%%%%%%%%%%%%%%%%%%%%%%%%%%%%%%%%%%%%%%%%%%%%%%%%%%%%%%%%%%%%%%%%%%%%
\subsection{Reverse discrete-time reduction: $R_n(t) = \sigma Q_{-n}(-t)$}
%%%%%%%%%%%%%%%%%%%%%%%%%%%%%%%%%%%%%%%%%%%%%%%%%%%%%%%%%%%%%%%%%%%%%%%%%%%%%%%%%%%%%%%%%
Interestingly, the system of discrete equations (\ref{PT-IDNLS1}) 
and (\ref{PT-IDNLS2}) are compatible under the symmetry reduction
\begin{equation}
\label{sym-reduction-a}
R_n(t) = \sigma Q_{-n}(-t)\;, \;\;\; \sigma = \mp 1\;,
\end{equation}
and gives rise to the reverse discrete-time nonlocal 
discrete NLS equation:
\begin{eqnarray}
\label{AL-new-eqn-1}
i\frac{dQ_n(t)}{dt} = \Delta_n Q_n
 - \sigma  Q_n(t) Q_{-n}(-t) \left[Q_{n+1}(t) + Q_{n-1}(t) \right].
\end{eqnarray}
The discrete symmetry constraint (\ref{sym-reduction-a}) is new and was not noticed in the literature. Since Eq.~(\ref{AL-new-eqn-1}) comes out of the Ablowitz-Ladik scattering problem, as such, it constitute an infinite dimensional integrable Hamiltonian dynamical system. The first two conserved quantities are given by
%%%%%%%%%%%%%%%%%%
\begin{equation}
\label{cons-quant-1}
\sum_{n=-\infty}^{+\infty}Q_n(t)~ Q_{1-n}(-t) = \text{constant}\;.
\end{equation}
%%%%%%%%%%%%%%%%%%%
\begin{equation}
\label{cons-quant-2}
\sum_{n=-\infty}^{+\infty}\left[\sigma Q_n(t)~ Q_{2-n}(-t) 
-\frac{1}{2}\left( Q_n(t)~ Q_{1-n}(-t) \right)^2 \right] = \text{constant}
\;.
\end{equation}
%%%%%%%%%%%%%%%%%%
\begin{equation}
\label{cons-quant-3}
\prod_{n=-\infty}^{+\infty}\left[1 -\sigma  Q_n(t)~ Q_{-n}(-t) \right] =  \text{constant} \;.
\end{equation}
Importantly, Eq.~(\ref{AL-new-eqn-1}) is a Hamiltonian dynamical system with $Q_n(t)$ 
and $Q_{-n}(-t)$ playing the role of coordinates and conjugate momenta respectively. The corresponding Hamiltonian and (the non canonical) brackets are given by
%%%%%%%%%%%%%%%%%%
\begin{eqnarray}
\label{Hamiltonian}
H &=& -\sigma \sum_{n=-\infty}^{+\infty}Q_{-n}(-t)\left(Q_{n+1}(t) + Q_{n-1}(t)\right)
\\ \nonumber
 &&  -2 \sum_{n=-\infty}^{+\infty} \log \left(1 -\sigma  Q_n(t)~ Q_{-n}(-t) \right)\;.
\end{eqnarray}
%%%%%%%%%%%%%%%%%%%
%%%%%%%%%%%%%%%%%%
\begin{equation}
\label{braket-1}
\left\{ Q_m(t) , Q_{-n}(-t) \right\} = i \left( 1 - \sigma  Q_n(t)~ Q_{-n}(-t) \right)\delta_{n,m}\;.
\end{equation}
%%%%%%%%%%%%%%%%%%%
%%%%%%%%%%%%%%%%%%
\begin{equation}
\label{braket-2}
\left\{ Q_n(t) , Q_{m}(t) \right\} = \left\{ Q_n(t) , Q_{-m}(-t) \right\} =0 \;.
\end{equation}
%%%%%%%%%%%%%%%%%%%
%%%%%%%%%%%%%%%%%%%%%%%%%%%%%%%%%%%%%%%%%%%%%%%%%%%%%%%%%%%%%%%%%%%%%%%%%%%%%%%%%%%%%%%%%%%%
\subsection{Reverse time discrete symmetry: $R_n(t) = \sigma Q_{n}(-t)$}
%%%%%%%%%%%%%%%%%%%%%%%%%%%%%%%%%%%%%%%%%%%%%%%%%%%%%%%%%%%%%%%%%%%%%%%%%%%%%%%%%%%%%%%%%
Equations (\ref{PT-IDNLS1}) and (\ref{PT-IDNLS2}) admit another important symmetry reduction given by
\begin{equation}
\label{sym-reduction-b}
R_n(t) = \sigma Q_{n}(-t)\;, \;\;\; \sigma = \mp 1\;.
\end{equation}
This symmetry reduction is called reverse time Ablowitz-Ladik symmetry and results in the following discrete reverse time nonlocal discrete NLS equation:
\begin{eqnarray}
\label{AL-new-eqn-1-a}
i\frac{dQ_n(t)}{dt} = \Delta_n Q_n
 - \sigma  Q_n(t) Q_{n}(-t) \left[Q_{n+1}(t) + Q_{n-1}(t) \right].
\end{eqnarray}
The discrete symmetry constraint (\ref{sym-reduction-b}) is also new and was not noticed in the literature so far. As is the case with the complex discrete-time symmetry, 
Eq.~(\ref{AL-new-eqn-1-a}) is also integrable and posses an infinite number of conservation laws. The first few conserved quantities are listed below
%%%%%%%%%%%%%%%%%%
\begin{equation}
\label{cons-quant-1-a}
\sum_{n=-\infty}^{+\infty}Q_n(t)~ Q_{n-1}(-t) = \text{constant}\;.
\end{equation}
%%%%%%%%%%%%%%%%%%%
\begin{equation}
\label{cons-quant-2-a}
\sum_{n=-\infty}^{+\infty}\left[\sigma Q_n(t)~ Q_{n-2}(-t) 
-\frac{1}{2}\left( Q_n(t)~ Q_{n-1}(-t) \right)^2 \right] = \text{constant}
\;.
\end{equation}
%%%%%%%%%%%%%%%%%%
\begin{equation}
\label{cons-quant-3-a}
\prod_{n=-\infty}^{+\infty}\left[1 -\sigma  Q_n(t)~ Q_{n}(-t) \right] =  \text{constant} \;.
\end{equation}
Importantly, Eq.~(\ref{AL-new-eqn-1-a}) is a Hamiltonian dynamical system with $Q_n(t)$ 
and $Q_{-n}(-t)$ playing the role of coordinates and conjugate momenta respectively. The corresponding Hamiltonian and (the non canonical) brackets are given by
%%%%%%%%%%%%%%%%%%
\begin{eqnarray}
\label{Hamiltonian-a}
H &=& -\sigma \sum_{n=-\infty}^{+\infty}Q_{n}(-t)\left(Q_{n+1}(t) + Q_{n-1}(t)\right)
\\ \nonumber
 &&  -2 \sum_{n=-\infty}^{+\infty} \log \left(1 - \sigma  Q_n(t)~ Q_{n}(-t) \right)\;.
\end{eqnarray}
%%%%%%%%%%%%%%%%%%%
%%%%%%%%%%%%%%%%%%
\begin{equation}
\label{braket-1-a}
\left\{ Q_m(t) , Q_{n}(-t) \right\} = i \sigma 
\left( 1 - \sigma  Q_n(t)~ Q_{-n}(-t) \right)\delta_{n,m}\;.
\end{equation}
%%%%%%%%%%%%%%%%%%%
%%%%%%%%%%%%%%%%%%
\begin{equation}
\label{braket-2-a}
\left\{ Q_n(t) , Q_{m}(t) \right\} = \left\{ Q_n(t) , Q_{m}(-t) \right\} =0 \;.
\end{equation}
%%%%%%%%%%%%%%%%%%%
In summary, the discrete systems (\ref{PT-IDNLS1}) and (\ref{PT-IDNLS1}) admit four different symmetry reduction:
\begin{enumerate}
\item Standard Ablowitz-Ladik symmetry 
\begin{equation}
\label{AL-sym}
R_n(t) = \sigma Q^*_{n}(t)\;, \;\;\; \sigma = \mp 1\;,
\end{equation}
giving rise to the so-called Ablowitz-Ladik model (\ref{AL-eqn}).\\
%%%%%%%%%%%%%%%%%%%
\item Reverse discrete-time symmetry
\begin{equation}
\label{discrete-time-sym-0}
R_n(t) = \sigma Q_{-n}(-t)\;, \;\;\; \sigma = \mp 1\;,
\end{equation}
giving rise to Eq.~(\ref{AL-new-eqn-1})\;,\\
%%%%%%%%%%%%%%%%%%%
\item Reverse time discrete symmetry
\begin{equation}
\label{discrete-time-sym-1}
R_n(t) = \sigma Q_{n}(-t)\;, \;\;\; \sigma = \mp 1\;,
\end{equation}
giving rise to Eq.~(\ref{AL-new-eqn-1-a})\;,\\
%%%%%%%%%%%%%%%%%%%
\item Discrete $PT$ preserved symmetry
\begin{equation}
\label{discrete-time-sym-2}
R_n(t) = \sigma Q^*_{-n}(t)\;, \;\;\; \sigma = \mp 1\;,
\end{equation}
giving rise to the discrete $PT$ symmetric integrable nonlocal discrete NLS equation first found in \cite{AblowitzMusslimani2}:
\begin{eqnarray}
\label{AL-new-eqn-1-a-PT}
i\frac{dQ_n(t)}{dt} = \Delta_n Q_n
 - \sigma  Q_n(t) Q^*_{-n}(t) \left[Q_{n+1}(t) + Q_{n-1}(t) \right].
\end{eqnarray}
%%%%%%%%%%%%%%%%%%%
\end{enumerate}
%%%%%%%%%%%%%%%%%%%%%%%%%%%%%%%%%%%%%%%%%%%%%%%
%%%%%%%%%%%%%%%%%%%%%%%%%%%%%%%%%%%%%%%%%%%%
%%%%%%%%%%%%%%%%%%%%%%%%%%%%%%%%%%%%%%%%%%%%
\section{IST: $2 \times 2$ AKNS type}
\label{ISTsoliton}
Many of the above reverse space-time nonlocal evolution equations introduced in this paper came out of crucial symmetry reductions of general AKNS scattering problem 
(\ref{AKNS}) -- (\ref{LAX-2}). As such, they constitute  infinite-dimensional integrable Hamiltonian dynamical systems which are solvable by the
 inverse scattering transform. The method of solution involves three major steps: (i) direct scattering problem which involves finding the associated eigenfunctions, scattering data and their symmetries, (ii) identifying the time evolution of the scattering data and (iii) solving the inverse problem using the Riemann-Hilbert approach or other inverse methods. In what follows we shall highlight the main results behind each step for the AKNS scattering problem given in (\ref{LAX-1}) subject to the new reversed space-time symmetry reductions. The full account of the inverse scattering theory for all evolution equations introduced in this paper is beyond the scope of this paper and will be discussed in future work.\\\\
%%%%%%%%%%%%%%%%%%%%%%%%%%
\subsection{Direct scattering problem} 
%%%%%%%%%%%%%%%%%%%%%%%%%%
The analysis presented in this paper assumes that the potential functions $q(x,t)$ and $r(x,t)$ decay to zero sufficiently fast at infinity. Thus, solutions of the scattering problem (\ref{AKNS}) are defined and satisfy the boundary conditions
\begin{equation}
\label{eig-fun-bc}
\begin{split}
\phi    & \sim
\left(\begin{array}{cc}
\!\! 1\!\!
\\
\!\!0\!\!
\end{array}\right)e^{-ikx}\;,\;\;
\overline{\phi}  \sim
\left(\begin{array}{cc}
\!\! 0\!\!
\\
\!\!1\!\!
\end{array}\right)e^{ikx}\;, ~~{\rm as}~ x\rightarrow -\infty
\\
\psi 
& \sim
\left(\begin{array}{cc}
\!\! 0\!\!
\\
\!\!1\!\!
\end{array}\right)e^{ikx}\;,\;\;
\overline{\psi}  \sim
\left(\begin{array}{cc}
\!\! 1\!\!
\\
\!\!0\!\!
\end{array}\right)e^{-ikx}\;, ~~{\rm as}~ x\rightarrow +\infty
\;.
\end{split}
\end{equation}
Note that bar does not denote complex conjugation; we use $*$ to denote complex conjugation. It is expedient to define new functions
\begin{equation}
\label{Jost-1}
M(x,t,k)=e^{ikx}\phi (x,t,k),\; \overline{M}(x,t,k)=e^{-ikx}\overline{\phi} (x,t,k)\;,
\end{equation}
\begin{equation}
\label{Jost-2}
N(x,t,k)=e^{-ikx}\psi (x,t,k),\; \overline{N}(x,t,k)=e^{ikx}\overline{\psi} (x,t,k)\;,
\end{equation}
with
\begin{equation}
\label{eig-fun-bc-Jf}
\begin{split}
M    & \sim
\left(\begin{array}{cc}
\!\! 1\!\!
\\
\!\!0\!\!
\end{array}\right)\;,\;\;
\overline{M}  \sim
\left(\begin{array}{cc}
\!\! 0\!\!
\\
\!\!1\!\!
\end{array}\right)\;, ~~{\rm as}~ x\rightarrow -\infty
\\
N
& \sim
\left(\begin{array}{cc}
\!\! 0\!\!
\\
\!\!1\!\!
\end{array}\right)\;,\;\;
\overline{N}  \sim
\left(\begin{array}{cc}
\!\! 1\!\!
\\
\!\!0\!\!
\end{array}\right)\;, ~~{\rm as}~ x\rightarrow +\infty
\;
\end{split}
\end{equation}
that satisfy constant boundary conditions at infinity and reformulate the direct scattering problem in terms of them. With this at hand, when the potentials $q,r$ are integrable 
(i.e. they are in class $L^1$) one can derive an integral equation for the above functions and use them to show that $M(x,t,k), N(x,t,k)$ are analytic functions in the upper half complex $k$ plane whereas
$\overline{M}(x,t,k), \overline{N}(x,t,k)$ are analytic functions in the lower half complex $k$ plane \cite{Ablowitz3}. \\\\
%%%%%%%%%%%%%%%%%%%%%%%%%%%%%%%%%%%%%%%%%%%%
The solutions $\phi (x,t,k)$ and $\overline{\phi} (x,t,k)$ of the scattering 
problem (\ref{AKNS}) with the boundary conditions (\ref{eig-fun-bc}) are linearly independent. The same hold for $\psi (x,t,k)$ and $\overline{\psi} (x,t,k).$ 
 We denote by $\Phi (x,t,k) \equiv (\phi (x,t,k),\; \overline{\phi}(x,t,k))$ and 
$\Psi (x,t,k) \equiv (\overline{\psi}(x,t,k),\; \psi (x,t,k)).$ Clearly, these two set of functions are linearly dependent and write
\begin{eqnarray}
\label{scat-I}
\Phi (x,t,k) = S(k,t) \Psi (x,t,k)\;,  
\end{eqnarray}
where $S(k,t)$ is the scattering matrix given by
\begin{eqnarray}
\label{S-matrix}
S(k)=
\left(\begin{array}{cr}
a(k,t)&b(k,t)
\\
\overline{b}(k,t)&\overline{a}(k,t)
\end{array}\right) \;.
\end{eqnarray}
The elements of the scattering matrix $S(k,t)$ are related to the Wronskian of the system via the relations 
\begin{equation}
\label{Wa}
a(k,t)=W(\phi (x,t,k),\psi (x,t,k)) 
\end{equation}

\begin{equation}
\label{Wabar}
\overline{a}(k,t)=W(\overline{\psi}(x,t,k) ,\overline{\phi}(x,t,k) ), 
\end{equation}
and 
\begin{equation}
\label{Wb}
b(k,t)=W(\overline{\psi}(x,t,k) ,\phi(x,t,k) ) 
\end{equation}
\begin{equation}
\label{Wbbar}
\overline{b}(k,t)=W(\overline{\phi}(x,t,k) ,\psi (x,t,k)) 
\end{equation}
where $W(u,v)$ is the Wronskian of the two solutions $u,v$ and is given by $W(u,v)=u_1v_2-v_1u_2$ where in terms of components  $u=[u_1,u_2]^T$ where $T$ represents the transpose. Moreover, it can be shown that $a(k), \overline{a}(k)$ are respectively analytic functions in the upper/lower half complex $k$ plane. However $b(k)$ and $\overline{b}(k)$ are generally not analytic anywhere. 
%%%%%%%%%%%%%%%%%%%%%%%%%%%%%%%%%%%%%%%%%%%%
%%%%%%%%%%%%%%%%%%%%%%%%%%%%%%%%%%%%%%%%%%%%
\subsection{Inverse scattering problem} 
%%%%%%%%%%%%%%%%%%%%%%%%%%%%%%%%%%%%%%%%%%%%
%%%%%%%%%%%%%%%%%%%%%%%%%%%%%%%%%%%%%%%%%%%%
The inverse problem consists of constructing the potential functions 
$r(x,t)$ and $q(x,t)$ from the scattering data (reflection coefficients), e.g. 
$\rho (k,t)=e^{-4ik^2t} b(k,0)/a(k,0)$ and $\overline{\rho} (k,t ) = e^{4ik^2t} \overline{b}(k,0)/\overline{a}(k,0)$ defined on ${\rm Im} k=0$ as well as the eigenvalues $k_j, \overline{k}_j$ and norming constants (in $x$) $C_j(t), \overline{C}_j(t).$ Using the Riemann-Hilbert approach, from equation (\ref{scat-I}) one can find equations governing the eigenfunctions $N(x, t,k), \overline{N}(x, t,k)$ \cite{Ablowitz3}
%%%%%%%%%%%%%%%%%%%%%%%%%%%
\begin{eqnarray}
\label{N-eq}
\overline{N}(x, t,k) &=&
\left(\begin{array}{cc}
\!\!1\!\!\\
\!\!0\!\!
\end{array}\right)
+\sum_{j=1}^{J}\frac{C_j(t)e^{2ik_jx}  N(x,t,k_j)}{k -k_j}     
\nonumber \\
&+&
\frac{1}{2\pi i}\int_{-\infty}^{+\infty}
\frac{\rho (\zeta,t )e^{2i\zeta x}N(x,t,\zeta)}{\zeta -(k-i0)}d\zeta \;,
\end{eqnarray}
%%%%%%%%%%%%%%%%%%%%%%%%%%%
\begin{eqnarray}
\label{Nbar-eq}
N(x, t,k) &=&
\left(\begin{array}{cc}
\!\!0\!\!\\
\!\!1\!\!
\end{array}\right)
+\sum_{j=1}^{\overline{J}}\frac{\overline{C}_j(t)e^{-2i\overline{k}_jx}  \overline{N}(x,t,\overline{k}_j)}{k -\overline{k}_j}
\nonumber \\
&-&
\frac{1}{2\pi i}\int_{-\infty}^{+\infty}
\frac{\overline{\rho}(t) (\zeta )e^{-2i\zeta x}\overline{N}(x,t,\zeta)}{\zeta -(k+i0)}d\zeta \;.
\end{eqnarray}
%%%%%%%%%%%%%%%%%%%%%%%%%%%
%%%%%%%%%%%%%%%%%%%%%%%%%%% 
To close the system we substitute $k=\overline{k}_\ell$ and $k=k_\ell$ in (\ref{Nbar-eq}) and (\ref{N-eq}) respectively and obtain a linear algebraic integral system of equations that solve the inverse problem for the eigenfunctions $N(x,t,k)$ and $\overline{N}(x,t,k).$ In the case with zero reflection coefficient, i.e., $\rho(t)= \overline{\rho}(t) =0$ the resulting algebraic system governing the soliton solution reads
%%%%%%%%%%%%%%%%%%%%%%%%%%%
\begin{eqnarray}
\label{Nbar-eq-sol}
\overline{N}(x, t,\overline{k}_\ell)
=
\left(\begin{array}{cc}
\!\!1\!\!\\
\!\!0\!\!
\end{array}\right)
+\sum_{j=1}^{J}\frac{C_j(t)e^{2ik_jx}  N(x,t,k_j)}{ \overline{k}_\ell -k_j}      \;,
\end{eqnarray}
%%%%%%%%%%%%%%%%%%%%%%%%%%%
\begin{eqnarray}
\label{N-eq-sol}
N(x, t,k_\ell)
=
\left(\begin{array}{cc}
\!\!0\!\!\\
\!\!1\!\!
\end{array}\right)
+\sum_{j=1}^{\overline{J}}\frac{\overline{C}_j(t)e^{-2i\overline{k}_jx} 
 \overline{N}(x,t,\overline{k}_j)}{k_\ell -\overline{k}_j} \;.
\end{eqnarray}

%%%%%%%%%%%%%%%%%%%%%%%%%%%
%%%%%%%%%%%%%%%%%%%%%%%%%%%%%%%%%%%%%%%%%%%%
%%%%%%%%%%%%%%%%%%%%%%%%%%%%%%%%%%%%%%%%%%%%
%%%%%%%%%%%%%%%%%%%%%%%%%%%%%%%%%%%%%%%%%%%%%
%%%%%%%%%%%%%%%%%%%%%%%%%%%%%%%%%%%%%%%%%%%%%
\subsection{Recovery of the potentials} 
%%%%%%%%%%%%%%%%%%%%%%%%%%%%%%%%%%%%%%%%%%%%%
%%%%%%%%%%%%%%%%%%%%%%%%%%%%%%%%%%%%%%%%%%%%%
To reconstruct the potentials for all time: $q(x,t),\; r(x,t)$ we compare the asymptotic 
expansions of Eq.~(\ref{Nbar-eq}) and  (\ref{N-eq})  to the Jost functions and find 
(for pure soliton solution only)
%%%%%%%%%%%%%%%%%%%%%%%%%%%%%%%%%%%%%%%%%%
%%%%%%%%%%%%%%%%%%%%%%%%%%%%%%%%%%%%%%%%%%
 \begin{eqnarray}
\label{q-of-x}
q(x,t) = 2i
\sum_{\ell =1}^{\bar{J}} \bar{C}_{\ell}(t) e^{-2i\bar{k}_{\ell}x} \bar{N}_1(x,\bar{k}_\ell ) \;.
\end{eqnarray}
%%%%%%%%%%%%%%%%%%%%%%%%%%%%%%%%%%%%%%%%%%%%
%%%%%%%%%%%%%%%%%%%%%%%%%%%%%%%%%%%%%%%%%%%%
\begin{eqnarray}
\label{r-of-x}
r(x,t) = -2i
\sum_{\ell =1}^{J}C_{\ell}(t) e^{2ik_{\ell}x} N_2(x,k_\ell ) \;.
\end{eqnarray}
Once all the symmetries of the scattering data are known, we can obtain the solution $q$ which satisfies the spatial symmetries by solving the above equations. 
%%%%%%%%%%%%%%%%%%%%%%%%%%%%%%%%%%%%%%%%%%%%
%%%%%%%%%%%%%%%%%%%%%%%%%%%%%%%%%%%%%%%%%%%%
\subsection{Evolution of the scattering data}
 %%%%%%%%%%%%%%%%%%%%%%%%%%%%%%%%%%%%%%%%%%%%
%%%%%%%%%%%%%%%%%%%%%%%%%%%%%%%%%%%%%%%%%%%%
The time dependence of the potentials $q$ and $r$ in Eq.~(\ref{q-of-x}) and (\ref{r-of-x}) is encoded in the eigenvalues and norming constants $C_j$ 
and $\overline{C}_j.$ Their time evolution is derived from Eq.~ (\ref{time-AKNS})
and (\ref{LAX-2}). The space, time and space-time nonlocal NLS, mKdV and SG equations belong to the same hierarchy, i.e., they all originate from the same scattering problem
(\ref{AKNS}) with different $\mathsf{A}, \mathsf{B}$ and $\mathsf{C}$ which in turn determines the time evolution of the scattering data and norming constants. 
For the problems we will be studying in detail here, following the derivation outlined in \cite{Ablowitz3} for the temporal evolution one finds the following:
In all cases we have
%%%%%%%%%%%%%%%%%
\[ a(k,t) = a(k,0), ~\bar{a}(k,t)= \bar{a}(k,0),\]
so that the zero's of $a(k)$ and $\bar{a}(k)$, denoted by,  $k_j, \bar{k}_j,~j=1,2...J$ respectively are constant in time. For NLS and nonlocal NLS problems
\[ b^{{\rm NLS}}(k,t)=b(k,0)e^{-4ik^2t},\]
\[
  \bar{b}^{{\rm NLS}}   (k,t)= \bar{b} (k,0)e^{4ik^2t}, k \in \mathbb{R} \],
%%%%%%%%%%%%%%%%%%%%%%%%%
%%%%%%%%%%%%%%%%%%%%%%%%%
\begin{equation}
\label{Cj-evolv-nls}
C^{{\rm NLS}}_j(t)=C_j(0)e^{-4ik_j^2t}\;,
\end{equation}
\begin{equation}
\label{Cj-bar-evolv-nls}
 \overline{C}^{{\rm NLS}}_j(t)=\overline{C}_j(0)e^{4i\overline{k}_j^2t}.
 \end{equation}
Here,  $k_j$ and $\overline{k}_j$ are often called the soliton eigenvalues and $C_j(0), \bar{C}_j(0)$ are termed norming constants. 
 %%%%%%%%%%%%%%%%%%%%%
 For  mKdV and nonlocal mKdV problems
\[ b^{{\rm mKdV}}(k,t)=b(k,0)e^{8ik^3t}, \]
\[
\bar{b}^{{\rm mKdV}}  (k,t)= \bar{b} (k,0)e^{-8ik^3t}, k \in \mathbb{R} \]
 %%%%%%%%%%%%%%%%%
 \begin{equation}
\label{Cj-evolv-mkdv}
C^{{\rm mKdV}}_j(t)=C_j(0)e^{8ik_j^3t}\;,
\end{equation}
\begin{equation}
\label{Cj-bar-evolv-mkdv}
 \overline{C}^{{\rm mKdV}}_j(t)=\overline{C}_j(0)e^{-8i\overline{k}_j^3t},
 \end{equation}
 %%%%%%%%%%%%%%%%%%
and for the sine-Gordon (sG) equation we have
%%%%%%%%%%%%%%%%%
 \begin{equation}
\label{Cj-evolv-SG}
C^{{\rm sG}}_j(t)=C_j(0)e^{-it/(2k_j)}\;,
\end{equation}
\begin{equation}
\label{Cj-bar-evolv-sG}
 \overline{C}^{{\rm sG}}_j(t)=\overline{C}_j(0)e^{it/(2\overline{k}_j)}.
 \end{equation}
In the latter equations we used the boundary condition (\ref{sGinfty}). 
 %%%%%%%%%%%%%%%%%%
%%%%%%%%%%%%%%%%%%%%%%%%%%%%%%%%%%%%%%%%%%%%
%%%%%%%%%%%%%%%%%%%%%%%%%%%%%%%%%%%%%%%%%%%%%
\section{Symmetries and soliton solutions}
\label{solitons-sol}
In this section we construct soliton solutions to the time and space-time nonlocal NLS  as well as the mKdV and sine-Gordon (sG) equations. 
Pure soliton solutions correspond to zero reflection coefficients, i.e., $\rho (\xi,t)=0$ and $\bar{\rho}(\xi,t)=0$ for all real $\xi.$ In this case the system (\ref{N-eq}), (\ref{Nbar-eq}) reduces to an algebraic equations (\ref{N-eq-sol}) and (\ref{Nbar-eq-sol}) supplemented by the time dependence (\ref{Cj-evolv-nls}-\ref{Cj-bar-evolv-mkdv}) that determine the functional form of the solitons for the nonlocal NLS, mKdV and sG equations.\\\\
 %%%%%%%%%%%%%%%%%%%%%%%%%%%%%%%%%%%%%%%%%%%%%%%
Next, we obtain a one-soliton solution of 
the $N,\bar{N}$ equations (\ref{Nbar-eq-sol}) and (\ref{N-eq-sol}) by 
taking $J=\bar{J}=1$ to find 
%%%%%%%%%%%%%%%%
\begin{equation}
\label{SolN}
N_2(x,t)=\bar{N}_1(x,t)=  \frac{1}{1+\frac{C_1(t)\bar{C}_1(t)}{(k_1-\bar{k}_1)^2
e^{2i(k_1-\bar{k}_1)x}}} \;.
\end{equation}
The corresponding potentials (\ref{q-of-x}) -- (\ref{r-of-x}) are given by
%%%%%%%%%%%%%%%%%
\begin{equation}
\label{Solq}
q(x,t)=  \frac{2i e^{-2i\bar{k}_1x}\bar{C}_1(t)}{1+\frac{C_1(t)\bar{C}_1(t)}{(k_1-\bar{k}_1)^2e^{2i(k_1-\bar{k}_1)x}}} \;,
\end{equation}
%%%%%%%%%%%%%%%%
\begin{equation}
\label{Solr}
r(x,t)=  -\frac{2i e^{2i k_1x} C_1(t)}{1+\frac{C_1(t)\bar{C}_1(t)}{(k_1-\bar{k}_1)^2
e^{2i(k_1-\bar{k}_1)x}}} \;.
\end{equation}
%%%%%%%%%%%%%%%%%%%%%%%%%%%%%%%%%%%%%%%%%%%%%
Below, for the $2 \times 2$ AKNS scattering problem we will give the relevant symmetries and (for simplicity) their associated one-soliton solutions considered in this paper.
%%%%%%%%%%%%%%%%%%%%%%%%%%%%%%%%%%%%%%%%%%%%
\subsection{Standard AKNS Symmetry: $r(x,t) = \sigma q^*(x,t) $}
%%%%%%%%%%%%%%%%%%%%%%%%%%%%%%%%%%%%%%%%%%%
The original symmetry (associated with solitons) considered in \cite{AKNS} was
\begin{equation}
\label{AKNSorigsym1}
r(x,t) = \sigma q^*(x,t) \;,
\end{equation}
where we recall $\sigma= \mp 1$. The (additional) time dependence of the scattering data associated with the classical NLS equation is 
\begin{equation*}
\bar{b}(k,t)=\sigma b^*(k,t),  ~ k \in \mathbb{R} \;,
\end{equation*}
and
\begin{equation*}
 \bar{C_j}(t)= -C_j^*(t), ~j=1,2...,J \;.
 \end{equation*} 
The corresponding continuous and discrete symmetries in scattering space, at the initial time, are given by
\begin{equation*}
\label{AKNSorigsymCont}
%\bar{a}(k)=a^*(k^*), ~
\bar{a}(k,0)=a^*(k^*,0), ~~\bar{b}(k,0)=\sigma b^*(k,0),  ~ k \in \mathbb{R}
\end{equation*}
\begin{equation}
\label{AKNSorigsymDisc}
%\bar{a}(k)=a^*(k^*), ~
\sigma=-1: ~\bar{k}_j=k_j^*, ~\bar{C_j}(0)= -C_j^*(0), ~j=1,2...,J 
\end{equation}
The above symmetries allow us to formulate the general linearization of the classical NLS equation (\ref{NLS}) with the reduction (\ref{AKNSorigsym1}) given above. Then the corresponding well-known  one soliton solution of the classical NLS Eq.~(\ref{NLS}) is obtained from Eq.~(\ref{Solq}-\ref{Solr}) with $J=1, k_1=\xi+i \eta$; it is given by
\begin{equation}
\label{AKNSorig-1sol}
q_{NLS}(x,t)= 2 \eta \text{sech}( 2 \eta(x-4 \xi t- x_0)) e^{-2i \xi x+4i(\xi^2-\eta^2)t-i \psi_0}\;,
\end{equation}
where $e^{2 \eta x_0}= |C_1(0)|/(2 \eta), \psi_0 = arg(C_1(0))-\pi/2$. 
We also note that the above symmetries in scattering space imply that $r(x,t)$ given 
by Eq.~(\ref{Solr}) automatically satisfy the physical symmetry (\ref{AKNSorigsym1}).
%%%%%%%%%%%%%%%%%%%%%%%%%%%%%%%%%%%%%%%%%%
\subsection{Reverse time AKNS symmetry: $r(x,t) = \sigma q(x,-t), ~q \in \mathbb{C}$}
%%%%%%%%%%%%%%%%%%%%%%%%%%%%%%%%%%%%%%%%%%%
The  solution corresponding to the physical symmetry
\begin{equation}
\label{AKNSsym1}
r(x,t) = \sigma q(x, -t) \;,
\end{equation}
of the corresponding nonlocal in time NLS Eq.~(\ref{complex-time-nonloc-NLS}) can be obtained by employing the following temporal symmetries in scattering space
\begin{equation*}
 \bar{b}(k,t)=-\sigma b(-k,-t) \;,
\end{equation*}
\begin{equation*}
\bar{C}(\bar{k}_j, t)=  C(k_j,-t), ~\sigma=-1 \;,
\end{equation*}
and we denote 
\begin{equation*}
\bar{C}(\bar{k}_j, t)= \bar{C}_j(t) ~\text{and} ~ C(k_j,t)=C_j(t) \;.
\end{equation*}
The symmetries at $t=0$ satisfy
\begin{equation}
\label{AKNSsymcontA}
\bar{a}(k,0)=-a^*(-k,0), \bar{b}(k,0)=-\sigma b(-k,0), ~k \in \mathbb{R}
\end{equation}
\begin{equation}
\label{AKNSsymdiscA}
\sigma=-1: \bar{k}_j=-k_j,  \bar{C_j}(0)= C_j(0), ~j=1,2,...J
\end{equation}
Further details of how to obtain these this symmetries are given in the 
appendix (see also \cite{Ablowitz2}). With the symmetries: $\bar{k}_1=-k_1$ 
and $ \bar{C}_1(0)= C_1(0)$  and using  the above time dependence 
for $C_1(t), \bar{C}_1(t)$ the nonlocal in time NLS 
equation (\ref{complex-time-nonloc-NLS}) has the following one soliton solution
\begin{equation}
\label{1solTDNLSq}
q_{TNLS}(x,t)=  \frac{ 2i C_1(0) e^{2i k_1x}e^{4ik_1^2t}} {1+\frac{C_1^2(0)}{4k_1^2} e^{4ik_1x} } \;,
\end{equation}

\begin{equation}
\label{1solTDNLSr}
r_{TNLS}(x,t)=  -\frac{ 2i C_1(0) e^{2i k_1x}e^{-4ik_1^2t}} {1+\frac{C_1^2(0)}{4k_1^2} e^{4ik_1x} } \;.
\end{equation}
One can see that the symmetry condition $r(x,t)=-q(x,-t)$ is automatically satisfied.
With $k_1=\xi+i\eta$ another form of the solution is 
\begin{equation}
\label{1solTDNLSqB}
q_{TNLS}(x,t)=  \frac{ 2i C_1(0) e^{2i \xi x} e^{4i(\xi^2-\eta^2)t} e^{-2\eta x}  e^{-8\xi \eta t} } {1+\frac{C_1^2(0)}{4k_1^2} e^{4i\xi x} e^{-4 \eta x } } \;.
\end{equation}
Note that as $|x| \rightarrow \infty $, $q_{TNLS}(x,t) \rightarrow 0$, but as $\xi t \rightarrow -\infty, q_{TNLS}(x,t)  \rightarrow \infty$ so in general it is an unstable solution.  If we write
\begin{equation*}
\frac{C_1(0)}{2k_1}= e^{2 \eta x_0}e^{-2i\psi_0}  \;,
\end{equation*}
then a singularity can occur when 
\begin{equation*}
 1+ e^{4i(\xi x-\psi_0)}e^{-4\eta(x-x_0)}=0 \;,
 \end{equation*}
or when 
\begin{equation*}
 x=x_0, ~~4(\xi x_0-\psi_0)= (2n+1) \pi, ~n  \in \mathbb{Z} \;.
\end{equation*}
When we take a special case: $\xi=0$  the solution is stable; it can be singular depending on $C_1(0)$; but if we further take $C_1(0)= |C_1(0)|$ so that $\psi_0=0$, and call
$|C_1(0)|/(2 \eta)= e^{-2 \eta x_0}$ we find
\begin{equation}
\label{1solTDNLSqBeta}
q_{TNLSR}(x,t)= 2 \eta \text{sech} [2\eta(x-x_0)] e^{4i \eta^2 t} \;,
\end{equation}
which is not singular. We note that from Eq.~(\ref{AKNSorig-1sol}) the one soliton solution of NLS with $\xi=0$ is given by
\begin{equation}
\label{AKNS-1solTNLS}
q_{TNLS}(x,t)= 2 \eta \text{sech}( 2 \eta(x-x_0)) e^{-4i\eta^2t-i \psi_0} \;,
\end{equation}
which is the same solution as given above in Eq.~(\ref{1solTDNLSqBeta}) but with $\psi_0=0.$ Indeed, $\psi_0=0$ is necessary for this to be a solution 
of the (\ref{complex-time-nonloc-NLS}) equation.  
Indeed any solution to the classical NLS (\ref{NLS}) that satisfies the property
 \begin{equation}
\label{property}
q^*(x,t) = q(x,-t)\;,
\end{equation}
automatically satisfies the corresponding nonlocal (in time) NLS 
equation (\ref{complex-time-nonloc-NLS}). This holds when the solution (\ref{AKNS-1solTNLS}) obeys $\psi_0=0.$ In this regard, we also note that the solution
 \begin{equation}
\label{sol-2}
q(x,t) = \eta \text{tanh} (\eta x) e^{2i\eta^2 t}\;,
\end{equation}
with nonzero boundary conditions $q(x,t) \sim \pm \eta e^{2i\eta^2 t}$ as $x \rightarrow \pm \infty,$ which 
is a ``dark" soliton solution of the classical NLS Eq.~(\ref{NLS}),  solves 
Eq.~(\ref{complex-time-nonloc-NLS}) with $\sigma= 1$. 
%%%%%%%%%%%%%%%%%%%%%%%%%%%%%%%%%%%%%%%%%%%
\subsection{$PT$ Symmetry: $r(x,t) = \sigma q^*(-x,t) $}
%%%%%%%%%%%%%%%%%%%%%%%%%%%%%%%%%%%%%%%%%%%%
The physical $PT$ symmetry (associated with solitons) considered in \cite{AblowitzMusslimani,AblowitzMusslimaniNonlinearity} was
\begin{equation}
\label{PTsym-complex1}
r(x,t) = \sigma q^*(-x,t) \;.
\end{equation}
The corresponding continuous and discrete symmetries in scattering space are given by
\begin{equation}
\label{PTsymCont-a-abar}
a(k,t)=a^*(-k^*,t)=a(k,0), ~\bar{a}(k,t)=\bar{a}^*(-k^*,t) =\bar{a}(k,0)\;,
\end{equation}
%%%%%%%%%%%%%
\begin{equation}
\label{PTsymCont-b-bbar}
\bar{b}(k,t)=\sigma b^*(-k,t),~ k \in \mathbb{R}\;.
\end{equation}
When $\sigma= -1$ there are soliton eigenvalues
\begin{equation*}
\label{PTsymDisck*}
 ~k_j=-k_j^*, ~\bar{k}_j= -\bar{k}_j^*,  ~ j=1,2...,J\;.
\end{equation*}
We calculate the norming constants from
\begin{equation*}
\label{PTsymDisckb*}
%\bar{a}(k)=a^*(k^*), ~
C_j(0)=b_j/a'(kj), b_j=e^{i\theta_j}, \theta_j \in \mathbb{R},  ~j=1,2...,J 
\end{equation*}
\begin{equation}
\label{PTsymDiscCj}
%\bar{a}(k)=a^*(k^*), ~
~\bar{C}_j(0)= \bar{b}_j/\bar{a}'(\bar{k}_j), \bar{b}_j=e^{i\bar{\theta}_j},\bar{\theta}_j \in \mathbb{R}, ~j=1,2...,J 
\end{equation}
and  the terms   $a'(kj),\bar{a}'(\bar{k}_j)$ are computed via the trace formulae \cite{AblowitzMusslimaniNonlinearity}.
When $J=1$ the eigenvalues are on the imaginary axis: $k_1= i \eta, \bar{k}_1= -i \bar{\eta}, \eta>0,\bar{\eta}>0$; then the trace formulae gives
\begin{equation}
\label{PTsym}
%a(k)=a^*(-k^*), \bar{a}(k)=\bar{a}^*(-k^*), ~
C_1(0)= i(\eta+\bar{\eta})e^{i \theta}, \bar{C}_1(0)= -i(\eta+\bar{\eta})e^{i \bar{\theta}}\;,
\end{equation}
the 1 soliton solution of the $PT$ symmetric nonlocal NLS Eq.~(\ref{PTNLS-1}) with the reduction
\begin{equation*}
r(x,t) = \sigma q^*(-x,t) \;,
\end{equation*} 
is found to be
\begin{equation}
\label{PTsym1sol}
q_{PT}(x,t)= \frac{2(\eta+\bar{\eta})e^{i \bar{\theta}} e^{-2\bar{\eta}x-4i\bar{\eta}^2t}}{1-e^{i (\theta+ \bar{\theta})}e^{-2(\eta+ \bar{\eta})x+4i(\eta^2-\bar{\eta}^2t)}   }\;.
\end{equation}
An alternative form of writing the above 1-soliton solution (\ref{PTsym1sol}) is 
\begin{eqnarray}
\label{one-soliton}
q(x,t)=\frac{(\eta + \bar{\eta})  e^{i( \bar{\theta}-\theta -\pi)/2}  e^{-( \bar{\eta}- \eta  )x}e^{-2i(\eta^2 + \bar{\eta}^2)t}}
%%%%%%%
{\cosh\left[(\eta + \bar{\eta})x - 2i(\eta^2 - \bar{\eta}^2)t -i(\theta + \bar{\theta}+\pi)/2
\right]}\;.
\end{eqnarray}
Next, some remarks are in order.
%%%%%%%%%%%%%%%%%%%%%%%%%%%%%%%%%%%%%%
\begin{itemize}
%%%%%%%%%%%%%
\item The solution $q(x,t)$ given in (\ref{PTsym1sol}) is doubly periodic in time with periods given by $T_1= \frac{\pi }{2\bar{\eta}^2}$ and 
$T_2= \frac{\pi }{2(\eta^2 - \bar{\eta}^2)}.$
%%%%%%%%%%%%
\item The intensity $|q(x,t)|^2$ breathes in time with period given by 
$T= \frac{\pi}{2(\eta^2 - \bar{\eta}^2)}$
%%%%%%%%%%%%
\item The solution (\ref{PTsym1sol}) can develop a singularity in finite time.
Indeed, at the origin ($x=0$) the solution (\ref{one-soliton}) becomes singular when
 \begin{eqnarray}
t_n = \frac{2n\pi -(\theta+\overline{\theta})}{4(\eta^2 - {\overline{\eta}}^2)}\;,
n\in\mathbb{Z}\;.
  \end{eqnarray}
%%%%%%%%%%%%
\item The solution (\ref{PTsym1sol}) is characterized by two important time scales: the 
  singularity time scale and the  periodicity of breathing.
  %%%%%%%%%%%%
  \item A  feature of this solution of  (\ref{PTsym1sol}) (and other singular solutions discussed in this paper) is that it can be defined after singularity has developed; i.e. 
it has a pole in time and it can be avoided in the complex time plane; i.e. the solution is of Painlev\'e type.
%%%%%%%%%%%%%  
\item
We recall that not all members of the one-soliton family develop a singularity at finite time. Indeed, if one let $\eta = \bar{\eta} \equiv \eta$ in (\ref{PTsym1sol})
then we arrive at the well behaved soliton solution of the nonlocal $PT$ symmetric NLS Eq.~(\ref{PTNLS-1})  
\begin{equation}
\label{non-sing}
q(x,t) = 2\eta {\rm sech} [2\eta x - i\theta ] e^{-4 i \eta^2 t}\;,
\end{equation}
where $\eta$ and $\theta$ are arbitrary real constants. 
\end{itemize}
%%%%%%%%%%%%%%%%%%%%%%%%%%%%%%%%%%%%
Note that when $\theta \ne 0$
the soliton given (\ref{non-sing}) is not a solution to the classical (local) NLS Eq.~(\ref{NLS}). The $PT$ symmetric induced potential is given by (see Eq.~(\ref{PTNLS-1V}))
\begin{equation}
\label{non-singV1}
V \equiv q(x,t)q^*(-x,t) = 4\eta^2 {\rm sech}^2 [2\eta x - i\theta ]\;.
\end{equation}
The real and imaginary parts of the induced potential are respectively given by
%%%%%%%%%%%%%%%%%%%%%%%
\begin{equation*}
\label{non-sing-real}
V_R = \frac{4\eta^2\left[\cos^2 \theta \cosh^2 (2\eta x) 
- \sin^2 \theta \sinh^2 (2\eta x)\right]}
{\left[\cos^2 \theta \cosh^2 (2\eta x) 
+ \sin^2 \theta \sinh^2 (2\eta x)\right]^2}
\end{equation*}
%%%%%%%%%%%%%%%%%%%%%%%%%%
\begin{equation*}
\label{non-sing-Img}
V_I = \frac{\sin (2 \theta) \sinh (4\eta x)}
{2\left[\cos^2 \theta \cosh^2 (2\eta x) 
+ \sin^2 \theta \sinh^2 (2\eta x)\right]^2}
\end{equation*}
%%%%%%%%%%%%%%%%%%%%%%%%%
%%%%%%%%%%%%%%%%%%%%%%%%%%%%%%%%%%%%%%%%%%%%
\subsection{Reverse space-time symmetry: $r(x,t) = \sigma q(-x,-t), q \in \mathbb{C} $}
%%%%%%%%%%%%%%%%%%%%%%%%%%%%%%%%%%%%%%%%%%%%
The corresponding continuous and discrete symmetries in scattering space are given by
\begin{equation}
\label{PTsymCont-new}
%\bar{a}(k)=a^*(k^*), ~
\bar{b}(k,t)=\sigma b(k,-t),~ k \in \mathbb{R} \;.
\end{equation}
When $\sigma= -1$ we calculate the norming constants from
\begin{equation*}
\label{PTsymDisckb*-new}
C_j(0)=b_j/a'(kj), b_j=e^{i\theta_j}, \theta_j \in \mathbb{R},  ~j=1,2...,J \;,
\end{equation*}
where the terms $a'(kj),\bar{a}'(\bar{k}_j)$ are computed via the trace formulae \cite{AblowitzMusslimaniNonlinearity}. Following the same procedure as in \cite{AblowitzMusslimaniNonlinearity} we also find
\begin{equation}
\label{sym-x-t,k,b2}
b(k_j,-t)b(k_j,t)=1 \Rightarrow b(k_j,0)= \pm1 \;,
\end{equation}
and 
\begin{equation}
\label{sym-x-t,k,b22}
\bar{b}(k_j,-t)\bar{b}(k_j,t)=1  \Rightarrow \bar{b}(k_j,0)= \pm1 \;.
\end{equation}
For a one soliton solution, $\sigma=-1, J=1$, the trace formulae yield
\begin{equation}
\label{sym-x-t,k,b2-trace}
a'(k_1)=  \frac{1}{k_1-\bar{k}_1},~~ \bar{a}'(\bar{k}_1)=  \frac{1}{k_1-\bar{k}_1} ~  \Rightarrow
\bar{a}'(\bar{k}_1)=-a'(k_1) \;.
\end{equation}
Thus
\begin{equation*}
C_1(0)=  2(k_1-\bar{k}_1)b(k_1,0), \bar{C}_1(0)= -2(k_1-\bar{k}_1)\bar{b}(k_1,0) \;.
\end{equation*}
This implies that
\begin{equation*}
C^2_1(0) =  \bar{C}^2_j(0) \;.
\end{equation*}
The one soliton solution of the complex space-time nonlocal 
NLS equation (\ref{complex-space-time-nonloc-NLS}) is again found using the above 
method with time evolution of the scattering data. We have 
\begin{equation}
\label{1solAq2}
q(x,t)=  \frac{2i \bar{C}_1(0) e^{-2i\bar{k}_1x} e^{-4i \bar{k}_1^2 t} }{1+\frac{C_1(0)\bar{C}_1(0)}{(k_1-\bar{k}_1)^2}e^{2i(k_1-\bar{k}_1)x} e^{4i( \bar{k}_1^2-k_1^2) t} }\;,
\end{equation}
and 
\begin{equation}
\label{1solAr2}
r(x,t)= - \frac{2i C_1(0) e^{2ik_1x}e^{4i k_1^2 t} } {1+\frac{C_1(0)\bar{C}_1(0)}{(k_1-\bar{k}_1)^2} e^{2i(k_1-\bar{k}_1)x}  e^{4i( \bar{k}_1^2-k_1^2) t}  }\;.
\end{equation}
With $C^2_1(0) =  \bar{C}^2_j(0)$ it follows that $r(x,t)=-q(-x,-t)$.
Calling $k_1= \xi_1+i\eta_1, \bar{k_1}= \bar{\xi}_1 -i\bar{\eta_1}, ~\eta_1>0, \bar{\eta}_1>0$ and the above time dependence for $C_1(t), \bar{C}_1(t)$ leads to the one soliton solution for Eq.~(\ref{complex-space-time-nonloc-NLS})
\begin{equation}
\label{CSTNLSsol}
q_{CSTNLS}(x,t)= 
 \frac{2i C_1(0) e^{-2i \bar{\xi}_1x-2 \bar{\eta}_1x} e^{-4i (\bar{\xi}_1^2 - \bar{\eta}_1^2)t}  e^{8\bar{\xi}_1\bar{\eta}_1t}  }{1+\Gamma_1 \Delta} \;,
\end{equation}
where 
\begin{equation*}
 \Delta= 
 e^{-4i(\xi_1^2-\eta_1^2)t +4i (\bar{\xi}_1^2 - \bar{\eta}_1^2)t} e^{8\xi_1\eta_1t+8\bar{\xi}_1\bar{\eta}_1t} 
e^{-2i(\xi_1-\bar{\xi}_1)x} e^{-2(\eta_1+\bar{\eta}_1)x} \;,
\end{equation*}
and $ \Gamma_1=C_1(0)\bar{C}_1(0)/[k_1-\bar{k}_1]^2=\gamma_1=\pm 1$.The above soliton is stable in the sense that as $\bar{\xi}_1\bar{\eta}_1 \rightarrow \infty$ find $q_{CSTNLS}(x,t) \rightarrow 0$. It also appears that if we let $\Gamma_1= e^{2(\eta_1+\bar{\eta}_1)x_0)} e^{2 i \psi_0}$ we can have a singularity when 
\begin{equation*}
 -2(\eta_1+\bar{\eta}_1)(x-x_0) +8(\xi_1\eta_1+\bar{\xi}_1\bar{\eta}_1)t =0 \;,
\end{equation*}
and
\begin{equation*}
 4((\bar{\xi}_1^2 - \bar{\eta}_1^2) -(\xi_1^2-\eta_1^2) )t +2  \psi_0 = (2n+1)\pi, ~ n\in \mathbb{Z} \;.
\end{equation*}
The singularity can be eliminated by taking $(\bar{\xi}_1^2 - \bar{\eta}_1^2) -(\xi_1^2-\eta_1^2) =0$ and $2  \psi_0 \neq (2n+1)\pi, ~ n\in \mathbb{Z}.$
As shown, the above symmetries yield solutions of NLS and nonlocal NLS type  equations.
%%%%%%%%%%%%%%%%%%%%%%%%%%%%%%%%%%%%%%%%%%
\subsection{Complex reverse time symmetry:  $r(x,t) = \sigma q^*(-x,-t)$}
%%%%%%%%%%%%%%%%%%%%%%%%%%%%%%%%%%%%%%%%%%
This symmetry yields a solution of the complex space-time nonlocal mKdV 
equation (\ref{complex-space-time-nonloc-mkdv}). The symmetries needed for this case are
\[ a(k,t)=a^*(-k^*,-t)=a(k,0), \]
\[ \bar{a}(k,t)=\bar{a}^*(-k^*,-t)=\bar{a}^*(-k^*,0),\] 
\[  \bar{b}(k,t)=\sigma b^*(-k,-t), ~k \in \mathbb{R} \]
When $\sigma=-1$
\begin{equation}
\label{symcmkdvk}
k_1=i \eta, \eta>0, ~\bar{k}_1=-i \bar{\eta}, \bar{\eta}>0\;,
\end{equation}
\begin{equation}
\label{symcmkdvC}
~C_1(t)=   C_1(0)  e^{8\eta^3 t}\;,
\end{equation}
\begin{equation}
\label{symcmkdv1Cb}
 \bar{C}_1(t)= \bar{C}_1(0)e^{8\bar{\eta}^3 t}\;,
 \end{equation}
\begin{equation*}
 C_1(0)= i(\eta+\bar{\eta})b_1, b_1=e^{i \theta}, ~~\theta \in \mathbb{R} \;,
\end{equation*}
\begin{equation*}
 \bar{C}_1(0)= - i (\eta+\bar{\eta})\bar{b}_1, \bar{b}_1=e^{i \bar{\theta}}, ~~\bar{\theta} \in \mathbb{R} \;.
\end{equation*}
Substituting into Eq.~(\ref{Solq}) yields the one soliton solution of the complex nonlocal mKdV equation
\begin{equation}
\label{solcmkdv}
q(x,t)= -\frac{2(\eta+\bar{\eta})e^{i\bar{\theta}} e^{-2 \bar{\eta}x+8\bar{\eta}^3t}}{ 1+e^{i(\theta +\bar{\theta})} e^{-2\eta x + 8\eta^3t -2 \bar{\eta}x +8\bar{\eta}^3t }} \;.
  \end{equation}
  We see that there are four real parameters in the above solution: $\eta,\bar{\eta},\theta, \bar{\theta}.$ Another way to write this solution is as follows
\begin{eqnarray}
\label{one-solitoncmkdv}
q(x,t)=\frac{(\eta + \bar{\eta})e^{-i(\theta/2+\bar{\theta}/2 +\pi)}   e^{ \eta (x-4\eta^2 t) }  e^{-\bar{ \eta} (x-4\bar{\eta}^2 t)  } }
%%%%%%%
{\cosh\left[(\eta(x - 4(\eta^2t) + \bar{\eta}(x- 4 \bar{\eta}^2)t -i(\theta + \bar{\theta})/2
\right]}\;.
\end{eqnarray}
We see that this solution can be singular if  $\theta +\bar{\theta}= (2n+1) \pi, ~n \in \mathbb{Z}$.
%%%%%%%%%%%%%%%%%%%%%%%%%%%%%%%%%%%%%%%%%%%%
\subsection{Real reverse space-time symmetry: $r(x,t) = \sigma q(-x,-t), ~q 
\in \mathbb{R}$}
%%%%%%%%%%%%%%%%%%%%%%%%%%%%%%%%%%%%%%%%%%%%%
There is only one change from the complex $PT$ time reversal symmetry case,
\begin{equation}
\label{PTsym-x1}
C_1(0)= i(\eta+\bar{\eta})b_1, \bar{C_1}(0)= -i(\eta+\bar{\eta})\bar{b}_1 \;,
\end{equation}
but now with
\[b_1=\pm1, \bar{b}_1=\pm1. \] 
Thus the only difference from the complex $PT$ time reversal symmetry case is that in the prior case we require $\theta,\bar{\theta}=0,\pi$.
Therefore, in this case there are only two free real parameters $\eta, \bar{\eta}$ and the
real nonlocal mKdV Eq.~(\ref{real-space-time-nonloc-mkdv}) the one soliton solution is given by
\begin{equation}
\label{solrmkdv}
q(x,t)= \frac{2\gamma_1(\eta+\bar{\eta}) e^{-2 \bar{\eta}x+8\bar{\eta}^3t} }{ 1+\gamma_2 e^{-2\eta x + 8\eta^3t -2 \bar{\eta}x +8\bar{\eta}^3t }}\;,
\end{equation}
where $\gamma_j= \pm 1, j=1,2$. If, say $\gamma_1=\gamma_2=1$ then the solution can be written in the following form
\begin{eqnarray}
\label{one-solitonrmkdv}
q(x,t)=\frac{(\eta + \bar{\eta})   e^{ \eta (x-4\eta^2 t) }  e^{-\bar{ \eta} (x-4\bar{\eta}^2 t)  } }
%%%%%%%
{\cosh\left[(\eta(x - 4\eta^2t) + \bar{\eta}(x- 4 \bar{\eta}^2)t
\right]}\;.
\end{eqnarray}
This solution is not singular. When $\eta=\bar{\eta}$ the solution reduces to the well-known solution of the real mKdV equation
\begin{eqnarray}
\label{one-solitonrmkdv-2}
q(x,t)=\frac{2 \eta}  
%%%%%%%
{\cosh\left[(2\eta(x - 4\eta^2t)
\right]}\;.
\end{eqnarray}
Finally, we construct soliton solution for the (real) space-time nonlocal sine-Gordon 
Eq.~(\ref{q-sG-final}). The sG equation belongs to the same symmetry class as the 
space-time nonlocal mKdV equation. As such, for the one soliton solution, the eigenvalues are given by $k_1=i\eta_1$ and $\bar{k}_1 = -i\bar{\eta}_1$ with $\eta_1>0$ 
and $\bar{\eta}_1>0.$ Furthermore,  the evolution of the norming constants is given by
Eqns.~(\ref{Cj-evolv-SG}) and (\ref{Cj-bar-evolv-sG}):
 %%%%%%%%%%%%%%%%%
 \begin{equation}
\label{Cj-evolv-SG-1}
C^{{\rm sG}}_1(t) = C_1(0)e^{-t/(2\eta_1)}\;,
\end{equation}
\begin{equation}
\label{Cj-bar-evolv-sG-1}
 \overline{C}^{{\rm sG}}_1(t) = \overline{C}_1(0)e^{-t/(2\overline{\eta}_1)}.
 \end{equation}
%%%%%%%%%%%%%%%%%
The solution is thus found from Eq.~(\ref{Solq}) to be\\
\begin{equation}
\label{Solq-sG-final}
q(x,t)=  \frac{2i e^{2i\bar{\eta}_1x}\overline{C}_1(0)e^{-t/(2\overline{\eta}_1)}}
{1- \frac{C_1(0)\bar{C}_1(0)e^{-t/(2\eta)}}
{(\eta_1+\bar{\eta}_1)^2e^{-2(\eta_1+\bar{\eta}_1)x}}} \;,
\end{equation}
%%%%%%%%%%%%%%%%
where $C_1(0)= i(\eta+\bar{\eta})b_1, \bar{C_1}(0)= -i(\eta+\bar{\eta})\bar{b}_1, ~b_1= \pm 1,  \bar{b}_1= \pm 1$ and 
\[
\frac{1}{\eta} = \frac{1}{\eta_1} + \frac{1}{\bar{\eta}_1}\] 
%%%%%%%%%%%%%%%%%%%%%%%%%%%%%%%%%%%%%%%%%%%%%
\section{Conclusion and outlook} 
\label{conclusion}
%%%%%%%%%%%%%%%%%%%%%%%%%%%%%%%%%%%%%%%%%%%%%
More than forty years has passed since AKNS published their paper:
``Inverse scattering transform - Fourier analysis for nonlinear problems" which appeared 
in this journal in1974. Until recently, it was thought that all ``simple" and physically relevant symmetry reductions of the ``classical" AKNS scattering problem had been identified. 
But, in 2013, the authors discovered a new ``hidden" reduction of the $PT$ symmetric type which leads to a nonlocal NLS equation that admits a novel soliton solution. 
Surprisingly enough, the AKNS symmetry reduction found in \cite{AblowitzMusslimani} is not the end of the story.  In this paper we unveil many new ``hidden" symmetry reductions that are nonlocal both in space and time and, in some cases, nonlocal in time-only. Each new symmetry condition give rise to its own new nonlocal nonlinear integrable evolution equation. These include the reverse time NLS equation, reverse space-time nonlocal forms of the NLS equation, derivative NLS equation (which includes the reverse space-time nonlocal derivative NLS equation as a special case), loop soliton, modified Korteweg-deVries (mKdV), sine-Gordon, 
(1+1) and (2+1) dimensional multi-wave/three-wave interaction, reverse discrete-time nonlocal discrete integrable NLS models and Davey-Stewartson equations. Linear Lax pairs and an infinite number of conservation laws are discussed along with explicit soliton solutions in some cases. All equations arise from remarkably simple symmetry reductions of AKNS and related scattering problems.  For convenience, below we list some of the symmetries associated with the AKNS scattering problem (\ref{AKNS}-\ref{LAX-2}). 
%%%%%%%%%%%%%%%
\begin{equation}
\label{AKNSorigsym}
r(x,t) = \sigma q^*(x,t) \;,
\end{equation}
%%%%%%%%%%%%%%%
\begin{equation}
\label{PTsym-complex}
r(x,t) = \sigma q^*(-x,t) \;,
\end{equation}
%%%%%%%%%%%%%%%
\begin{equation}
\label{sym-reduction-intro-3c}
r(x,t)=\sigma q(-x,-t), ~q \in \mathbb{C} \;,
\end{equation}
%%%%%%%%%%%%%%%
%%%%%%%%%%%%%%%
\begin{equation}
\label{AKNSsym}
r(x,t) = \sigma q(x,-t), ~q \in \mathbb{C} \;,
\end{equation}
%%%%%%%%%%%%%%%
\begin{equation}
\label{PTsymt-complex-2}
r(x,t)=\sigma q^*(-x,-t) \;,
\end{equation}
%%%%%%%%%%%%%%%
\begin{equation}
\label{sym-reduction-intro-3r}
r(x,t)=\sigma q(-x,-t), ~~~q \in \mathbb{R} \;,
\end{equation}
%%%%%%%%%%%%%%%
where $\sigma \mp 1$. 
%%%%%%%%%%%%%%%
In future work these symmetries will be extended to other vector, matrix AKNS and $2+1$ dimensional AKNS type systems. The symmetry (\ref{AKNSorigsym}) was discussed in \cite{AKNS} along with the subcase $r(x,t) = \sigma q(x,t), q \in \mathbb{R}$. The symmetry (\ref{PTsym-complex}) was first discussed in \cite{AblowitzMusslimani}, particularly with application to the $PT$ symmetric NLS equation and related hierarchies. The symmetry (\ref{sym-reduction-intro-3c}) was first noted in \cite{AblowitzMusslimaniNonlinearity} with regard to the nonlocal mKdV and SG equations, though the IST and one soliton solutions were not given there. \\\\
We show here that the symmetries (\ref{AKNSorigsym}), (\ref{PTsym-complex}), 
(\ref{sym-reduction-intro-3c}) and (\ref{AKNSsym}) are all associated with the  IST and solutions of the NLS and nonlocal NLS equations while the symmetries (\ref{AKNSorigsym}, 
(\ref{PTsymt-complex-2}) and (\ref{sym-reduction-intro-3r}) are associated with the IST and solutions of the mKdV and nonlocal SG equation.
\\\\
%%%%%%%%%%%%%%%%%%%%%%%%%%%%%%%%%%%%%%%%%%%%%
%%%%%%%%%%%%%%%%%%%%%%%%%%%%%%%%%%%%%%%%%%%%%
We close this section with an outlook towards future research direction pertaining to the emerging field of integrable nonlocal equations including what we here term here as {\it reverse space-time and reverse time systems}.
%%%%%%%%%%%%%%%%%%%%%%%%%%%%%%%%%%%%%%%%%%%%%
\begin{enumerate}
%%%%%%%%%%%%%%%%%%%%%%%%%%%%%%%%%%%%%%%%%%%%%
\item {\it Inverse scattering transform and left-right Riemann-Hilbert (RH) problems for reverse space-time and inverse scattering for the reverse time-only nonlocal NLS type equations}. \\
%%%%%%%%%%%%%%%%%%%%%%%%%%%%%%%%%%%%%%%%%%%%%
In~\cite{AblowitzMusslimani,AblowitzMusslimaniNonlinearity}, it was shown that a ``natural" approach to solve the inverse problem associated with the nonlocal NLS equation (\ref{PTNLS-1}) is to formulate two separate RH problems: one for $x<0$ (left) and one at $x>0$ (right) then use the appropriate (nonlocal) symmetries between the eigenfunctions to reduce the number of independent equations and recover the 
potentials $q$ and $r$. The left-right RH approach has the advantage of reducing the integral equations on the inverse side to integral equations for one function.  It will be valuable  to develop the left-right RH equations for the reverse space-time nonlocal equations and thereby develop a more complete inverse scattering theory. Indeed, inverse scattering is an important field of mathematics and physics independent of solving nonlinear equations.
%%%%%%%%%%%%%%%%%%%%%%%%%%%%%%%%%%%%%%%%%%%%%
\\
\item {\it Nonlocal Painlev\'e type equations}. \\\\
%%%%%%%%%%%%%%%%%%%%%%%%%%%%%%%%%%%%%%%%%%%%%
The Painlev\'e equations are certain class of nonlinear second-order complex ordinary differential equations that usually arise as reductions of the ``soliton evolution equations" which are solvable by IST cf. \cite{Ablowitz1}.  They are particularly interesting due to their properties in the complex plane and their associated integrability properties. The first nonlocal (in space) Painlev\'e type equation was obtained in \cite{AblowitzMusslimani} and came out of a reduction of Eq.~(\ref{PTNLS-1}). Using the ansatz
\begin{equation}
\label{simantz}
q(x,t) = \frac{1}{(2t)^{1/2}}f(z)e^{i \nu \log t /2 }, \;\;\;\;\; z = \frac{x}{(2t)^{1/2}}\;,
\end{equation}
one can show that $f(z)$ satisfies
\begin{equation}
\label{P-space-nonloc}
f_{zz}(z) + izf_z(z)  + (\nu+ i) f(z)  - 2 \sigma f^2(z)f^*(-z) = 0\;,
\end{equation}
where $\sigma = \mp 1$. Since Eq.~(\ref{P-space-nonloc}) comes out of Eq.~(\ref{PTNLS-1}) which, in turn arose using the so-called $PT$ preserving symmetry reduction 
$r(x,t) = \sigma q^*(-x,t),$ we thus refer to (\ref{P-space-nonloc}) as a $PT$ preserving 
Painlev\'e equation. The situation for the reverse space-time and reverse time only nonlocal NLS cases 
is different. Here, the proper ansatz we use for the reduction to ODE is of the form
\begin{equation}
\label{simantz-2}
q(x,t) = \frac{1}{(2t)^{1/2}}f(z), \;\;\;\;\; z = \frac{x}{(2t)^{1/2}}.
\end{equation}
Substituting this ansatz into Eq.~(\ref{complex-time-nonloc-NLS}) gives
\begin{equation}
\label{P-time-nonloc}
f_{zz}(z) + izf_z(z) + i f(z) - 2 \sigma \kappa f^2(z)f(\kappa z) = 0, 
\end{equation}
where $ \sigma = \mp 1$ and $\kappa = (-1)^{-1/2}$. 
In this case, $\kappa = i $ if one chooses 
$-1=e^{-i \pi}$ and $(-1)^{-1/2}= e^{i \pi/2}$ but $\kappa = -i $ if one chooses 
$-1=e^{i \pi}$ and $(-1)^{-1/2}= e^{-i \pi/2}$, i.e., it is branch dependent. 
Since the number $\kappa$  is branch dependent, it can wait to be defined when one 
does an application. On the other hand, from equation (\ref{complex-space-time-nonloc-NLS}) one obtains the following ODE reduction
\begin{equation}
\label{P-space-time-nonloc}
f_{zz}(z) + izf_z(z) + i f(z)  - 2\sigma \kappa f^2(z) f(-\kappa z) = 0, 
\end{equation}
with $\sigma = \mp 1$. Equations (\ref{P-time-nonloc}) and (\ref{P-space-time-nonloc})  are nonlocal Painlev\'e type equations.
As a future research direction, it would be interesting to study the behavior of solutions to the above new nonlocal Painlev\'e equations.\\
%%%%%%%%%%%%%%%%%%%%%%%%%%%%%%%%%%%%%%%%%%%%%
\item {\it Inverse scattering transform for the reverse time discrete and the reverse discrete-time nonlocal nonlinear Schr\"odinger equations}.\\\\
 In Sec.~\ref{discrete-sys} we used various discrete symmetry reductions based on the Ablowitz-Ladik scattering problem to obtain two new discrete nonlocal in both ``space" and time  nonlinear Schr\"odinger equations. A future research direction would be to develop the full inverse scattering transform and obtain soliton 
solutions of these equations.
\end{enumerate}
%%%%%%%%%%%%%%%%%%%%%%%%%%%%%%%%%%%%%%%%%%%%%
\section{Acknowledgements} 
The research of M.J.A. was partially supported by NSF 
under Grant No. DMS-1310200. We thank Dr. Xudan Luo for helpful interactions.  \
%%%%%%%%%%%%%%%%%%%%%%%%%%%%%%%%%%%%%%%%%%%
%%%%%%%%%%%%%%%%%%%%%%%%%%%%%%%%%%%%%%%%%%%
\section{Appendix} 
In this appendix for the physical space symmetries discussed in this paper we will provide the symmetries associated with the AKNS eigenfunctions. 
To do so, we call $v (x,k)\equiv (v_1 (x,k),\; v_2 (x,k))^{{\rm T}}$ a solution to system (\ref{AKNS}). Note: $\sigma=\mp 1$.
%%%%%%%%%%%%%%%%%%%%%%%%%%%%%%%%%%%%%%%%%%%
%%%%%%%%%%%%%%%%%%%%%%%%%%%%%%%%%%%%%%%%%%%
\begin{enumerate}
%%%%%%%%%%%%%%%
\item 
For the standard AKNS symmetry (\ref{AKNSsym}), i.e.,  $r(x,t)=  \sigma q^*(x,t)$ we have
\begin{eqnarray}
\label{sym-barpsi-barphi}
\overline{\psi} (x,t,k)=
\left(\begin{array}{cr}
0&1
\\
\sigma &0
\end{array}\right) \psi^*(x,t,k^*) \;,
\end{eqnarray}
and 
\begin{eqnarray}
\label{sym-barpsi-barphi-2a}
\overline{\phi} (x,t,k)=
\left(\begin{array}{cr}
0&\sigma
\\
1 &0
\end{array}\right) \phi^*(x,t,k^*) \;.
\end{eqnarray}
%%%%%%%%%%%%%%
\item 
For the reverse time AKNS symmetry (\ref{AKNSsym}), i.e., 
$r(x,t)=  \sigma q(x,-t)$ we have
\begin{eqnarray}
\label{sym-barpsi-barphi2}
\overline{\psi} (x,t,k)=
\left(\begin{array}{cr}
0&1
\\
\sigma &0
\end{array}\right) \psi (x,-t,-k) \;,
\end{eqnarray}

and 

\begin{eqnarray}
\label{sym-barpsi-barphi2-2}
\overline{\phi} (x,t,k)=
\left(\begin{array}{cr}
0&\sigma
\\
1 &0
\end{array}\right) \phi (x,-t,-k)  \;.
\end{eqnarray}
%%%%%%%%%%%%%%%%
\item
For the $PT$ symmetry (\ref{PTsym-complex}), i.e.,  $r(x,t)=  \sigma q^*(-x,t)$ we have
\begin{eqnarray}
\label{sym-psi-phi}
\psi (x,t,k)=
\left(\begin{array}{cr}
0&-\sigma
\\
1&0
\end{array}\right)\phi^* (-x,t,-k^*)\;,
\end{eqnarray}
\begin{eqnarray}
\label{sym-barpsi-barphi-a}
\overline{\psi} (x,t,k)=
\left(\begin{array}{cr}
0&1
\\
-\sigma &0
\end{array}\right)\overline{\phi}^* (-x,t,-k^*)\;.
\end{eqnarray}
%%%%%%%%%%%%%%%%%%
\item For the reverse space-time symmetry (\ref{sym-reduction-intro-3c}), i.e., \\
$r(x,t)=  \sigma q(-x,-t), ~q \in \mathbb{C}$ we have
\begin{eqnarray}
\label{sym-psi-phi-2}
\psi (x,t,k)=
\left(\begin{array}{cr}
0&-\sigma
\\
1&0
\end{array}\right)\phi (-x,-t,-k)\;,
\end{eqnarray}
\begin{eqnarray}
\label{sym-barpsi-barphi-2}
\overline{\psi} (x,t,k)=
\left(\begin{array}{cr}
0&1
\\
-\sigma &0
\end{array}\right)\overline{\phi}(-x,-t,-k)\;.
\end{eqnarray}
%%%%%%%%%%%%%%%%%%%
\item
For the complex reverse space-time symmetry (\ref{sym-reduction-intro-3c}), i.e.,\\
$r(x,t)=  \sigma q^*(-x,-t)$ we have
\begin{eqnarray}
\label{sym-psi-phi-3}
\psi (x,t,k)=
\left(\begin{array}{cr}
0&-\sigma
\\
1&0
\end{array}\right)\phi^* (-x,-t,-k^*)\;,
\end{eqnarray}
\begin{eqnarray}
\label{sym-barpsi-barphi-3}
\overline{\psi} (x,t,k)=
\left(\begin{array}{cr}
0&1
\\
-\sigma &0
\end{array}\right)\overline{\phi}^*(-x,-t,-k^*)\;.
\end{eqnarray}
%%%%%%%%%%%%%%%%
\item
For the real reverse space-time symmetry (\ref{sym-reduction-intro-3c}), i.e.,  \\
$r(x,t)=  \sigma q(-x,-t), ~q \in \mathbb{R}$ we have the above symmetry given in item (5) associated with $r(x,t)=  \sigma q^*(-x,-t)$ and 
\begin{eqnarray}
\label{sym-psi-phi-4}
\psi (x,t,k)=
\left(\begin{array}{cr}
0&-\sigma
\\
1&0
\end{array}\right)\phi (-x,-t,-k)\;,
\end{eqnarray}
\begin{eqnarray}
\label{sym-barpsi-barphi-4}
\overline{\psi} (x,t,k)=
\left(\begin{array}{cr}
0&1
\\
-\sigma &0
\end{array}\right)\overline{\phi}(-x,-t,-k)\;.
\end{eqnarray}
\end{enumerate}
The above symmetry relations can be turned into symmetry relations for the scattering data $a(k), b(k)$ and eigenvalues $k_j,\bar{k}_j,~j=1,2...J$ from the Wronskian relations 
(\ref{Wa}), (\ref{Wabar}), (\ref{Wb}) and (\ref{Wbbar}). Finally symmetries for the normalization coefficients $C_j,\bar{C}_j,~ j=1,2...J$ can be found either directly from the above by analytic continuation  or by individually finding $b_j$ and $a'(k_j)$  associated with $C_j=b_j/a'(k_j)$ and  $\bar{b}_j$ and $\bar{a}'(k_j)$  associated with $\bar{C}_j=\bar{b}_j/\bar{a}'(k_j)$ as was done in \cite{AblowitzMusslimaniNonlinearity}.
%%%%%%%%%%%%%%%%%%%%%%%%%%%%%%%%%%%%%%%%%%%
%%%%%%%%%%%%%%%%%%%%%%%%%%%%%%%%%%%%%%%%%%%


\begin{thebibliography}{99}
%%%%%%%%%%%%%%%%%%%%%%%%%%%%%%%%%%%%%%%%%%%
\bibitem{Chen_et.al}
Z. Chen, D.N. Christodoulides and M. Segev. 
Optical spatial solitons: historic overview and recent advances, 
Reports on Progress in Physics 75: 086401 (2012).
%%%%%%%%%%%%%%%%%%%%%%%%%%%%%%%%%%%%%%%%%%%
\bibitem{Lederer_et.al}
F. Lederer, G.I.. Stegeman, D. N. Christodoulides, G. Assanto, M. Segev, and Y. Silberberg, Discrete solitons in optics, 
Physics Reports, 463: 1-126 (2008). 
%%%%%%%%%%%%%%%%%%%%%%%%%%%%%%%%%%%%%%%%%%%
\bibitem{kivshar_et.al}
Y. S. Kivshar and B. Luther-Davies.
Dark optical solitons: physics and applications,
Phys. Rep. 298: 81-197 (1998).
%%%%%%%%%%%%%%%%%%%%%%%%%%%%%%%%%%%%%%%%%%%
\bibitem{kev_et.al}
P. G. Kevrekidis, D. J. Frantzeskakis, and R. Carretero-González.
Emergent Nonlinear Phenomena in Bose-Einstein Condensates: Theory and Experiment,
Springer Verlag, (2007).
%%%%%%%%%%%%%%%%%%%%%%%%%%%%%%%%%%%%%%%%%%%
\bibitem{yangbook}
J. Yang, Nonlinear Waves in Integrable and Non integrable Systems, SIAM Mathematical Modeling and Computation (2010).
%%%%%%%%%%%%%%%%%%%%%%%%%%%%%%%%%%%%%%%%%%%
\bibitem{Ablowitz2} 
M. J. Ablowitz and H. Segur, Solitons and Inverse
Scattering Transform, SIAM Studies in Applied Mathematics Vol. 4, SIAM, 
Philadelphia, PA, (1981).
%%%%%%%%%%%%%%%%%%%%%%%%%%%%%%%%%%%%%%%%%%%%
\bibitem{NOVIKOV} 
S. P. Novikov, S. V. Manakov, L. P. Pitaevskii and V. E. Zakharov. 
Theory of Solitons: The inverse Scattering Method, Plenum (1984).
%%%%%%%%%%%%%%%%%%%%%%%%%%%%%%%%%%%%%%%%%%
\bibitem{Cal-Deg} 
F. Calogero and A. Degasperis,
Spectral transform and solitons I, North Holand (1982).
%%%%%%%%%%%%%%%%%%%%%%%%%%%%%%%%%%%%%%%%%%%%
 \bibitem{GGKM} 
 C. S. Gardner, J. M. Greene, M. D.  Kruskal and R. M. Miura.
Method for Solving the Korteweg-deVries Equation,
Phys. Rev. Lett. 19: 1095-1098 (1967). 
%%%%%%%%%%%%%%%%%%%%%%%%%%%%%%%%%%%%%%%%%%%%
\bibitem{Lax} 
P. D. Lax.
Integrals of nonlinear equations of evolution and solitary waves,
Comm. Pure. Appl. Math. 21: 467-490 (1968).
%%%%%%%%%%%%%%%%%%%%%%%%%%%%%%%%%%%%%%%%%%%
\bibitem{ZS} V. E. Zakharov and A. B. Shabat. 
Exact theory of two-dimensional self-focusing and one-dimensional self-modulation of waves in nonlinear media,
Sov. Phys. JETP 34: 62-69 (1972).
%%%%%%%%%%%%%%%%%%%%%%%%%%%%%%%%%%%%%%%%%%
\bibitem{AKNS} 
M. J. Ablowitz, D. J. Kaup, A. C. Newell and H. Segur.
Inverse scattering transform - Fourier analysis for nonlinear problems,
Stud. Appl. Math. 53: 249-315 (1974). 
%%%%%%%%%%%%%%%%%%%%%%%%%%%%%%%%%%%%%%%%%%%
\bibitem{Ablowitz1} 
M. J. Ablowitz and P. A. Clarkson,
Solitons, Nonlinear Evolution Equations and Inverse Scattering, 
Cambridge University Press, Cambridge, (1991).
%%%%%%%%%%%%%%%%%%%%%%%%%%%%%%%%%%%%%%%%%%%
\bibitem{AblowitzMusslimani} 
M. J. Ablowitz and Z. H. Musslimani.
Integrable nonlocal nonlinear Schr\"odinger equation,
Phys. Rev. Lett. 110: 064105 (2013). 
%%%%%%%%%%%%%%%%%%%%%%%%%%%%%%%%%%%%%%%%%%
\bibitem{Bender} 
C. M. Bender and S. Boettcher.
Real spectra in non-Hermitian Hamiltonians having $PT$ symmetry,
Phys. Rev. Lett. 80: 5243-5246 (1998).
%%%%%%%%%%%%%%%%%%%%%%%%%%%%%%%%%%%%%%%%%%%
\bibitem{RKDM2} 
K. G. Makris, R. El Ganainy, D. N. Christodoulides and Z. H. Musslimani.
 Beam dynamics in $PT$ symmetric optical lattices,
Phys. Rev. Lett. 100: 103904 (2008).
%%%%%%%%%%%%%%%%%%%%%%%%%%%%%%%%%%%%%%%%%%
\bibitem{RKDM3} 
Z. H. Musslimani, K. G. Makris, R. El Ganainy and D. N. Christodoulides.
Optical solitons in $PT$ periodic potentials,
Phys. Rev. Lett. 100: 030402 (2008). 
%%%%%%%%%%%%%%%%%%%%%%%%%%%%%%%%%%%%%%%%%%
\bibitem{yang_review}
V.V. Konotop, J. Yang and D.A. Zezyulin.
Nonlinear waves in $PT$-symmetric systems, 
Rev. Mod. Phys. 88: 035002 (2016)
%%%%%%%%%%%%%%%%%%%%%%%%%%%%%%%%%%%%%%%%%%%
\bibitem{ApplicNlclNLS16} 
T. A. Gadzhimuradov and A. M. Agalarov. 
Towards a gauge-equivalent magnetic structure of the
nonlocal nonlinear Schr\"odinger equation, 
Phys Rev A 93: 062124 (2016).
%%%%%%%%%%%%%%%%%%%%%%%%%%%%%%%%%%%%%%%%%%%%
\bibitem{AblowitzMusslimani2} 
M. J. Ablowitz and Z. H. Musslimani. 
Integrable discrete $PT$ symmetric model,
Phys. Rev. E 90: 032912 (2014). 
%%%%%%%%%%%%%%%%%%%%%%%%%%%%%%%%%%%%%%%%%%%
\bibitem{Abl-ladik}
M.J.  Ablowitz  and J.F.  Ladik.
Nonlinear  differential-difference  equations,
J. Math. Phys.,  16: 598-603 (1975).
%%%%%%%%%%%%%%%%%%%%%%%%%%%%%%%%%%%%%%%%%%%
\bibitem{AblowitzMusslimaniNonlinearity} 
M. J. Ablowitz and Z. H. Musslimani.  
Inverse scattering transform for the integrable nonlocal nonlinear Schr\"odinger equation, 
Nonlinearity 29: 915-946 (2016). 
%%%%%%%%%%%%%%%%%%%%%%%%%%%%%%%%%%%%%%%%%
\bibitem{yang-partialPT}
J. Yang.
Partially $PT$ symmetric optical potentials with all-real spectra and
soliton families in multidimensions,
Optics Letters 39: 113-1136 (2014).
%%%%%%%%%%%%%%%%%%%%%%%%%%%%%%%%%%%%%%%%
\bibitem{Fokas2016} 
A. S. Fokas.
Integrable multidimensional versions of the nonlocal nonlinear Schr\"odinger equation,
Nonlinearity 29: 319-324 (2016).
%%%%%%%%%%%%%%%%%%%%%%%%%%%%%%%%%%%%%%%%%
\bibitem{He} 
J. S. He and D. Q. Qiu. 
Mirror symmetrical nonlocality of a parity-time symmetry system, 
Private Communication (2016).
%%%%%%%%%%%%%%%%%%%%%%%%%%%%%%%%%%%%%%%%%%
\bibitem{He2} 
Y. Zhang, D, Qiu, Y. Cheng, J. He.
Rational solution of the nonlocal nonlinear Schr\"odinger equation and its application, Private Communication (2016).
%%%%%%%%%%%%%%%%%%%%%%%%%%%%%%%%%%%%%%%%%%
\bibitem{Lou1}
S. Y. Lou.
Alice-Bob systems, $P_s-T_d-C$ principles and multi-soliton solutions,
https://arxiv.org/abs/1603.03975 (2016).
%%%%%%%%%%%%%%%%%%%%%%%%%%%%%%%%%%%%%%%%%%%
\bibitem{Lou2}
S. Y. Lou.
Alice-Bob Physics: Coherent Solutions of Nonlocal KdV Systems,
https://arxiv.org/abs/1606.03154 (2016).
%%%%%%%%%%%%%%%%%%%%%%%%%%%%%%%%%%%%%%%%%%%
\bibitem{Ablowitz3} 
M. J. Ablowitz, B. Prinari and A. D. Trubatch, 
Discrete and Continuous Nonlinear Schr\"odinger Systems, Cambridge University Press, Cambridge, (2004).
%%%%%%%%%%%%%%%%%%%%%%%%%%%%%%%%%%%%%%%%%%
\bibitem{KaupNewell} D. J. Kaup and A. C. Newell.  
An exact solution for a derivative nonlinear Schr\"odinger equation,
J. Math. Phys. 19: 798-801 (1978). 
%%%%%%%%%%%%%%%%%%%%%%%%%%%%%%%%%%%%%%%%%%%
%%%%%%%%%%%%%%%%%%%%%%%%%%%%%%%%%%%%%%%%%%%
\bibitem{KJ84} 
K. Konno and A. Jeffrey.
The loop soliton, 
Advances in Nonlinear Waves, Ed. L. Debnath, Research Notes Math. 95: 162-183, Pitman, London
%%%%%%%%%%%%%%%%%%%%%%%%%%%%%%%%%%%%%%%%%%%
%%%%%%%%%%%%%%%%%%%%%%%%%%%%%%%%%%%%%%%%%
\bibitem{KSI74} 
K. Konno, H. Sanuki, and Y.H. Ichikawa.
Conservation laws of nonlinear-evolution equations, 
Prog. Theor., Phys. 52: 886-889 (1974).
%%%%%%%%%%%%%%%%%%%%%%%%%%%%%%%%%%%%%%%%%%
\bibitem{WSK75} 
M. Wadati, H. Sanuki, and K. Konno.
Relationships among inverse method, B\"acklund transformation 
and an infinite number of conservation laws,
Prog. Theor. Phys. 53: 419-436 (1975).
%%%%%%%%%%%%%%%%%%%%%%%%%%%%%%%%%%%%%%%%%
\bibitem{AblHa75} 
M. J. Ablowitz and R. Haberman. 
Nonlinear evolution equations -- two and three dimensions, 
Phys. Rev. Lett., 35: 1185-1188 (1975).
%%%%%%%%%%%%%%%%%%%%%%%%%%%%%%%%%%%%%%%%%%
\bibitem{AblHbmn} 
M. J. Ablowitz and R. Haberman.
Resonantly coupled nonlinear evolution equations, 
J. Math. Phys. 16: 2301-2305 (1975).
%%%%%%%%%%%%%%%%%%%%%%%%%%%%%%%%%%%%%%%%%%
%%%%%%%%%%%%%%%%%%%%%%%%%%%%%%%%%%%%%%%%
\bibitem{Hbmn75} 
R Haberman.
An infinite number of conservation laws for coupled nonlinear evolution equations,
J. Math. Phys. 18: 1137-1139 (1977).
%%%%%%%%%%%%%%%%%%%%%%%%%%%%%%%%%%%%%%%%%
\bibitem{AL1}  
M. J. Ablowitz and J. F. Ladik.
Nonlinear  differential-difference  equations,
J. Math. Phys. 16: 598-603 (1975).
%%%%%%%%%%%%%%%%%%%%%%%%%%%%%%%%%%%%%%%%%%%
\bibitem{AL2} M. J. Ablowitz and J. F.  Ladik.
Nonlinear differential?difference  equations and Fourier-analysis
J. Math. Phys. 17: 1011-1018, (1976).
%%%%%%%%%%%%%%%%%%%%%%%%%%%%%%%%%%%%%%%%%%%
%%%%%%%%%%%%%%%%%%%%%%%%%%%%%%%%%%%%%%%%%%%%%
\end{thebibliography}
\end{document}